\documentclass[12pt]{article}
\usepackage[scheme=plain]{ctex}
\usepackage{amscd}
\usepackage{bm}
\usepackage{bbm}
\usepackage{mathrsfs}
\usepackage{amssymb}
\usepackage{graphics}
\usepackage{graphicx}
\usepackage{amsfonts}
\usepackage{amsmath}
\usepackage{array}
\usepackage{multirow}
\usepackage{booktabs}
\usepackage{subfigure}
\usepackage{float} % 引入float宏包

\UseRawInputEncoding
\usepackage[numbers,sort&compress]{natbib}
\begin{document}
\date{}
\title{Optical dispersive shock waves of initial pulses in optical fibers with high-order dispersion and quintic nonlinearity effects}
\author{Hua-Ying Ren, Rui Guo$\thanks{Corresponding author:
gr81@sina.com}$\ , Jian-Wen Zhang
\\
\\{\em
School of Mathematics, Taiyuan University  of} \\
{\em Technology, Taiyuan 030024, China}
} \maketitle
\begin{abstract}
This paper probes the dispersive shock waves (DSWs) theory in nonlinear optical systems through Whitham modulation theory for the high-order Chen-Lee-Liu (HOCLL) equation. We systematically derived the one-phase periodic solutions and the corresponding Whitham equations. For all feasible initial discontinuous conditions, we delineate a classification of wave structures during the evolutionary process by virtue of the initial distribution of Riemann invariants, including both convex and non-convex cases. In this classification, we find that the evolved wave structures in non-convex cases incorporate an additional and more complex region-contact DSW region, compared with those in convex cases. So we analyze the propagation behavior of this region. Additionally, we also take into account the problem that the initial wave at the instant of wave breaking approximated by the cubic root function of $x^\frac{1}{3}$, and elaborately analyze the propagation of photons in optical fibers under the effects of high-order dispersion and quintic nonlinearity.

\vspace{5mm}\noindent\emph{Keywords}: dispersive shocks waves; higher-order Chen-Lee-Liu equation; Whitham modulation theory; initial discontinuous conditions; cubic root function;

\end{abstract}

\maketitle

\vspace{7mm}\noindent\textbf{1 Introduction}
\hspace*{\parindent}
\renewcommand{\theequation}{1.\arabic{equation}}\setcounter{equation}{0}\\

Dispersive shock waves (DSWs)~\cite{ck1,ck2,ck3} are a representative class of phenomena in nonlinear wave theory, widely observed across various fields such as classical mechanics, quantum physics, and nonlinear optics~\cite{ck4,ck5,ck6,ck7,ck8}. It is generated by the interaction between dispersive effects and nonlinear effects in nonlinear wave systems. As we all know, if the dissipative and dispersive effects of the medium are neglected, nonlinear effects will give rise to the gradual steepening of the initial pulse due to discrepancies in its propagation velocity. Eventually, a breaking singularity-with a slope approaching infinity-emerges at a certain point and the solution to the wave equation exhibits multi-valuedness. But such a solution which disregards dispersive effects, carries no physical significance.\ Naturally, introducing dispersive effects suppresses this nonphysical behavior and the originally multi-valued region is replaced by a domain of rapid nonlinear oscillations. At the front edge, a series of solitons will appear as a result of the balance between dispersive effects and nonlinear effects, while the amplitude of oscillations at the trailing edge will gradually weaken. This domain of rapid oscillations is referred to as a dispersive shock wave.

Regarding research on the theory of DSW, Gurevich and Pitaevskii made foundational contributions in 1974. Building upon the nonlinear wave modulation theory proposed by Whitham~\cite{ck9,ck10}-where the ideology of averaging was employed to derive Whitham equations governing the slow evolution of parameters that regulate the oscillation of nonlinear modulated waves-they developed this theory to describe the structures of DSWs~\cite{ck11,ck12,ck13}. Subsequently, the theoretical framework has been instrumental in addressing the evolution of a myriad of wave dynamics, with the most canonical and extensively studied being the Riemann problem, which concerns the evolution of an initial discontinuity.

Within the ambit of Whitham's theory, the Riemann problem is also designated as the Gurevich-Pitaevskii (GP) problem where the initial conditions are assumed to be piecewise constant functions, with a discontinuity manifesting at the origin. The simplest instantiation is the Korteweg-de Vries (KdV) equation triggered by a monotonic step at the origin, which ultimately evolves into either a rarefaction wave (RW) or a harmonically DSW~\cite{ck12}. In Reference~\cite{ck14}, the evolution of initial discontinuity of the Nonlinear Schrödinger (NLS) equation was investigated. It was demonstrated that the nonlinear evolution of the NLS equation with initial discontinuity yields combination of several fundamental wave structures, and six potential classification scenarios were deliberated. With the advancement of the finite-band integration method~\cite{ck15}, the GP problems corresponding to a broader range of integrable equations have been successfully resolved~\cite{ck16,ck17,ck18,ck19}.

For classical convex dispersive hydrodynamic equations of the KdV or NLS type, their corresponding Whitham modulation equations exhibit strict hyperbolicity and genuine nonlinearity~\cite{ck20,ck21}. However, for non-convex equations, like the modified KdV (mKdV) equation~\cite{ck22,ck23}, Gardner equation~\cite{ck24,ck25} and derivative NLS equation~\cite{ck26}, the Whitham modulation equations fail to satisfy these conditions.\ Alternatively, they only fulfill the criterion of strict hyperbolicity and genuine nonlinearity within specific parameter ranges.\ Consequently, beyond the classical wave structures, additional structures emerge, including kinks, contact DSWs and combined DSWs.

In the realm of nonlinear optics, the nonlinear Schrödinger (NLS) equation is commonly utilized to characterize the dynamics of pulses propagating in optical fibers. It is a lowest-order approximate model encompassing quadratic normal dispersion and Kerr nonlinearity~\cite{ck27,ck28,ck29,ck30,ck31}. While besides these, a host of other factors in practical problems exert a substantial impact on pulse propagation. The derivative nonlinear Schrödinger (DNLS) equation~\cite{ck32} that incorporates self-steepening effect is widely applied to describe Alfvén waves in plasmas as well as the propagation of ultrashort pulses in single-mode optical fibers~\cite{ck33,ck34}. There are three types in total: the Kaup-Newell (KN) equation~\cite{ck35,ck36}, the Chen-Lee-Liu (CLL) equation~\cite{ck37,ck38}, and the Gerdjikov-Ivanov (GI) equation~\cite{ck39,ck40,ck41}. These are commonly referred to collectively as the DNLSI, DNLSII, DNLSIII equations. Notably, the three types of equations can be transformed into one another through gauge transformations. With the advancement of high-speed optical fiber technology, high-order nonlinear terms have become increasingly significant and cannot be neglected when investigating pulse propagation in nonlinear optical fiber. In this paper, we consider the propagation of nonlinear optical pulses in optical fibers, which is described by the high-order CLL (HOCLL) equation 
\begin{equation}
u_t + u_{xxx} + \frac{3}{2}i|u|^2 u_{xx} - \frac{3}{4}|u|^4 u_x + \frac{3}{2}i u_x^2 u^* = 0,
\end{equation}
that includes the three-order dispersion and quintic nonlinearity terms. Owing to the exact integrability of this equation, we are able to utilize Whitham's modulation theory to derive the DSW structure from the slow modulation of nonlinear waves, and predict the wave evolution under initial discontinuous conditions.

Beyond the highly specific scenario where an initial discontinuity gives rise to the formation of a DSW, we also consider another case. In Reference~\cite{ck42}, the wave breaking during the propagation in a Bose-Einstein condensate (BEC) to a stationary medium was investigated, with particular focus on the situation where the initial profile can be approximated by a power function of the form $-x^\frac{1}{n}$ at the instant of breaking~\cite{ck43,ck44,ck45}. The present study considers the case when $n=3$, i.e., when the pulse is approximated by a cubic root function at the moment of breaking and derives the motion laws of the two boundaries of the DSW.

The paper is organized as follows: Single-phase periodic solutions and the Whitham equations are discussed in Sec. II. The dispersionless limit of the equation is derived in Sec. III. In Sec. IV we describe the elementary wave structures in the general wave patterns. The full classification of the solutions of the Riemann problem is presented in Sec. V. The evolution of DSWs under a condition of cubic initial values formed at the wave breaking moment is considered and the laws of motion of DSWs at the small-amplitude and soliton edge is analyzed in Sec. VI. The Sec. VII is devoted to conclusions.\\

\vspace{7mm}\noindent\textbf{2 Single-phase periodic solution and the Whitham equations}
\hspace*{\parindent}
\renewcommand{\theequation}{2.\arabic{equation}}\setcounter{equation}{0}\\

This section elaborates on the derivation of the single-phase periodic solutions to Eq. (1.1) and the corresponding Whitham equations. The Lax pair of Eq.(1.1) is given from [46] as follows:
$$
\Psi_x = \begin{pmatrix} F & G \\ H & -F \end{pmatrix} \Psi, \quad \Psi_t = \begin{pmatrix} A & B \\ C & -A \end{pmatrix} \Psi,
$$
where,
$$
F=i\lambda ^2+\frac{1}{4}i|u|^2,G=i\lambda u,H=-i\lambda u^*,
$$
$$
A=A_1\lambda ^6+A_2\lambda ^4+A_3\lambda ^2+A_4,B=B_1\lambda ^5+B_2\lambda ^3+B_3\lambda ,C=C_1\lambda ^5+C_2\lambda ^3-C_3\lambda ,
$$
$$
A_1=4i,A_2=2i|u|^2,A_3=-uu_{x}^{*}+u_xu^*+\frac{1}{2}i|u|^4,
$$
$$
A_4=\frac{1}{16}i|u|^6+\frac{1}{4}i|u_x|^2+\frac{3}{8}|u|^2\left( u_xu^*-uu_{x}^{*} \right) -\frac{1}{4}i\left( uu_{xx}^{*}+u_{xx}u^* \right) ,
$$
$$
B_1=4iu,B_2=2u_x+i|u|^2u^*,B_3=\frac{1}{4}iu|u|^2+\frac{3}{2}|u|^2u_x-\frac{1}{2}u^2u_{x}^{*}-iu_{xx},
$$
$$
C_1=-4iu,C_2=2u_{x}^{*}-i|u|^2u^*,C_3=\frac{1}{4}i|u|^2u^*-\frac{3}{2}|u|^2u_{x}^{*}+\frac{1}{2}u_x\left( u^* \right) ^2-iu_{xx}^{*}.
$$
Assuming that $(\phi_1, \phi_2)^T$and $(\psi_1, \psi_2)^T$ represent two linearly independent fundamental solutions of the Lax pair (the superscript $T$ stands for matrix transpose). We construct the squared wave functions as follows:
\begin{equation}
f = -\frac{i}{2}(\phi_1 \psi_2 + \phi_2 \psi_1), \quad g = \phi_1 \psi_1, \quad h = -\phi_2 \psi_2,
\end{equation}
thus, $f$, $g$ and $h$ satisfy the following linear equations
\begin{equation}
\begin{aligned}
f_x &= -iHg + iGh,\ g_x = 2iGf + 2Fg,\ h_x = -2iHf - 2Fh, \\
f_t &= -iCg + iBh,\ g_t = 2iBf + 2Ag,\ h_t = -2iCf - 2Ah.
\end{aligned}
\end{equation}

It can be proved that the quantity $P(\lambda) = f^2 - gh = (-1/4) (\phi_1 \psi_2 - \phi_2 \psi_1)$ is independent of $x$ and $t$.
Generally, $f$, $g$ and $h$ are likewise supposed to be polynomials in $\lambda$ and the single-phase solution corresponds to the polynomials having the following forms:
\begin{equation}\label{}
f = f_2\lambda^4 - f_1\lambda^2 + f_0,\ g = -i\lambda u(\lambda^2 - \mu),\ h = i\lambda u^*(\lambda^2 - \mu^*),
\end{equation}
where the asterisk * denotes the complex conjugate. Substituting Eqs. (2.3) into Eqs. (2.2) and comparing the coefficients of the powers of $\lambda$ yields
\begin{equation}\label{}
\begin{aligned}
&f_2=1,\ f_{0x}=0,\ f_{1x}=i\rho \left( \mu -\mu ^* \right),\\
&f_{1t}=\rho \left( i\left( 4f_0-4f_{1}^{2} \right) \left( \mu -\mu ^* \right) \right) -\rho ^2\left( i\left( \mu -\mu ^* \right) \left( 6f_1-2\mu -2\mu ^* \right) \right) -\rho ^3\left( \frac{3i}{2}\left( \mu -\mu ^* \right) \right),\\
&u_x=\frac{i}{2}u\left( 4f_1-4\mu +\rho \right) ,\ \mu _x=2i\left( f_0-\mu \left( f_1-\mu \right) \right) ,\ \mu _{x}^{*}=-2i\left( f_0-\mu ^*\left( f_1-\mu ^* \right) \right),\quad \\
&u_t=iu\left( \frac{5}{4}\rho ^3+\left( 9f_1-6\mu -3\mu ^* \right) \rho ^2+\left( -8f_0+18f_{1}^{2}+6\mu \mu ^*+4\mu ^2-4f_1\left( 5\mu +2\mu ^* \right) \right) \rho \right) \\
&+iu\left( -16f_0f_1+8f_{1}^{3}+8f_0\mu -8f_{1}^{2}\mu ^* \right).
\end{aligned}
\end{equation}
The polynomial $P(\lambda)$ can be written as
\begin{equation}\label{}
\left( \lambda ^4+f_1\lambda ^2+f_0 \right) ^2-\lambda ^2\left| u \right|^2\left( \lambda ^2-\mu \right) \left( \lambda ^2-\mu ^* \right) =\lambda ^8-s_1\lambda ^6+s_2\lambda ^4-s_3\lambda ^2+s_4=\prod_{i=1}^4{\left( \lambda ^2-\lambda _{i}^{2} \right)},
\end{equation}
where,
\begin{equation}\label{}
s_1=\sum_i{\lambda}_{i}^{2},\ s_2=\sum_{i<j}{\lambda}_{i}^{2}\lambda _{j}^{2},\ s_3=\sum_{i<j<k}{\lambda}_{i}^{2}\lambda _{j}^{2}\lambda _{k}^{2},\ s_4=\lambda _{1}^{2}\lambda _{2}^{2}\lambda _{3}^{2}\lambda _{4}^{2}.
\end{equation}
According to Eq. (2.5), comparing the coefficients of $\lambda^i$ on both sides gives
\begin{equation}\label{}
s_1=2f_1+\rho ,\,\,s_2=2f_0+f_{1}^{2}+\rho \left( \mu +\mu ^* \right) ,\,\,s_3=2f_0f_1+\rho \mu \mu ^*,\,\,s_4=f_{0}^{2}.
\end{equation}
From Eq. (2.7), we obtain 
\begin{equation}\label{}
\mu=\frac{1}{8\rho}\bigl( s_{1}^{2}-4s_2\pm 8\sqrt{s_4}-2s_1\rho +\rho ^2-i\sqrt{-R\left( \rho \right)} \bigr) ,
\end{equation}
where,
\begin{equation}\label{}
\begin{aligned}
R\left( \rho \right) =&\rho ^4+4s_1\rho ^3+\left( 6s_{1}^{2}-8s_2\pm 48\sqrt{s_4} \right) \rho ^2+\left( 4s_{1}^{3}-16s_1s_2+64s_3\pm 32s_1\sqrt{s_4} \right) \rho \\
&+\left( -s_{1}^{2}+4s_2\pm 8\sqrt{s_4} \right) ^2.
\end{aligned}
\end{equation}
The roots of the polynomial $R(\rho)$ are related to the roots of $P(\lambda)$: the upper signs ($+$) in Eq. (2.9) corresponds to the relationships
\begin{equation}\label{}
\begin{aligned}
\rho _1=\left( \lambda _1+\lambda _2+\lambda _3-\lambda _4 \right) ^2,\ \rho _2=\left( \lambda _1+\lambda _2-\lambda _3+\lambda _4 \right) ^2,\\
\rho _3=\left( \lambda _1-\lambda _2+\lambda _3+\lambda _4 \right) ^2,\ \rho _4=\left( -\lambda _1+\lambda _2+\lambda _3+\lambda _4 \right) ^2,
\end{aligned}
\end{equation}
and the lower signs ($-$) correspond to the relationships
\begin{equation}\label{}
\begin{aligned}
\rho _1=\left( -\lambda _1+\lambda _2+\lambda _3-\lambda _4 \right) ^2,\ \rho _2=\left( \lambda _1-\lambda _2+\lambda _3-\lambda _4 \right) ^2,\\
\rho _3=\left( \lambda _1+\lambda _2-\lambda _3-\lambda _4 \right) ^2,\ \rho _4=\left( \lambda _1+\lambda _2+\lambda _3+\lambda _4 \right) ^2.
\end{aligned}
\end{equation}

From Eqs. (2.7), the following equations can be obtained 
\begin{equation}\label{}
\begin{aligned}
\rho _x=-2i\rho \left( \mu -\mu ^* \right) =\frac{1}{2}\sqrt{-R\left( \rho \right)},\ \rho _t=\rho _x\left( \frac{3}{2}s_{1}^{2}-2s_2 \right) ,
\end{aligned}
\end{equation}
which indicates $\rho$ depends on the phase $\xi = x-Vt$, we obtain
\begin{equation}\label{}
\begin{aligned}
\frac{d\rho}{d\xi}=\frac{1}{2}\sqrt{-R\left( \rho \right)},\ V=-\frac{3s_{1}^{2}}{2}+2s_2.
\end{aligned}
\end{equation}
Since $-R\left( \rho \right)>0$ for $\rho\in [\rho_1, \rho_2]\cup[\rho_3, \rho_4]$, the $\rho$ oscillates within the intervals.

Firstly, for $\rho_1<\rho<\rho_2$, after a calculation yields the one-phase periodic solution
\begin{equation}\label{}
\begin{aligned}
\rho =\frac{\rho _2\left( \rho _4-\rho _1 \right) -\rho _4\left( \rho _2-\rho _1 \right) \operatorname{cn}^2\left( \omega ,m \right)}{\rho _4-\rho _2+\left( \rho _2-\rho _1 \right) \operatorname{sn}^2\left( \omega ,m \right)},
\end{aligned}
\end{equation}
where,
\begin{equation}\label{}
\begin{aligned}
L=\frac{8K\left( m \right)}{\sqrt{\left( \rho _3-\rho _1 \right) \left( \rho _4-\rho _2 \right)}}=\frac{2K\left( m \right)}{\sqrt{\left( \lambda_{3}-\lambda_{1} \right) \left( \lambda_{4}-\lambda_{2} \right)}},\\
\omega =\frac{1}{4}\sqrt{\left( \rho _3-\rho _1 \right) \left( \rho _4-\rho _2 \right)}\xi ,\ m=\frac{\left( \rho _4-\rho _3 \right) \left( \rho _2-\rho _1 \right)}{\left( \rho _4-\rho _2 \right) \left( \rho _3-\rho _1 \right)},
\end{aligned}
\end{equation}
where the functions $cn$ and $sn$ are Jacobi elliptic functions, and $K(m)$ is the complete elliptic integral of the first kind.
When $\rho _3 \rightarrow \rho _2$ $(i.e., m \rightarrow 1)$, the solution (2.14) becomes a bright soliton
\begin{equation}\label{}
\begin{aligned}
\rho =\frac{\rho _1\left( \rho _4-\rho _2 \right) +\rho _4\left( \rho _2-\rho _1 \right) \operatorname{sech}^2\left( \omega \right)}{\rho _4-\rho _2+\left( \rho _2-\rho _1 \right) \operatorname{tanh}^2\left( \omega \right)}.
\end{aligned}
\end{equation}
When $m \rightarrow 0$, we can consider the degeneration of the one-phase periodic solution in two ways:
If $\rho _2 \rightarrow \rho _1$, the solution reduces into a constant $\rho=\rho_1$.
If $\rho _3 \rightarrow \rho _4$, we arrive at the nonlinear trigonometric solution
\begin{equation}\label{}
\begin{aligned}
\rho =\frac{\rho _1\left( \rho _4-\rho _2 \right) +\rho _4\left( \rho _2-\rho _1 \right) \cos ^2\left( \omega \right)}{\rho _4-\rho _2+\left( \rho _2-\rho _1 \right) \sin ^2\left( \omega \right)}.
\end{aligned}
\end{equation}

Secondly, for $\rho_3<\rho<\rho_4$, the one-phase periodic solution has the following form
\begin{equation}\label{}
\begin{aligned}
\rho = \frac{\rho_3(\rho_1 - \rho_4) + \rho_1(\rho_4 - \rho_3)\operatorname{cn}^2(\omega, m)}{\rho_1 - \rho_3 + (\rho_3 - \rho_4)\operatorname{sn}^2(\omega, m)}.
\end{aligned}
\end{equation}
In the soliton limit $\rho _2 \rightarrow \rho _3$ $(i.e., m \rightarrow 1)$ we get
\begin{equation}\label{}
\begin{aligned}
\rho =\frac{\rho _3\left( \rho _1-\rho _4 \right) -\rho _1\left( \rho _3-\rho _4 \right) \operatorname{sech}^2\left( \omega \right)}{\rho _1-\rho _4+\left( \rho _4-\rho _3 \right) \operatorname{tanh}^2\left( \omega \right)}.
\end{aligned}
\end{equation}
If $\rho _4 \rightarrow \rho _3$, the solution turns into a constant $\rho=\rho_3$.
If $\rho _2 \rightarrow \rho _1$, we obtain the nonlinear trigonometric solution
\begin{equation}\label{}
\begin{aligned}
\rho =\frac{\rho _3\left( \rho _1-\rho _4 \right) -\rho _1\left( \rho _3-\rho _4 \right) \cos ^2\left( \omega \right)}{\rho _1-\rho _4+\left( \rho _4-\rho _3 \right) \sin ^2\left( \omega \right)}.
\end{aligned}
\end{equation}

Utilizing Eqs. (2.2) yields the following equations
\begin{equation}\label{}
\left( \frac{G}{g} \right) _t=\left( \frac{B}{g} \right) _x,\ \left( \frac{H}{h} \right) _t=\left( \frac{C}{h} \right) _x.
\end{equation}
To average Eqs.(2.21), normalizing the function $f, g$ and $h$ gives
\begin{equation}\label{}
\left( \frac{f}{\sqrt{P\left( \lambda \right)}} \right) ^2-\left( \frac{g}{\sqrt{P\left( \lambda \right)}} \right) \left( \frac{h}{\sqrt{P\left( \lambda \right)}} \right)=1.
\end{equation} 
Substitution of $B$ and $G$ in Lax pair into Eq. (2.21), we obtain
\begin{equation}\label{}
\frac{\partial}{\partial t}\left( \sqrt{P\left( \lambda \right)}\frac{1}{\lambda ^2-\mu _1} \right) -\frac{\partial}{\partial x}\left( \sqrt{P\left( \lambda \right)}\left( 2s_1+4\lambda ^2+\frac{\frac{3}{2}s_{1}^{2}-2s_2}{\lambda ^2-\mu _1} \right) \right) =0.
\end{equation}
Averaging this equation over one wavelength
\begin{equation}\label{}
L=\oint{\frac{d\mu}{-2\sqrt{-P\left( \sqrt{\mu} \right)}}}
\end{equation}
according to the formula
\begin{equation}\label{}
\tilde{\mathcal{F}}=\frac{1}{L}\int_0^L{\mathcal{F}}\,dx,
\end{equation}
we can obtain the averaged conservation law
\begin{equation}\label{}
\begin{aligned} 
&\frac{\partial}{\partial t} \left[ \frac{\sqrt{P\left( \lambda \right)}}{L}\oint{\frac{d\mu}{-2\left( \lambda ^2-\mu \right) \sqrt{-P\left( \sqrt{\mu} \right)}}} \right] -\\
\frac{\partial}{\partial x}&\left[ \sqrt{P\left( \lambda \right)} \left( 2s_1+4\lambda ^2+\frac{\frac{3}{2}s_{1}^{2}-2s_2}{L}\oint{\frac{d\mu}{-2\left( \lambda ^2-\mu \right) \sqrt{-P\left( \sqrt{\mu} \right)}}} \right) \right] =0.
\end{aligned}
\end{equation}
Taking the limit $\lambda \rightarrow \lambda_i\ (i = 1, 2, 3, 4)$ gives
\begin{equation}\label{}
\begin{aligned}
\small & \frac{\partial \lambda _{i}^{2}}{\partial t} \oint{\frac{d\mu}{-2\left( \lambda _{i}^{2}-\mu \right) \sqrt{-P\left( \sqrt{\mu} \right)}}}-\\
\frac{\partial \lambda _{i}^{2}}{\partial x}& \left[ \left( 2s_1+4\lambda _{i}^{2} \right) L+ \left( \frac{3}{2}s_{1}^{2}-2s_2 \right) \oint{\frac{d\mu}{-2\left( \lambda _{i}^{2}-\mu \right) \sqrt{-P\left( \sqrt{\mu} \right)}}} \right] =0.
\end{aligned}
\end{equation}
Following simplification, Eqs. (2.27) become the following forms
\begin{equation}\label{}
\frac{\partial \lambda _{i}^{2}}{\partial t}+v_i\frac{\partial \lambda _{i}^{2}}{\partial x}=0,
\end{equation}
where,
\begin{equation}\label{}
v_i=-\left( \frac{\left( s_1+2\lambda _{i}^{2} \right) L}{-\frac{\partial L}{\partial \lambda _{i}^{2}}}+\frac{3}{2}s_{1}^{2}-2s_2 \right) , i=1,2,3,4.
\end{equation}
We define a new set of Riemann invariants $l_i = \lambda_i^2 ~(i = 1, 2, 3, 4)$ and they are ordered by $0\le l_1\le l_2\le l_3\le l_4\,\,\text{for\,\,} 0\le\lambda _{1}^{2}\le \lambda _{2}^{2}\le \lambda _{3}^{2}\le \lambda _{4}^{2}.$
Therefore, the parameters $\rho _i$ in periodic solutions are expressed in terms of $l_i$
\begin{equation}\label{}
\begin{aligned}
	\rho _1&=\left( \sqrt{l_1}+\sqrt{l_2}+\sqrt{l_3}-\sqrt{l_4} \right) ^2,\ \rho _2=\left( \sqrt{l_1}+\sqrt{l_2}-\sqrt{l_3}+\sqrt{l_4} \right) ^2,\\
	\rho _3&=\left( \sqrt{l_1}-\sqrt{l_2}+\sqrt{l_3}+\sqrt{l_4} \right) ^2,\ \rho _4=\left( -\sqrt{l_1}+\sqrt{l_2}+\sqrt{l_3}+\sqrt{l_4} \right) ^2,\\
\end{aligned}
\end{equation}
and
\begin{equation}\label{}
\begin{aligned}
	\rho _1&=\left( -\sqrt{l_1}+\sqrt{l_2}+\sqrt{l_3}-\sqrt{l_4} \right) ^2,\ \rho _2=\left( \sqrt{l_1}-\sqrt{l_2}+\sqrt{l_3}-\sqrt{l_4} \right) ^2,\\
	\rho _3&=\left( \sqrt{l_1}+\sqrt{l_2}-\sqrt{l_3}-\sqrt{l_4} \right) ^2,\ \rho _4=\left( \sqrt{l_1}+\sqrt{l_2}+\sqrt{l_3}+\sqrt{l_4} \right) ^2.\\
\end{aligned}
\end{equation}
Then the wavelength is given by
$$
L=\frac{2K\left( m \right)}{\sqrt{\left( l_2-l_4 \right) \left( l_1-l_3 \right)}},\quad m=\frac{\left( l_1-l_2 \right) \left( l_3-l_4 \right)}{\left( l_1-l_3 \right) \left( l_2-l_4 \right)},
$$
and the Whitham modulation equations can be expressed by $l_i$ 
\begin{equation}\label{}
\frac{\partial l_i}{\partial t}+v_i\frac{\partial l_i}{\partial x}=0,\quad i=1,2,3,4,
\end{equation}
where the Whitham velocities $v_i$ have the explicit expressions
\begin{equation}\label{}
\begin{aligned}
	v_1&=2\sum_{i<j}^4{l_i}l_j-\frac{3}{2}\left( \sum_{i=1}^4{l_i} \right) ^2-\frac{2\left( l_1-l_2 \right) \left( l_1-l_4 \right) \left( 3l_1+l_2+l_3+l_4 \right) K\left( m \right)}{\left( l_4-l_2 \right) E\left( m \right) +\left( l_1-l_4 \right) K\left( m \right)},\\
	v_2&=2\sum_{i<j}^4{l_i}l_j-\frac{3}{2}\left( \sum_{i=1}^4{l_i} \right) ^2-\frac{2\left( l_1-l_2 \right) \left( l_2-l_3 \right) \left( l_1+3l_2+l_3+l_4 \right) K\left( m \right)}{\left( l_1-l_3 \right) E\left( m \right) +\left( l_3-l_2 \right) K\left( m \right)},\\
	v_3&=2\sum_{i<j}^4{l_i}l_j-\frac{3}{2}\left( \sum_{i=1}^4{l_i} \right) ^2-\frac{2\left( l_3-l_4 \right) \left( l_2-l_3 \right) \left( l_1+l_2+3l_3+l_4 \right) K\left( m \right)}{\left( l_4-l_2 \right) E\left( m \right) +\left( l_2-l_3 \right) K\left( m \right)},\\
	v_4&=2\sum_{i<j}^4{l_i}l_j-\frac{3}{2}\left( \sum_{i=1}^4{l_i} \right) ^2-\frac{2\left( l_1-l_4 \right) \left( l_4-l_3 \right) \left( l_1+l_2+l_3+3l_4 \right) K\left( m \right)}{\left( l_3-l_1 \right) E\left( m \right) +\left( l_1-l_4 \right) K\left( m \right)}.\\
\end{aligned}
\end{equation}
For the case of $m \rightarrow 1\ (l_2 \rightarrow l_3)$, the Whitham velocities reduce into
$$
v_2=v_3=\frac{1}{2}\left( -3l_{1}^{2}-8l_{3}^{2}-4l_3l_4-3l_{4}^{2}-2l_1\left( 2l_3+l_4 \right) \right) ,
$$
$$
v_1=-\frac{3}{2}\left( 5l_{1}^{2}+2l_1l_4+l_{4}^{2} \right) ,\ v_4=-\frac{3}{2}\left( 5l_{4}^{2}+2l_1l_4+l_{1}^{2} \right).
$$
In the limit of $m \rightarrow 0\ (l_2 \rightarrow l_1)$, the Whitham velocities are given by
$$
v_1=v_2=-\frac{48l_{1}^{3}-6l_1\left( l_3-l_4 \right) ^2-24l_{1}^{2}\left( l_3+l_4 \right) -3\left( l_3-l_4 \right) ^2\left( l_3+l_4 \right)}{4l_1-2\left( l_3+l_4 \right)},
$$
$$
v_3=-\frac{3}{2}\left( 5l_{3}^{2}+2l_3l_4+l_{4}^{2} \right) ,\ v_4=-\frac{3}{2}\left( 5l_{4}^{2}+2l_3l_4+l_{3}^{2} \right).
$$
And in the other way where $m \rightarrow 0\ (l_3 \rightarrow l_4)$, we obtain
$$
v_3=v_4=\frac{1}{2}\left( -3l_{1}^{2}-3l_{2}^{2}-4l_2l_4-8l_{4}^{2}-2l_1\left( 2l_4+l_2 \right) \right) +\frac{4\left( l_1-l_4 \right) \left( l_4-l_2 \right) \left( l_1+l_2+4l_4 \right)}{l_1+l_2-2l_4},
$$
$$
v_1=-\frac{3}{2}\left( 5l_{1}^{2}+2l_1l_2+l_{2}^{2} \right) ,\ v_2=-\frac{3}{2}\left( 5l_{2}^{2}+2l_1l_2+l_{1}^{2} \right) .
$$
\\
\vspace{7mm}\noindent\textbf{3 Dispersionless limit}
\hspace*{\parindent}
\renewcommand{\theequation}{3.\arabic{equation}}\setcounter{equation}{0}

Using the Madelung transformation
\begin{equation}
u\left( x,t \right) =\sqrt{\rho \left( x,t \right)}e^{i\varphi \left( x,t \right)},\varphi _x\left( x,t \right) =\nu\left( x,t \right),
\end{equation}
where $\rho\left( x,t \right)$ and $\nu\left( x,t \right)$ are real functions and indicate the density and velocity in the hydrodynamics respectively. we obtain the hydrodynamic representation of Eq. (1.1)
\begin{equation}
\left\{ \begin{array}{l}
	\rho_t - \left( \dfrac{1}{4}\rho^3 + 3\rho^2 \nu + 3\rho \nu^2 \right)_x = \left( \rho_{xx} - \dfrac{3\rho_x^2}{4\rho} \right)_x,\\
	\nu_t - \left( \nu^3 + 3\rho \nu^2 + \dfrac{3}{4}\rho^2 \nu \right)_x = \left( -\dfrac{3\rho_x^2 \nu}{4\rho^2} + \dfrac{3\nu \rho_{xx}}{2\rho} + \dfrac{3\rho_x \nu_x}{2\rho} + \dfrac{3}{4}\rho_{xx} + \nu_{xx} \right)_x.\\
\end{array} \right.
\end{equation}

In the dispersionless limit, we can omit the dispersion terms in Eqs. (3.2) and the system takes the simple hydrodynamic form
\begin{equation}
\left( \begin{array}{c}
	\rho\\
	\nu\\
\end{array} \right) _t-\left( \begin{matrix}
	\frac{3}{4}\rho ^2+6\rho \nu +3\nu ^2&		3\rho ^2+6\rho \nu\\
	3\nu ^2+\frac{3}{2}\rho \nu&		\frac{3}{4}\rho ^2+6\rho \nu +3\nu ^2\\
\end{matrix} \right) \left( \begin{array}{c}
	\rho\\
	\nu\\
\end{array} \right) _x=0,
\end{equation}
which can be easily transformed into diagonal forms
\begin{equation}
\begin{aligned}
\frac{\partial l_+}{\partial t}+v_+\frac{\partial l_+}{\partial x}=0,\ 
\frac{\partial l_-}{\partial t}+v_-\frac{\partial l_-}{\partial x}=0,
\end{aligned}
\end{equation}
by introducing the Riemann invariants
\begin{equation}
l_{\pm}=\frac{1}{4}\left( \sqrt{\rho}\pm \sqrt{2\nu} \right) ^2,
\end{equation}
where the velocities $v_\pm$ in Eqs. (3.4) are expressed in terms of the Riemann invariants as
\begin{equation}
\begin{aligned}
v_+=-\frac{3}{2}\left( 5l_{+}^{2}+2l_+l_-+l_{-}^{2} \right),\ 
v_-=-\frac{3}{2}\left( 5l_{-}^{2}+2l_+l_-+l_{+}^{2} \right).
\end{aligned}
\end{equation}
Therefore, the functions of $\rho$ and $\nu$ are easily obtained
\begin{equation}
\rho =\left( \sqrt{l_+}\pm \sqrt{l_-} \right) ^2,\ \nu =\frac{1}{2}\left( \sqrt{l_+}\mp \sqrt{l_-} \right) ^2,
\end{equation}
where the Riemann invariants are positive satisfying $l_+>l_->0$.

Simple waves are characterized that only one of the Riemann invariants changes while the rest remains constant. If the invariant $l_- = constant$, the second equation in Eqs. (3.4) holds true naturally. Meanwhile, the first equation is converted into the Hopf equation.

One can notice that the Riemann invariants and the characteristic velocities are real for $\rho \ge 0$ and $\ \nu \ge 0$, that is, these conditions define the hyperbolicity domain in the plane $(\nu, \rho)$. Moreover, the Riemann invariant $l_-$ reaches its minimum value along the line $\rho = 2\nu$ which cuts the hyperbolicity domain into two monotonicity regions. It decreases monotonically in the region where $\rho > 2\nu$ and increases monotonically in that where $\rho < 2\nu$. If the solutions to the Eqs. (3.3) cross this curve, the dependence between physical variables and the Riemann invariants ceases to have a one-to-one correspondence, which significantly influences the wave structure arising from a discontinuous initial condition. Then, we are going to discuss the basic wave structures evolving from discontinuous initial conditions.

\vspace{7mm}\noindent\textbf{4 Basic wave structures}
\hspace*{\parindent}
\renewcommand{\theequation}{4.\arabic{equation}}\setcounter{equation}{0}\\

In this section, we explore elementary wave structures of HOCLL equation for the general discontinuous step initial value conditions
\begin{equation}
\begin{aligned}
\rho(x,0)=
\begin{cases} 
\rho^L, & x<0,\\
 \rho^R, & x>0, 
\end{cases}
\quad \text{and} \quad
v(x,0)=
\begin{cases} 
v^L, & x<0, \\
v^R, & x>0. 
\end{cases}
\end{aligned}
\end{equation}
where $\rho^R,\ \rho^L,\ v^R$ and $v^L$ are four arbitrary real constants and the values of Riemann invariants $l_-^R,\ l_-^L,\ l_+^R$ and $l_+^L$ can be matched on both sides of the discontinuity according to Eqs. (3.5).

\vspace{5mm}\noindent\textbf{4.1 Rarefaction waves}\\

In what follows, a detailed investigation will be conducted on the rarefaction wave solutions. Through introducing self-similar variables $z = x/t$, the Whitham Eqs. (3.4) are converted into 
\begin{equation}
\begin{aligned}
\left( v_+-z \right) \frac{\partial l_+}{\partial z}=0,\,\,\left( v_--z \right) \frac{\partial l_-}{\partial z}=0.
\end{aligned}
\end{equation}
Manifestly, these equations have solutions that one of the Riemann invariants is a constant,
i.e., $l_+ = const$ or $l_- = const$ and the other varies in accordance with Eqs. (4.2).
Below, two categories of rarefaction wave solutions are derived:
\begin{equation}
\begin{aligned}
\left( i \right) \ l_-=\text{const}\equiv l_{-}^{0},\ v_+=-\frac{3}{2}\left( 5l_{+}^{2}+2l_+l_-+l_{-}^{2} \right) =z,\\
\left( ii \right) \ l_+=\text{const}\equiv l_{+}^{0},\ v_-=-\frac{3}{2}\left( 5l_{-}^{2}+2l_+l_-+l_{+}^{2} \right) =z.
\end{aligned}
\end{equation}
\begin{figure}[htbp]
\centering
\setcounter{subfigure}{0}
\subfigure[]{\includegraphics[width=0.31\linewidth]{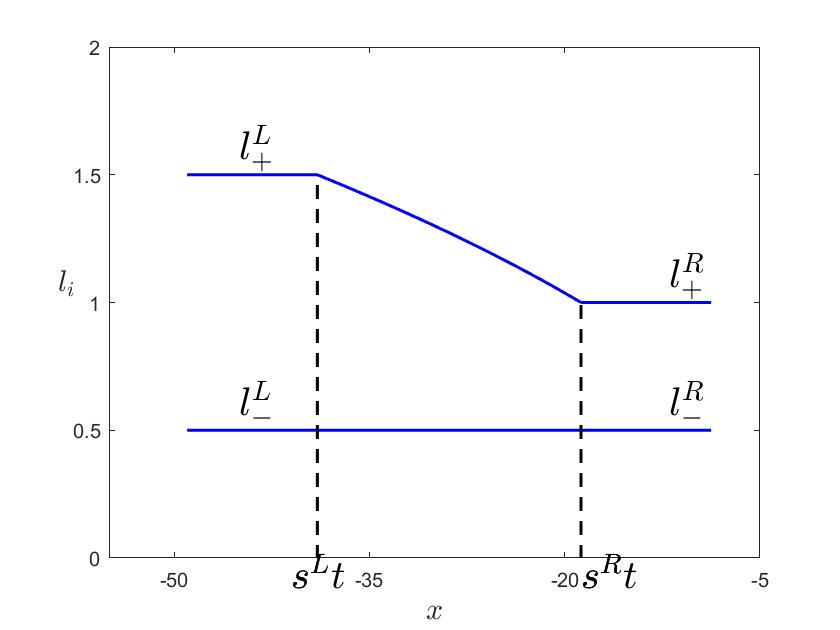}}
\quad
\subfigure[]{\includegraphics[width=0.31\linewidth]{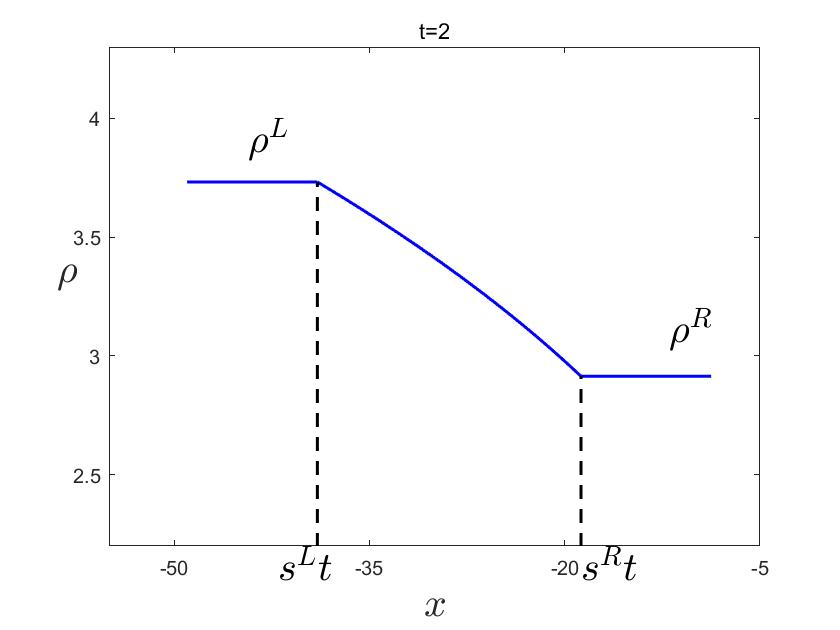}}
\quad
\subfigure[]{\includegraphics[width=0.31\linewidth]{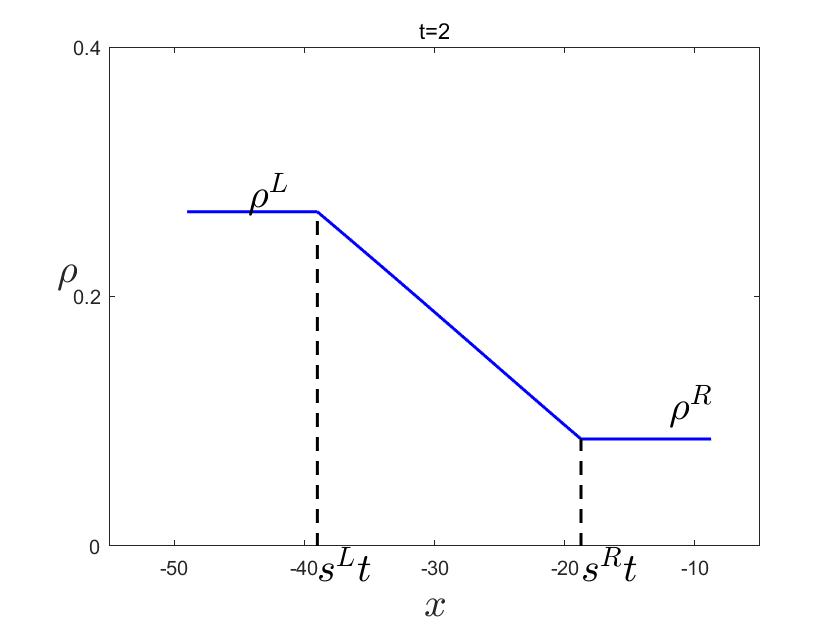}}
\quad
\subfigure[]{\includegraphics[width=0.31\linewidth]{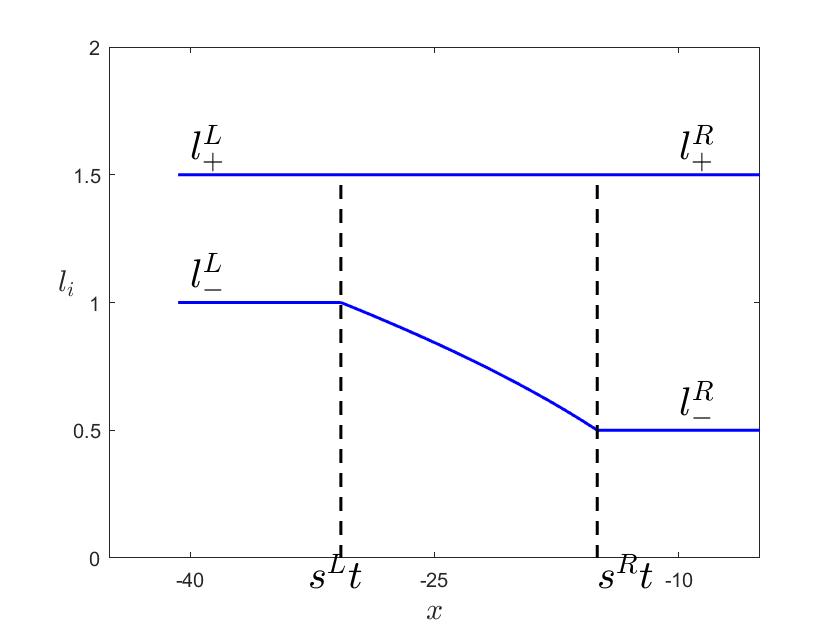}}
\quad
\subfigure[]{\includegraphics[width=0.31\linewidth]{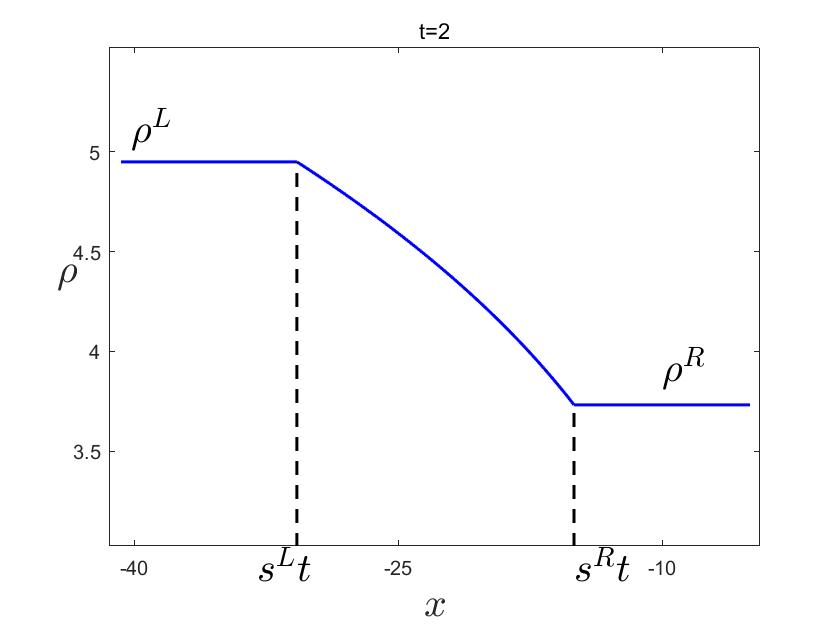}}
\quad
\subfigure[]{\includegraphics[width=0.31\linewidth]{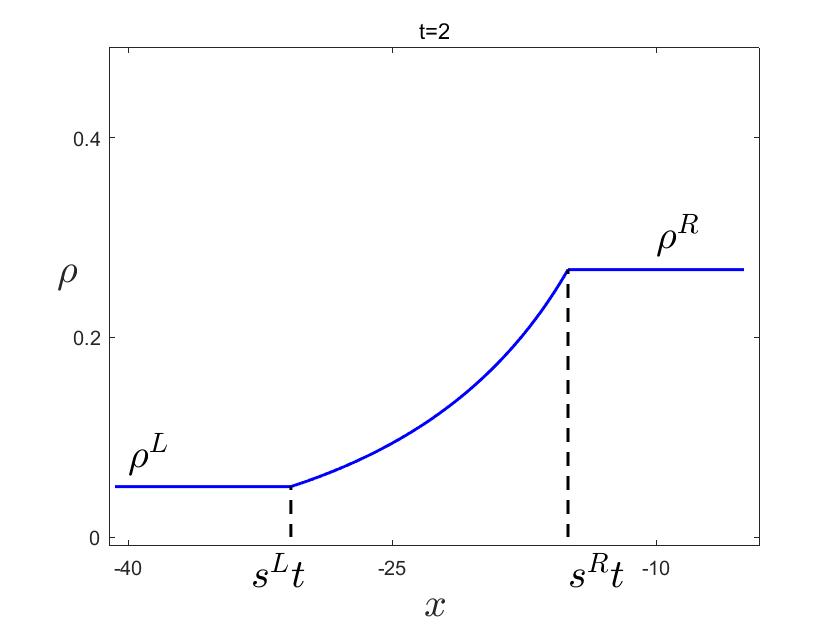}}
\quad
\flushleft{\footnotesize
\textbf{Fig.~$\bm{1}$.} The distributions of the Riemann invariants $l_{\pm}$ for the rare wave solutions of HOCLL equation and the functions of two sets of corresponding densities $\rho$ at $t = 2$.}
\end{figure}

For case $(i)$, the variables $\rho$ and $\nu$ can be represented as
\begin{equation}
\begin{aligned}
\rho =\left( \sqrt{-l_-^0+\sqrt{-4l_{-}^{02}+\frac{2z}{3}}}\pm \sqrt{l_-^0} \right) ^2,\ \nu =\frac{1}{2}\left( \sqrt{-l_-^0+\sqrt{-4l_{-}^{02}+\frac{2z}{3}}}\mp \sqrt{l_-^0} \right) ^2,
\end{aligned}
\end{equation}
and for case $(ii)$ can be obtained by
\begin{equation}
\begin{aligned}
\rho =\left( \sqrt{-l_+^0+\sqrt{-4l_{+}^{02}+\frac{2z}{3}}}\mp \sqrt{l_+^0} \right) ^2,\ \nu =\frac{1}{2}\left( \sqrt{-l_+^0+\sqrt{-4l_{+}^{02}+\frac{2z}{3}}}\pm \sqrt{l_+^0} \right) ^2.
\end{aligned}
\end{equation}

The distributions of Riemann invariants along with the basic structures of rarefaction waves are shown in Fig. 1, in which the edge velocities can be formulated as follows:

\begin{equation}
\begin{aligned}
\left( i \right) s^L=-\frac{3}{2}\left( 5l_{-}^{02}+2l_{+}^{L}l_{-}^{0}+l_{+}^{L2} \right) ,\ s^R=-\frac{3}{2}\left( 5l_{-}^{02}+2l_{+}^{R}l_{-}^{0}+l_{+}^{R2} \right) ,\\
\left( ii \right) s^L=-\frac{3}{2}\left( 5l_{+}^{02}+2l_{+}^{0}l_{-}^{L}+l_{-}^{L2} \right) ,\ s^R=-\frac{3}{2}\left( 5l_{+}^{02}+2l_{+}^{0}l_{-}^{R}+l_{-}^{R2} \right) .
\end{aligned}
\end{equation}

Fig.\ 2 illustrates the expressions of variables $\rho$ and $\nu$, and both their left and right endpoints must lie on the same side of $\rho = 2\nu$. As for the rarefaction wave characterized by $l_- = const$ (depicted by the red line in Fig.\ 2(a)), the states at its two edges can be identified as the intersection points of this red parabola with another two parabolas which respectively represent the curves corresponding to constant values $l_{+}^{R}$ and $l_{+}^{L}$. Then we obtain two pairs of points $P_1(\rho^L) \rightarrow P_2(\rho^R)$ and $Q_1(\rho^L) \rightarrow Q_2(\rho^R)$ corresponding to the two types of rarefaction waves described in Fig.\ 1(b) and Fig.\ 1(c), respectively. In a similar way, the scenarios corresponding to Fig.\ 1(e) and Fig.\ 1(f) which involve a constant Riemann invariant $l_+ = const$ are represented by the parabolas shown in Fig.\ 2(b).

\begin{figure}[htbp]
\centering
\setcounter{subfigure}{0}
\subfigure[]{\includegraphics[width=0.41\linewidth]{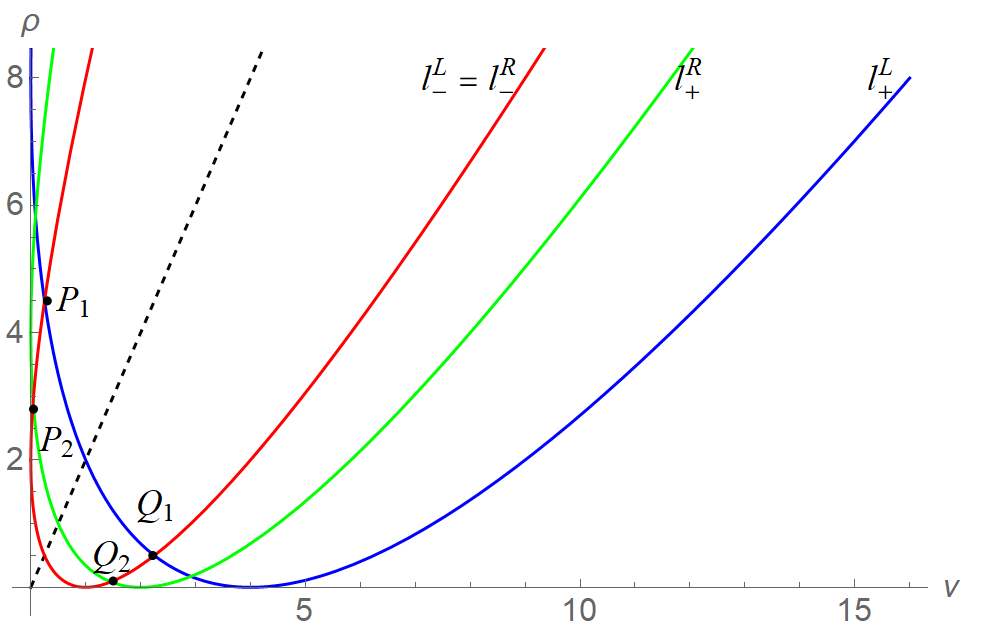}}
\quad
\subfigure[]{\includegraphics[width=0.41\linewidth]{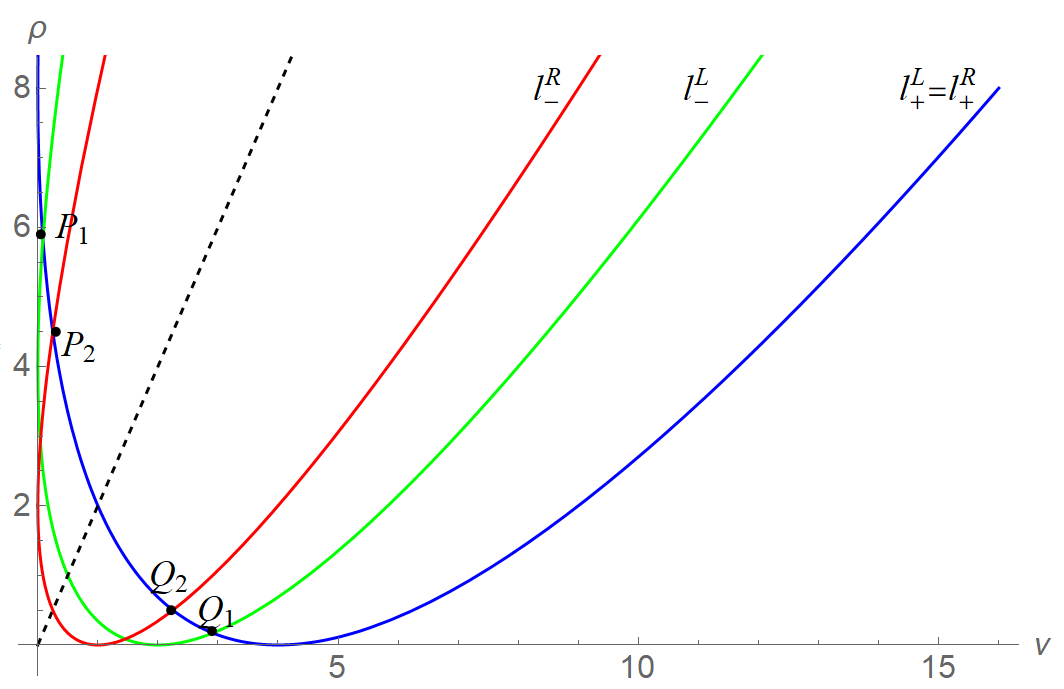}}
\quad
\flushleft{\footnotesize
\textbf{Fig.~$\bm{2}$.} The evolution of density $\rho$ with velocity $\nu$ under constant Riemann invariants, where: (a) corresponds to the case (i) where $l_-$ remains constant; (b) corresponds to the case (ii) where $l_+$ remains constant. The black dashed line in the middle is $\rho=2\nu$. The intersection points P and Q of the parabola represent the initial states of the Riemann invariants.}
\end{figure}

\vspace{5mm}\noindent\textbf{4.2 Cnoidal dispersive shock waves}\\

In this section, we focus on the study of the structures of DSWs which are governed by the Whitham system as presented in Eqs.\ (2.32). Introducing the self-similar variable $z = x/t$, the Whitham equations can be transformed into
\begin{equation}
\begin{aligned}
\left( v_i-z \right) \frac{\partial l_i}{\partial z}=0,\quad i=1,2,3,4,
\end{aligned}
\end{equation}
and the structures of DSWs satisfy the following boundary value conditions:
\begin{equation}
\begin{aligned}
\left( i \right) \ l_1=l_{-}^{L},\quad l_3=l_{-}^{R},\quad l_4=l_{+}^{L},\quad v_2\left( l_{-}^{L},l_2,l_{-}^{R},l_{+}^{L} \right) =z,\\
\left( ii \right) \ l_1=l_{-}^{R},\quad l_2=l_{+}^{L},\quad l_4=l_{+}^{R},\quad v_3\left( l_{-}^{R},l_{+}^{L},l_3,l_{+}^{R} \right) =z.
\end{aligned}
\end{equation}

These two cases of the structures of Riemann invariants are shown in Fig. 3(a) and Fig. 3(d). The motion of their edges (i.e., the leading and trailing edges of the oscillatory region) are given by
\begin{equation}
\begin{aligned}
&\left( i \right) \ s^L=-\frac{1}{2}\left( -3l_{-}^{L2}-8l_{-}^{R2}-4l_{-}^{R}l_{+}^{L}-3l_{+}^{L2}-2l_{-}^{L}\left( 2l_{-}^{R}+l_{+}^{L} \right) \right) ,\\
&s^R=-\frac{48l_{-}^{L3}-6l_{-}^{L}\left( l_{-}^{R}-l_{+}^{L} \right) ^2-24l_{-}^{L2}\left( l_{-}^{R}+l_{+}^{L} \right) -3\left( l_{-}^{R}-l_{+}^{L} \right) ^2\left( l_{-}^{R}+l_{+}^{L} \right)}{4l_{-}^{L}-2\left( l_{-}^{R}+l_{+}^{L} \right)},\\
\small \left( ii \right) \,\,s^L=&\frac{3\left( l_{-}^{R3}+l_{+}^{L3}-l_{-}^{R2}\left( l_{+}^{L}-2l_{+}^{R} \right)+2l_{+}^{L2}l_{+}^{R}+8l_{+}^{L}l_{+}^{R2}-16l_{+}^{R3}-l_{-}^{R}\left( l_{+}^{L2}+4l_{+}^{L}l_{+}^{R}-8l_{+}^{R2} \right) \right)}{2l_{-}^{R}+2l_{+}^{L}-4l_{+}^{R}},\\
&s^R=\frac{1}{2}\left( -3l_{-}^{R2}-8l_{+}^{L2}-4l_{+}^{L}l_{+}^{R}-3l_{+}^{R2}-2l_{-}^{R}\left( 2l_{+}^{L}+l_{+}^{R} \right) \right) .
\end{aligned}
\end{equation}
Substituting the solutions of Riemann invariants into periodic solutions (2.14) and (2.18), we obtain the DSWs on $x$ which are visually represented in Fig.\ 3. There exist two mappings from the Riemann invariants to the physical parameters, which is similar to the situation described above for the rarefaction waves. Therefore, each of the structures of Riemann invariants corresponds to two different cnoidal DSWs.
\begin{figure}[htbp]
\centering
\setcounter{subfigure}{0}
\subfigure[]{\includegraphics[width=0.31\linewidth]{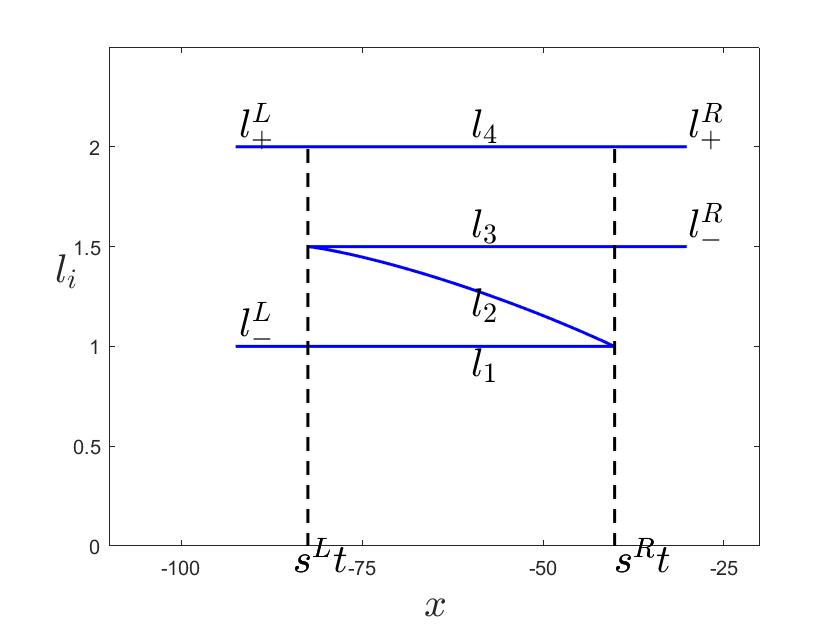}}
\quad
\subfigure[]{\includegraphics[width=0.31\linewidth]{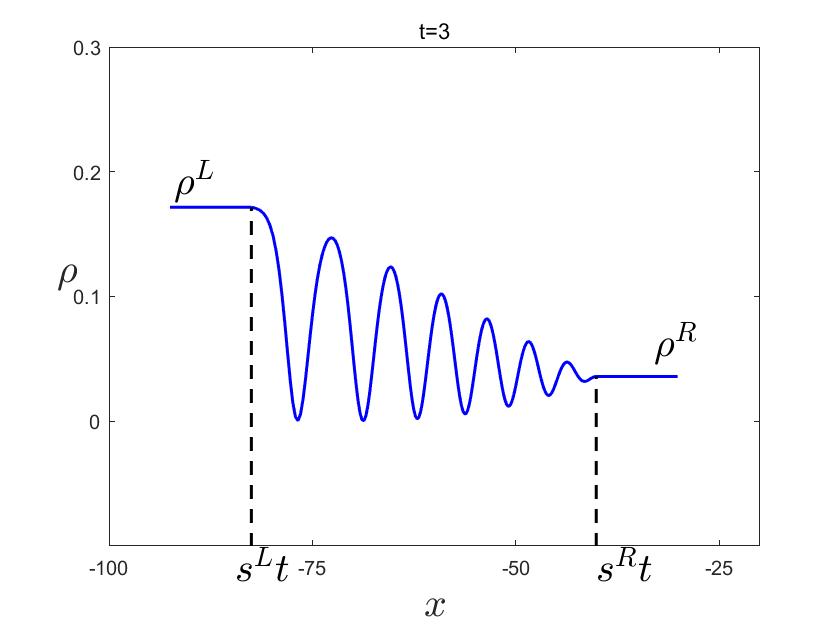}}
\quad
\subfigure[]{\includegraphics[width=0.31\linewidth]{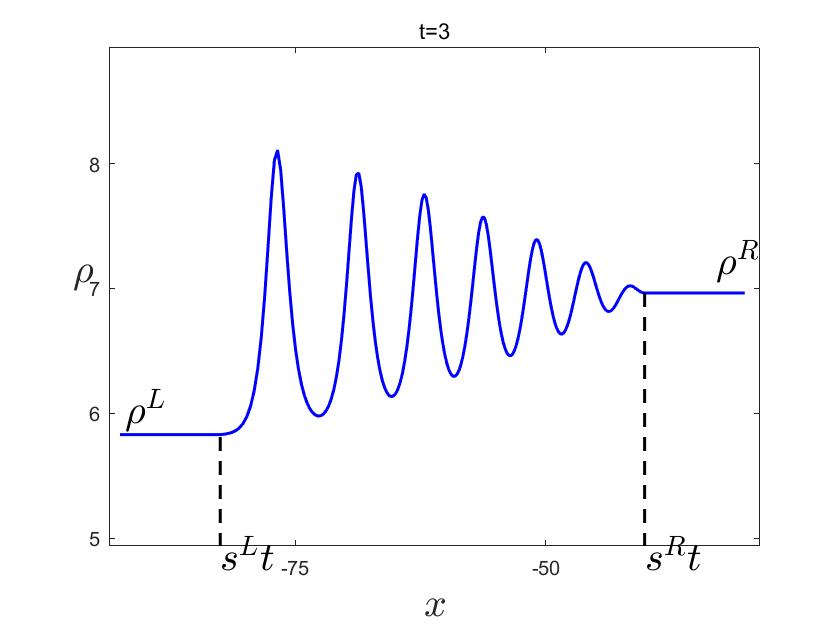}}
\quad
\subfigure[]{\includegraphics[width=0.31\linewidth]{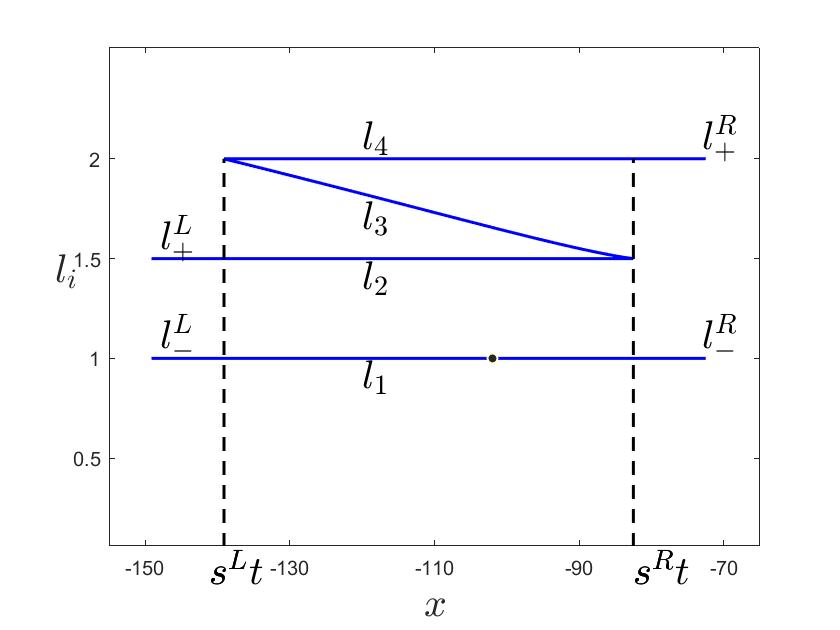}}
\quad
\subfigure[]{\includegraphics[width=0.31\linewidth]{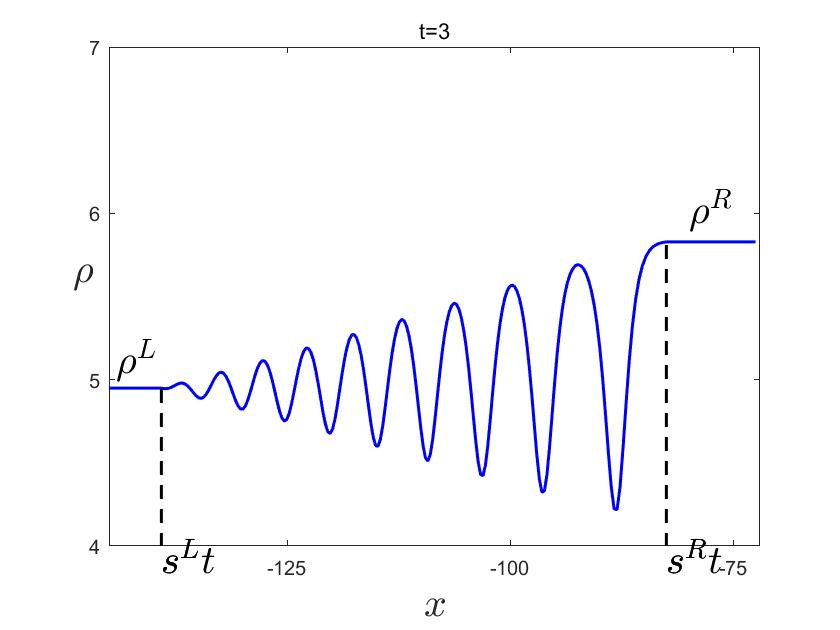}}
\quad
\subfigure[]{\includegraphics[width=0.31\linewidth]{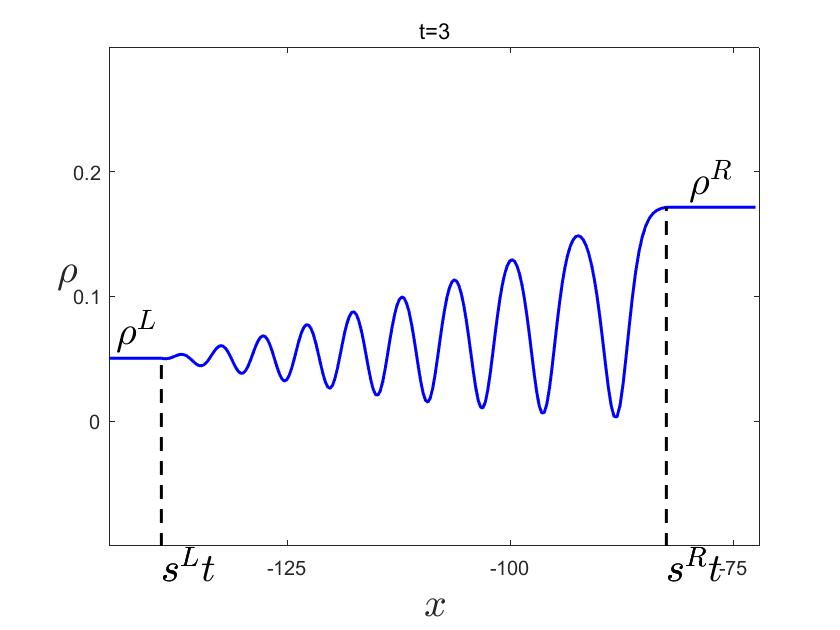}}
\quad
\flushleft{\footnotesize
\textbf{Fig.~$\bm{3}$.} The functions of the Riemann invariants $l_{\pm}$ for the cnoidal DSW solutions of HOCLL equation and the diagrams of two sets of corresponding densities $\rho$ at $t = 3$.}
\end{figure}

It is clear that the cnoidal DSWs also require that the dispersionless Riemann invariants belong to the regions with the same monotonicity. These can be described by the same diagrams of Fig.\ 2 as the rarefaction waves, but with inverted 'left' and 'right' states, i.e., $P_2(\rho^L) \rightarrow P_1(\rho^R)$ and $Q_2(\rho^L) \rightarrow Q_1(\rho^R)$. 

We consider a special case regarding the corresponding relationship of Eq. (2.31). The minimum value of $\rho$ can be known through the periodic solution (2.14), which is $\rho_1$. If the Riemann invariants of the initial condition (4.1) satisfy $l_{+}^{R}=l_{+}^{L}>l_{-}^{R}>l_{-}^{L}$ and $l_{+}^{R}-l_{-}^{R} \ge l_{-}^{R}-l_{-}^{L}$, there will exist a point with the minimum value of $\rho$ equal to 0 within the wave. We shall call this point a vacuum point and the phase velocity $V$ at this point is zero.
\begin{figure}[htbp]
\centering
\setcounter{subfigure}{0}
\subfigure[]{\includegraphics[width=0.45\linewidth]{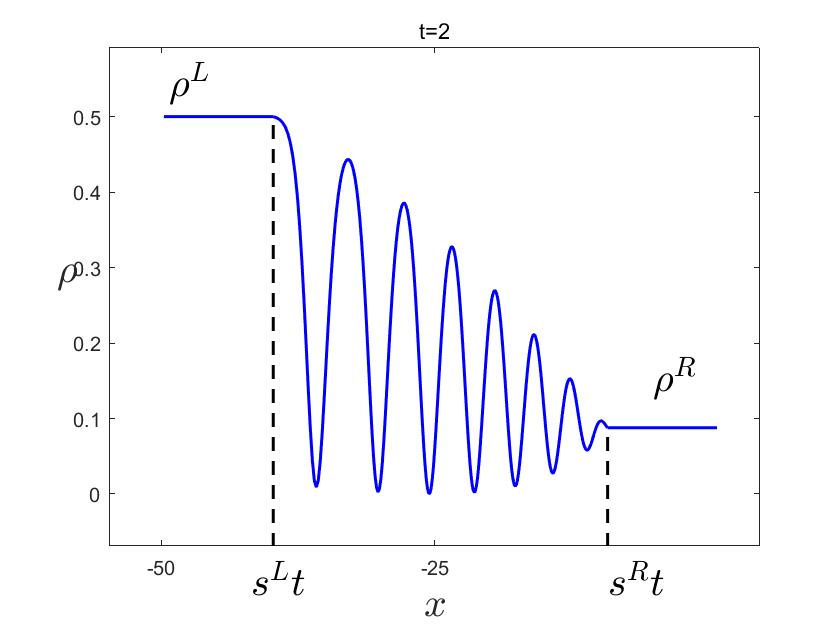}}
\quad
\subfigure[]{\includegraphics[width=0.45\linewidth]{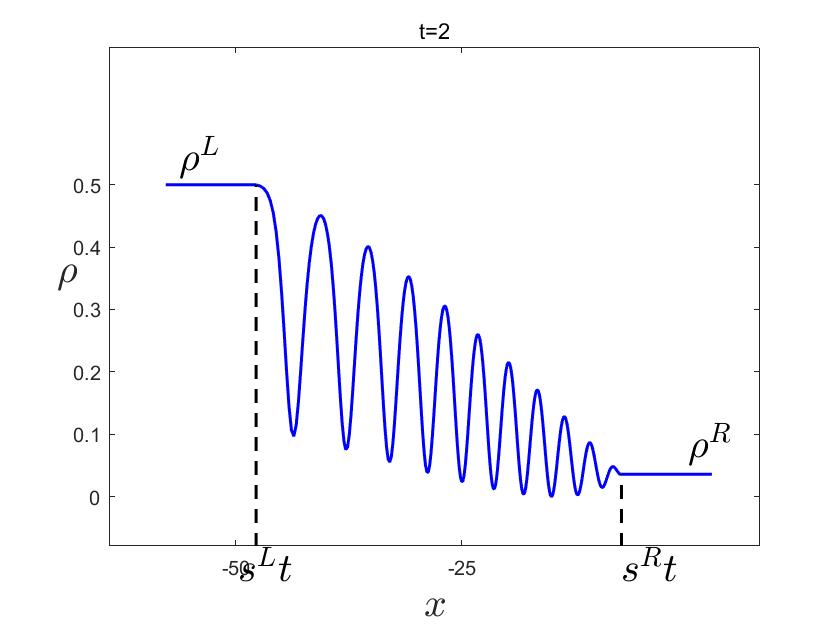}}
\quad
\subfigure[]{\includegraphics[width=0.45\linewidth]{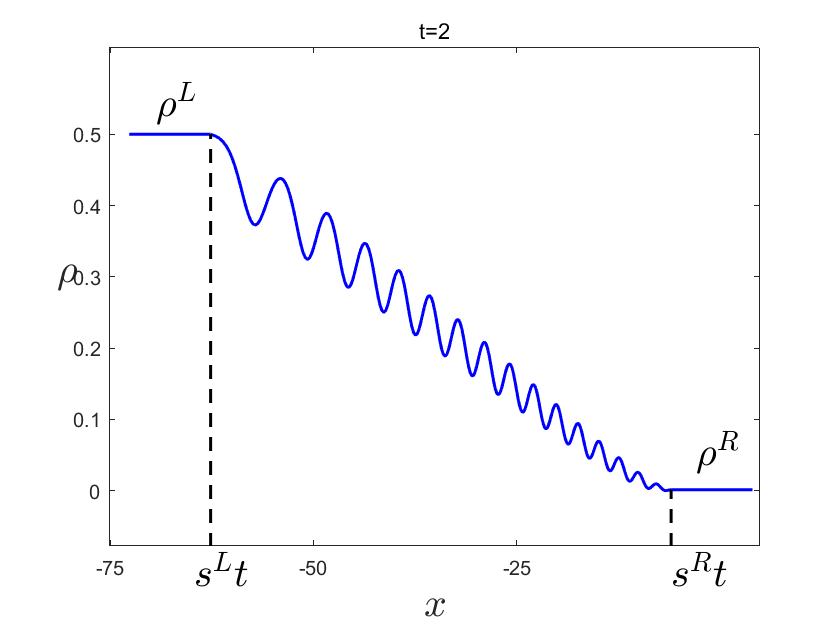}}
\quad
\subfigure[]{\includegraphics[width=0.45\linewidth]{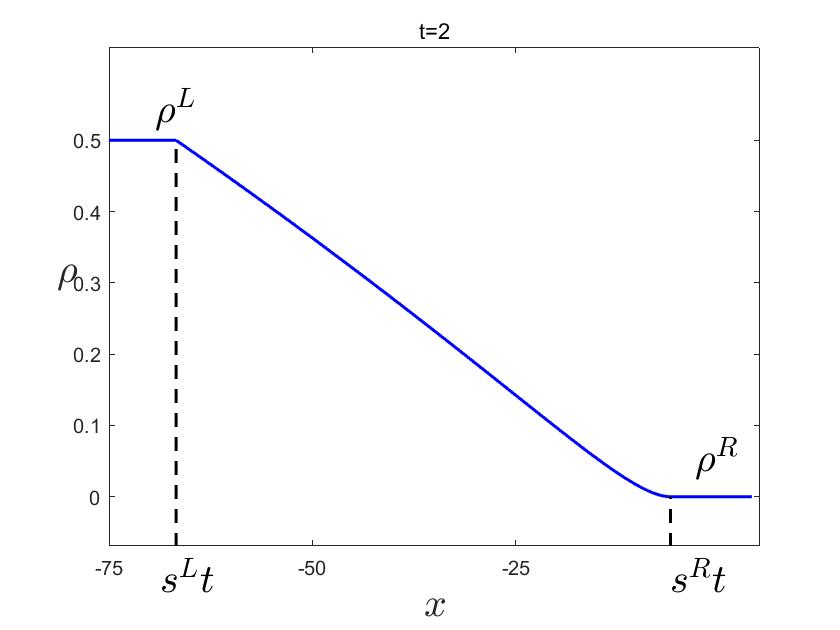}}
\quad
\flushleft{\footnotesize
\textbf{Fig.~$\bm{4}$.} The DSW solutions of HOCLL equation when a vacuum point exists. (a) Occurrence of a vacuum point at soliton edge. (b) Vacuum point inside the DSW. (c) DSW with a small amplitude. (d) Transformation of the DSW into a rarefaction wave at zero density on the right boundary. The boundary value conditions are followed by Eq.(4.8(i)).}
\end{figure}

Under the condition that the vacuum point exists, further, if $l_{+}^{R}-l_{-}^{R} = l_{-}^{R}-l_{-}^{L}$, the vacuum point will tend to the soliton edge, as shown in Fig.\ 4(a). If $l_{-}^{R} \rightarrow l_{+}^{R}$, the vacuum point will tend to the edge of small-amplitude harmonic wave as shown in Fig.\ 4(d), and the DSW converts into a RW. 

If the Riemann invariants corresponding to the initial conditions (4.1) satisfy $l_{+}^{R}>l_{+}^{L}>l_{-}^{R}=l_{-}^{L}$ and $l_{+}^{R}-l_{+}^{L} \ge l_{+}^{L}-l_{-}^{L}$, the vacuum point will also appear, and the DSWs are shown in Fig.\ 5.

Next section, we will consider the wave that connects the states of Riemann invariants at opposite sides of $\rho = 2\nu$ and crosses different regions of monotonicity.

\begin{figure}[htbp]
\centering
\setcounter{subfigure}{0}
\subfigure[]{\includegraphics[width=0.45\linewidth]{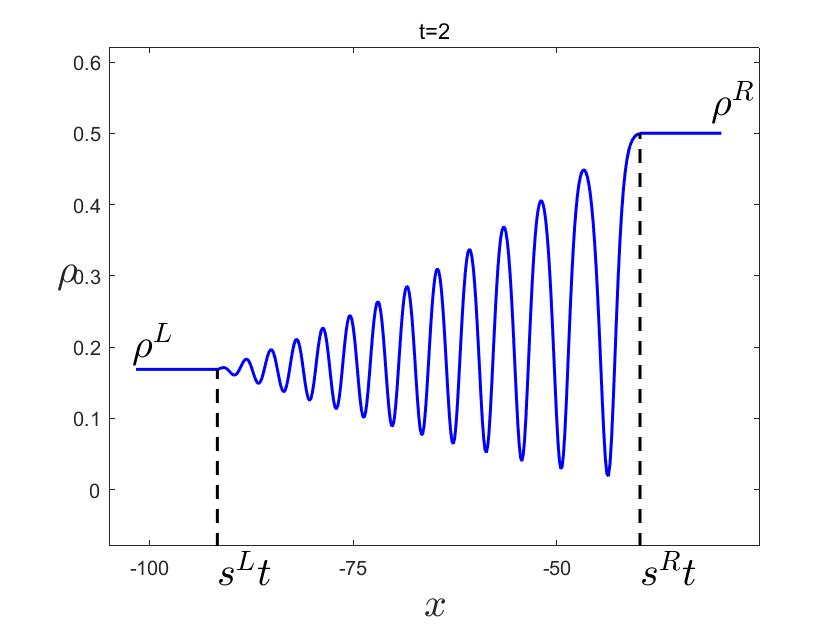}}
\quad
\subfigure[]{\includegraphics[width=0.45\linewidth]{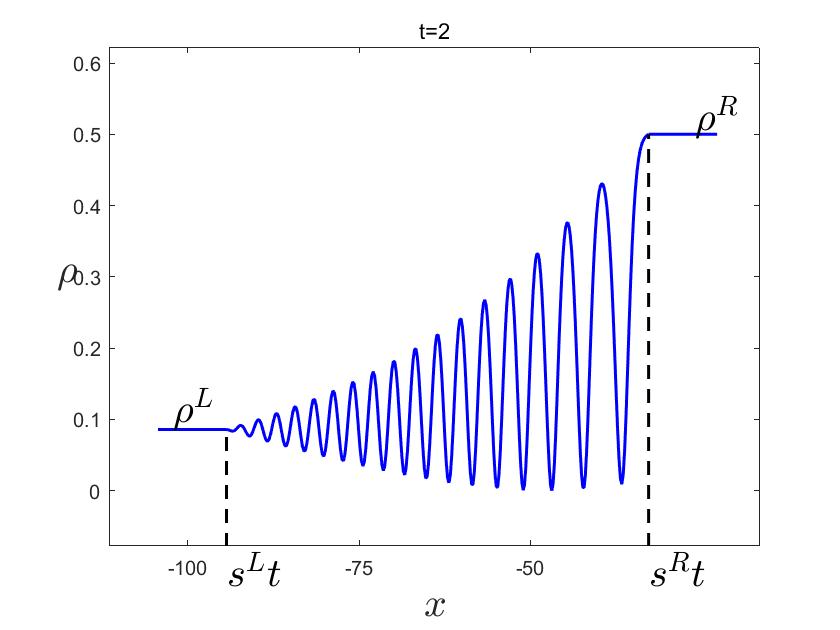}}
\quad
\subfigure[]{\includegraphics[width=0.45\linewidth]{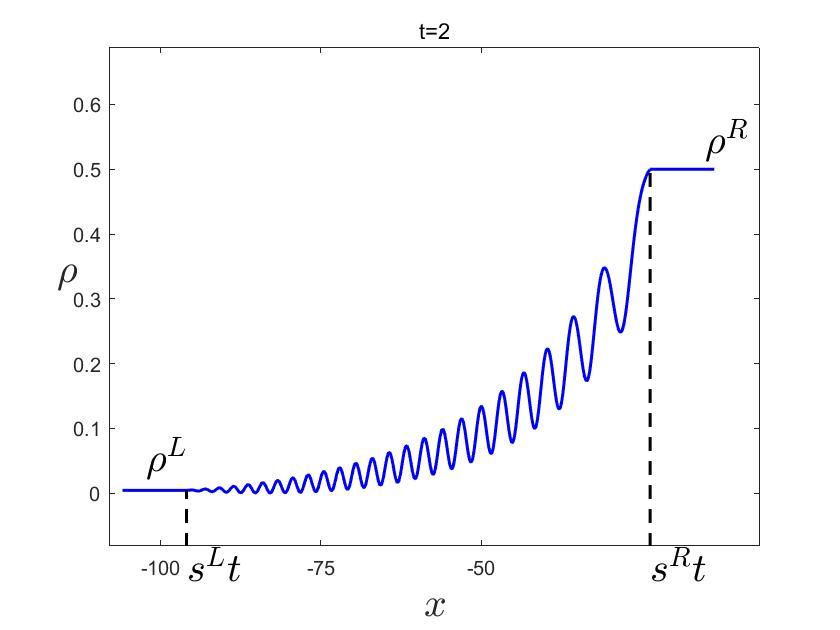}}
\quad
\subfigure[]{\includegraphics[width=0.45\linewidth]{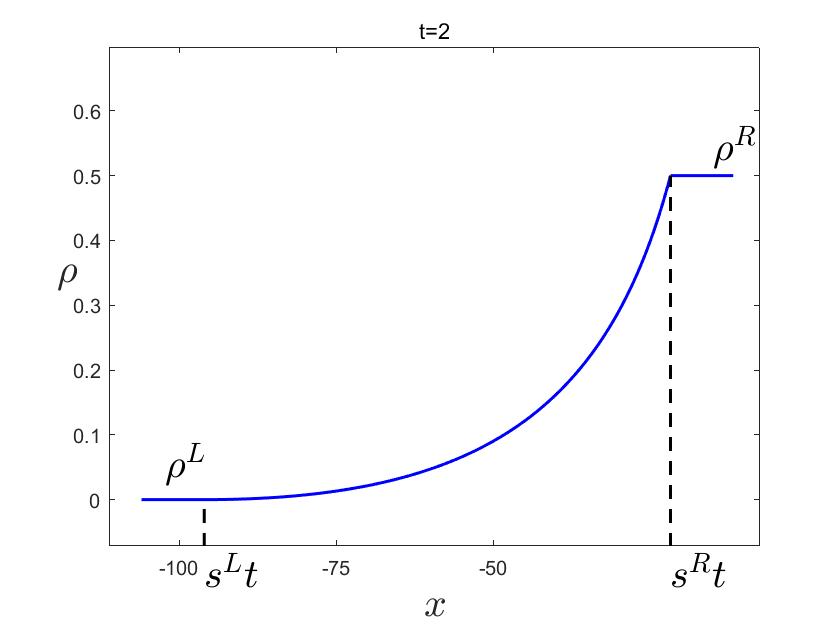}}
\quad
\flushleft{\footnotesize
\textbf{Fig.~$\bm{5}$.} The DSW solutions of HOCLL equation when a vacuum point exists where the boundary value conditions are followed by Eq.(4.8(ii)). (a) Occurrence of a vacuum point at soliton edge. (b) Vacuum point inside the DSW. (c) DSW with a small amplitude. (d) Transformation of the DSW into a rarefaction wave at zero density on the right boundary.}
\end{figure}

\vspace{5mm}\noindent\textbf{4.3 Contact dispersive shock waves}\\

In the subsequent analysis, we firstly consider that the Riemann invariants have the same values at the boundaries, that is, $l^L_+= l^R_+$ and $l^L_-= l^R_-$ as shown in Fig.\ 6(a). This equality of the Riemann invariants at the boundaries does not imply that the left and right boundary points share the same dynamical states. On the contrary, they are situated in regions with different monotonicity of the dispersionless Riemann invariants. Naturally, this produces a contact DSW. It is crucial for resolving the scenarios involving general initial discontinuity between different monotonicity regions.

\begin{figure}[H]
\centering
\setcounter{subfigure}{0}
\subfigure[]{\includegraphics[width=0.41\linewidth]{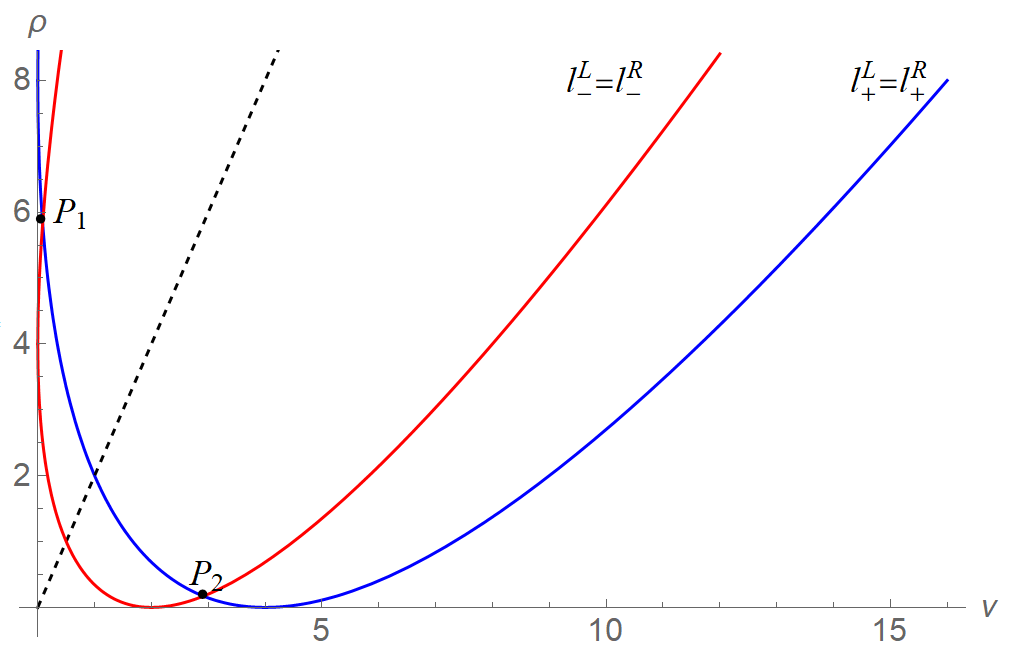}}
\quad
\subfigure[]{\includegraphics[width=0.41\linewidth]{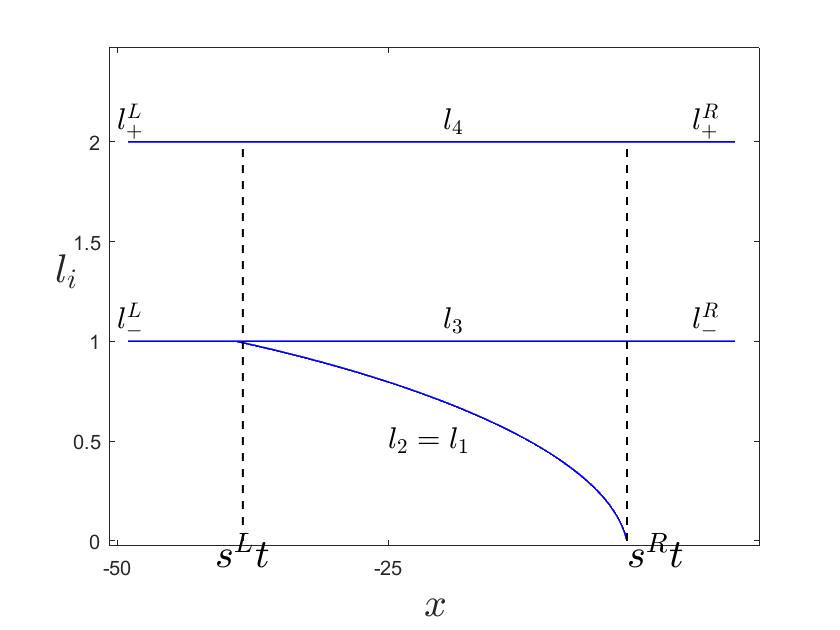}}
\quad
\subfigure[]{\includegraphics[width=0.41\linewidth]{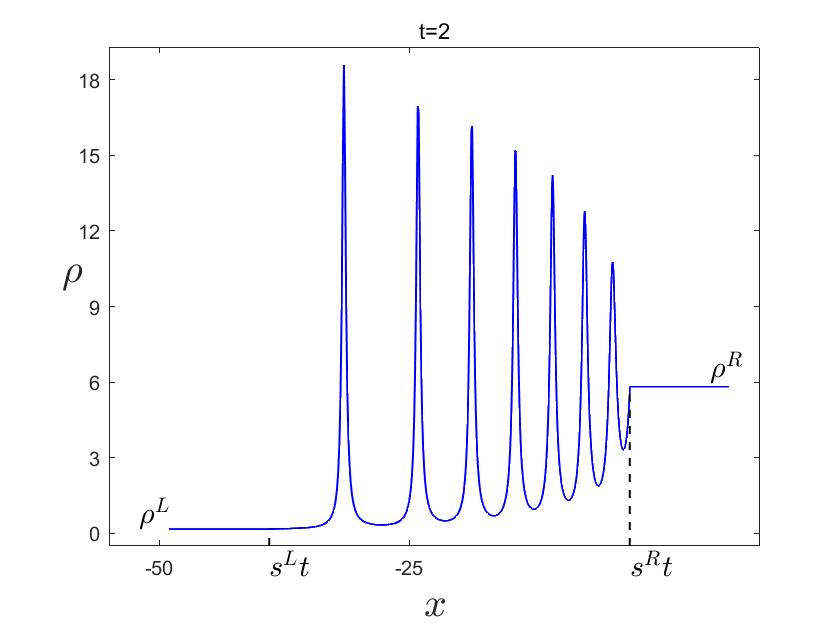}}
\quad
\subfigure[]{\includegraphics[width=0.41\linewidth]{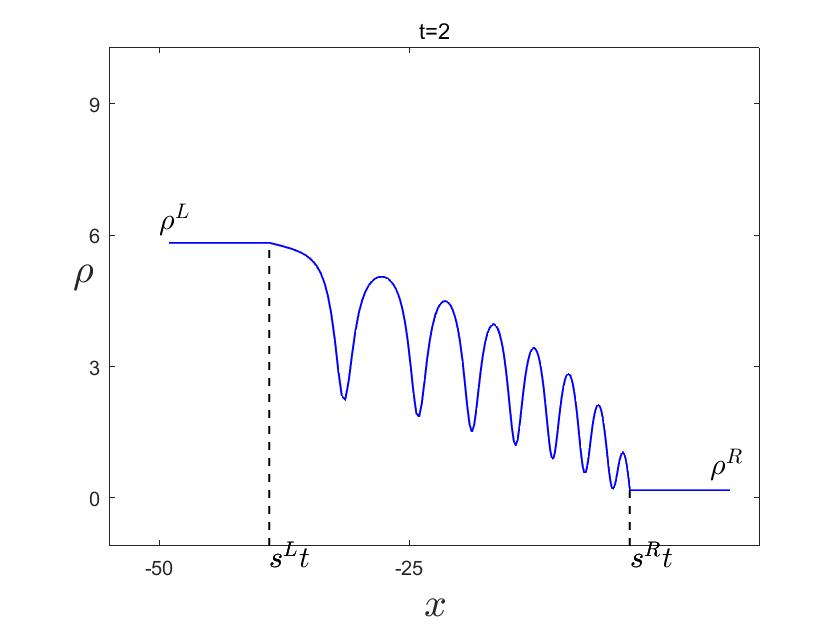}}
\quad
\flushleft{\footnotesize
\textbf{Fig.~$\bm{6}$.} Diagrams related to contact DSW. (a) The variation of density with velocity when Riemann invariants are constant. The marked intersections in the figure correspond to the initial conditions. (b) The Riemann invariant structures of the contact DSW and the values of Riemann invariants on both sides of the contact DSW in the figure remain equal. (c) The contact DSW at $t=2$ which shows the distribution between density $\rho$ and x. Herein, the density and Riemann invariants are correlated via Eq. (2.30). (d) The density and Riemann invariants are correlated via Eq. (2.31).}
\end{figure}

We arrive at the diagram of the Riemann invariants shown in Fig.\ 6(b) which the two smallest Riemann invariants are equal to each other, i.e., $l_1 = l_2$. Thus, one can obtain $m = 0$ and the solutions of the Whitham equations are determined by the following expressions
\begin{equation}
\begin{aligned}
v_1=v_2=-\frac{48l_{1}^{3}-6l_1\left( l_{-}^{L}-l_{+}^{L} \right) ^2-24l_{1}^{2}\left( l_{-}^{L}+l_{+}^{L} \right) -3\left( l_{-}^{L}-l_{+}^{L} \right) ^2\left( l_{-}^{L}+l_{+}^{L} \right)}{4l_1-2\left( l_{-}^{L}+l_{+}^{L} \right)}=z.
\end{aligned}
\end{equation}
The velocities of the edges are
\begin{equation}
\begin{aligned}
s^L=-\frac{3\left( 5l_{-}^{L2}+2l_{-}^{L}l_{+}^{L}+l_{+}^{L2} \right)}{2},~s^R=-\frac{3\left( l_{-}^{L}-l_{+}^{L} \right) ^2}{2}.
\end{aligned}
\end{equation}
Fig.\ 6(c) and Fig.\ 6(d) show two contact DSWs corresponding to the mappings (2.30) and (2.31), and the states of edges represent $P_1(\rho^L) \rightarrow P_2(\rho^R)$ and $P_2(\rho^L) \rightarrow P_1(\rho^R)$, respectively.

\vspace{5mm}\noindent\textbf{4.4 Combined shocks}\\

Now we consider the general elementary wave structures crossed different monotonicity regions of Riemann invariants. There exist two situations as illustrated in Fig.\ 7 which one of the Riemann invariants still remains constant while the boundary values of the other Riemann invariant are different: $l_-^L < l_-^R$ or $l_-^L > l_-^R$. 

\begin{figure}[htbp]
\centering
\setcounter{subfigure}{0}
\subfigure[]{\includegraphics[width=0.31\linewidth]{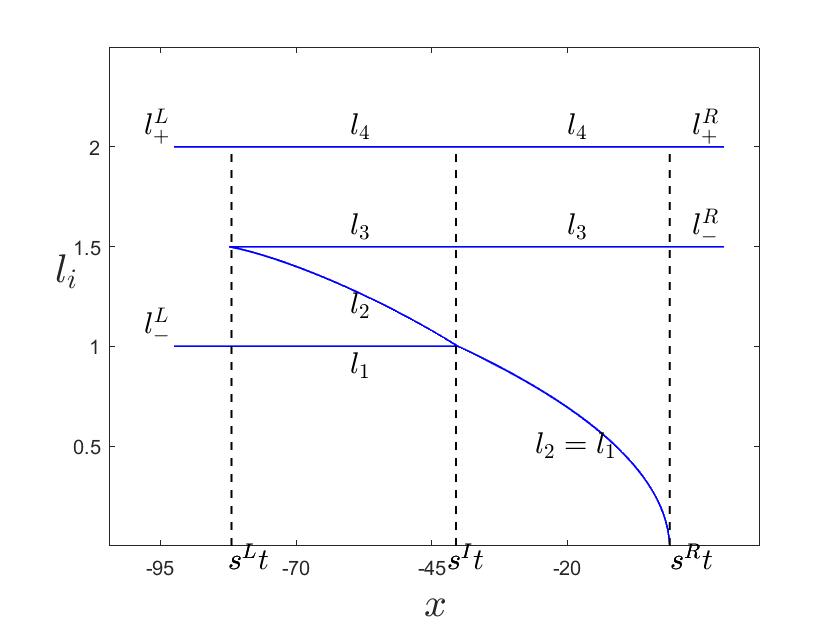}}
\quad
\subfigure[]{\includegraphics[width=0.31\linewidth]{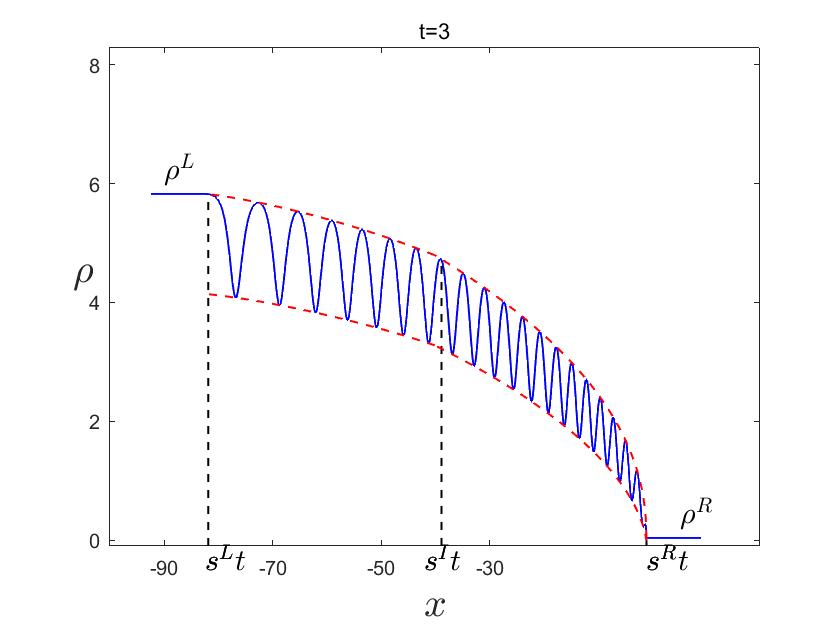}}
\quad
\subfigure[]{\includegraphics[width=0.31\linewidth]{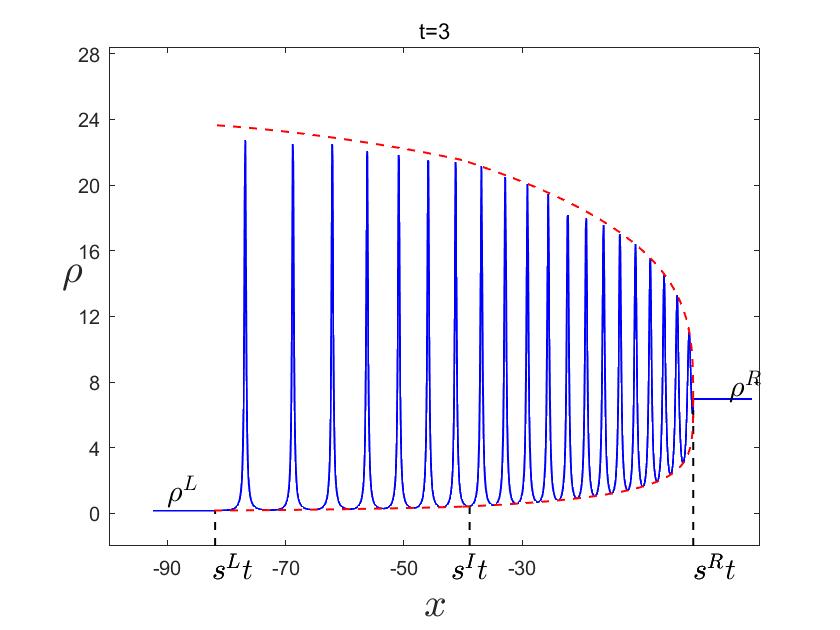}}
\quad
\subfigure[]{\includegraphics[width=0.31\linewidth]{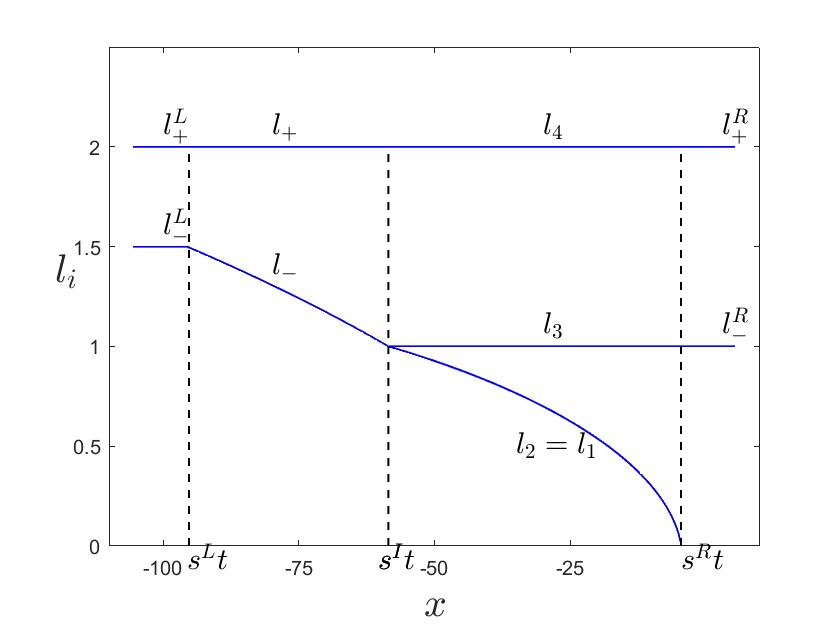}}
\quad
\subfigure[]{\includegraphics[width=0.31\linewidth]{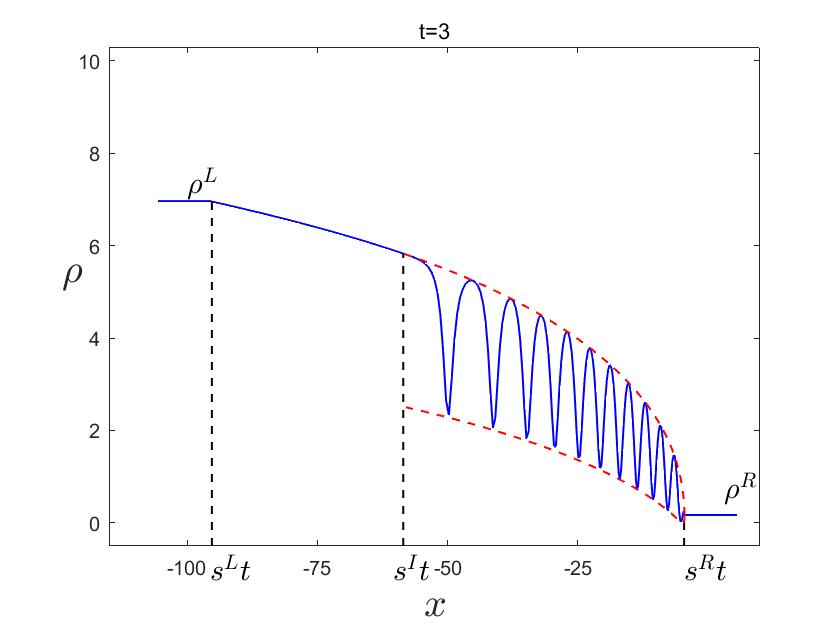}}
\quad
\subfigure[]{\includegraphics[width=0.31\linewidth]{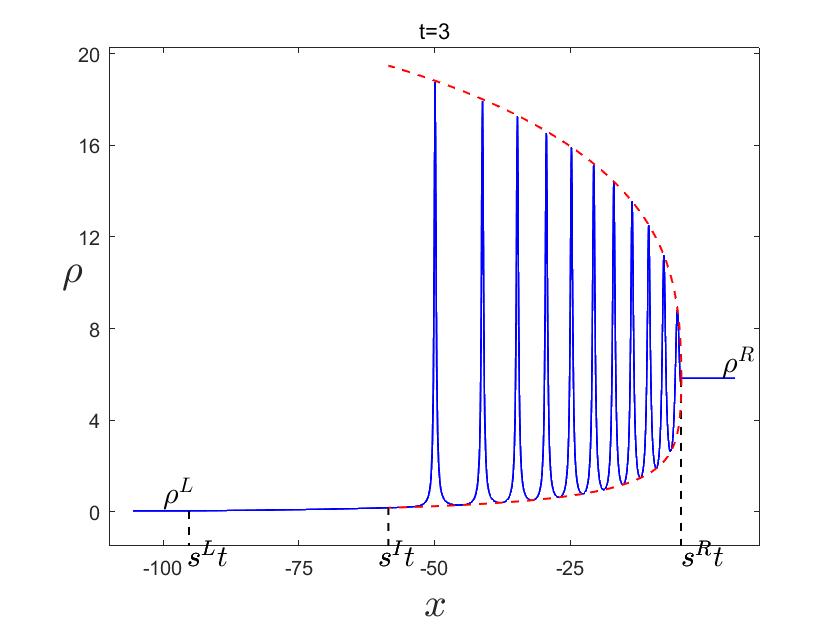}}
\quad
\flushleft{\footnotesize
\textbf{Fig.~$\bm{7}$.} Diagrams of combined wave structures. (a) represents the Riemann invariants structures of the DSW and contact DSW. (b) and (c) are the two corresponding diagrams of density $\rho$ versus $x$ corresponding to (a). (d) depicts the Riemann invariants structures of the RW and contact DSW. (e) and (f) are the diagrams of density $\rho$ versus $x$ corresponding to (d).}
\end{figure}

As we see, in the first case, the oscillating region consists of the combined wave structures-a cnoidal DSW and a contact DSW-through two mappings. The soliton edge of the cnoidal DSW matches with the left plateau and the edge with $m = 0$  degenerates into the contact DSW. The Whitham velocities can be given by
\begin{equation}
\begin{aligned}
s^L=-\frac{48l_{-}^{L3}-6l_{-}^{L}\left( l_{-}^{R}-l_{+}^{L} \right) ^2-24l_{-}^{L2}\left( l_{-}^{R}+l_{+}^{L} \right) -3\left( l_{-}^{R}-l_{+}^{L} \right) ^2\left( l_{-}^{R}+l_{+}^{L} \right)}{4l_{-}^{L}-2\left( l_{-}^{R}+l_{+}^{L} \right)},\\
s^I=-\frac{1}{2}\left( -3l_{-}^{L2}-8l_{-}^{R2}-4l_{-}^{R}l_{+}^{L}-3l_{+}^{L2}-2l_{-}^{L}\left( 2l_{-}^{R}+l_{+}^{L} \right) \right) ,
s^R=-\frac{3\left( l_{-}^{R}-l_{+}^{R} \right) ^2}{2}.
\end{aligned}
\end{equation}
In a similar way, in the second case we have a contact DSW region combined with a RW. The edge velocities can be expressed as
\begin{equation}
\begin{aligned}
s^L=-\frac{3}{2}\left( 5l_{-}^{L2}+2l_{+}^{L}l_{-}^{L}+l_{+}^{L2} \right) ,\ s^I=-\frac{3}{2}\left( 5l_{-}^{R2}+2l_{+}^{R}l_{-}^{R}+l_{+}^{R2} \right) ,
s^R=-\frac{3\left( l_{-}^{R}-l_{+}^{R} \right) ^2}{2}.
\end{aligned}
\end{equation}

\vspace{7mm}\noindent\textbf{5 Classification of all feasible initial discontinuous conditions}
\hspace*{\parindent}
\renewcommand{\theequation}{5.\arabic{equation}}\setcounter{equation}{0}\\

Following the preceding setting, the initial discontinuity conditions described by Eqs. (4.1) is parameterized by four parameters: $\nu_L,\ \rho_L,\ \nu_R$ and $\rho_R$, which correspond to the values of Riemann invariants $l_{-}^{L},\ l_{+}^{L},\ l_{-}^{R}$ and $l_{+}^{R}$ at the left and right boundary, respectively. As derived from Eq. (3.5), the relationship between the variables $\nu$ and $\rho$ is obtained. 

Next, we will present the complete classification to the Riemann problem of the HOCLL equation. Specifically, the line $\rho = 2\nu$ splits $(\nu ,\rho)$ plane into two regions with distinct monotonicity characteristics. We firstly investigate the classification problem from the case where both boundary points (corresponding to the left and right states) lie on the same side of the axis $\rho = 2\nu$. By fixing the values of the left boundary, it is seen that there are six regions in the $(\nu ,\rho)$ plane in Fig.\ 8, which are denoted as
\begin{equation}
\begin{aligned}
A:\ l_{-}^{R}<l_{+}^{R}<l_{-}^{L}<l_{+}^{L},\\
B:\ l_{-}^{R}<l_{-}^{L}<l_{+}^{R}<l_{+}^{L},\\
C:\ l_{-}^{L}<l_{-}^{R}<l_{+}^{R}<l_{+}^{L},\\
D:\ l_{-}^{R}<l_{-}^{L}<l_{+}^{L}<l_{+}^{R},\\
E:\ l_{-}^{L}<l_{-}^{R}<l_{+}^{L}<l_{+}^{R},\\
F:\ l_{-}^{L}<l_{+}^{L}<l_{-}^{R}<l_{+}^{R}.
\end{aligned}
\end{equation}
\begin{figure}[htbp]
\centering
\setcounter{subfigure}{0}
{\includegraphics[width=0.51\linewidth]{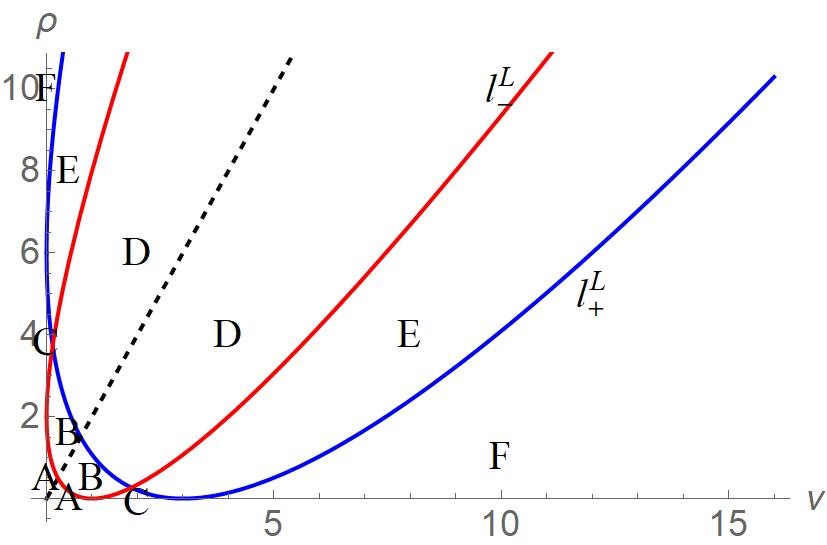}}
\flushleft{\footnotesize
\textbf{Fig.~$\bm{8}$.} Regional classification corresponding to evolution of different wave structures under initial discontinuous conditions (six categories). The dashed black line represents $\rho=2\nu$.}
\end{figure}
\begin{figure}[htbp]
\centering
\setcounter{subfigure}{0}
\subfigure[]{\includegraphics[width=0.31\linewidth]{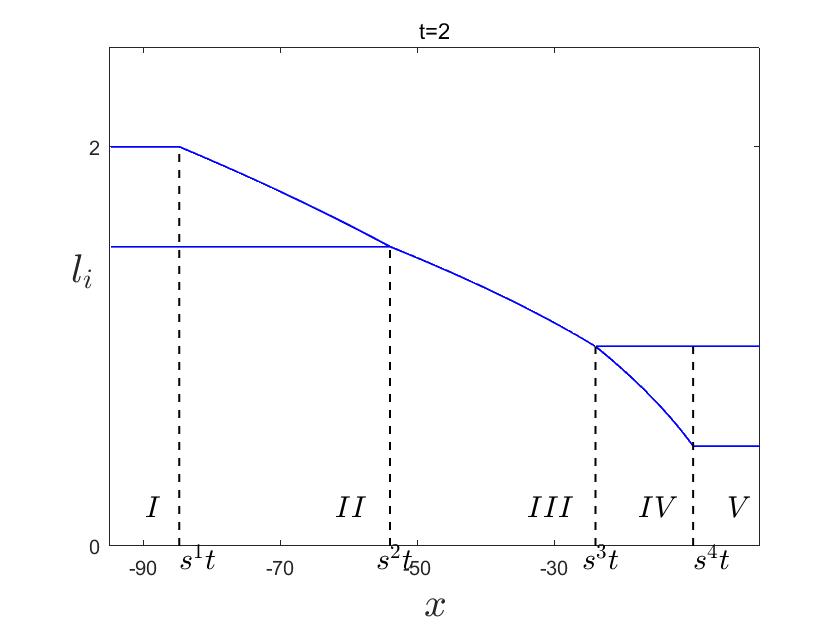}}
\quad
\subfigure[]{\includegraphics[width=0.31\linewidth]{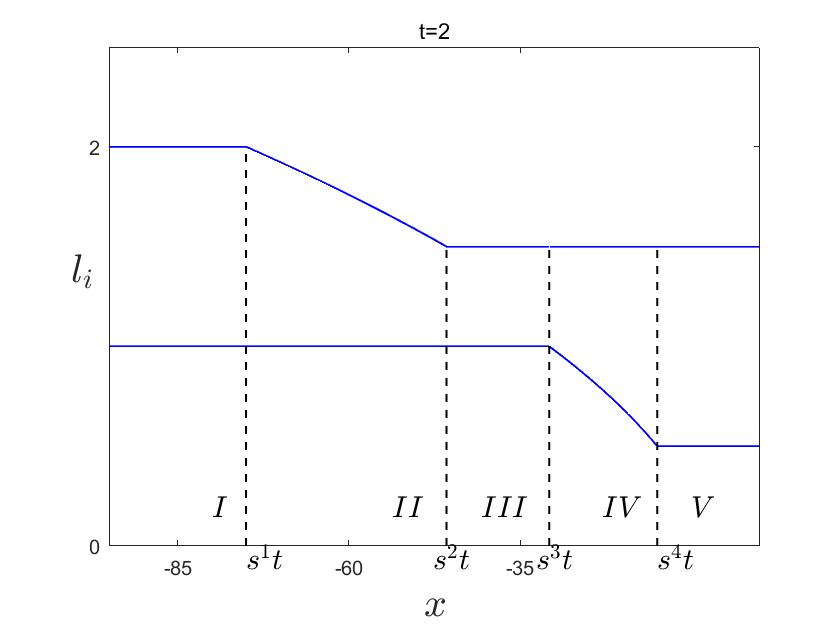}}
\quad
\subfigure[]{\includegraphics[width=0.31\linewidth]{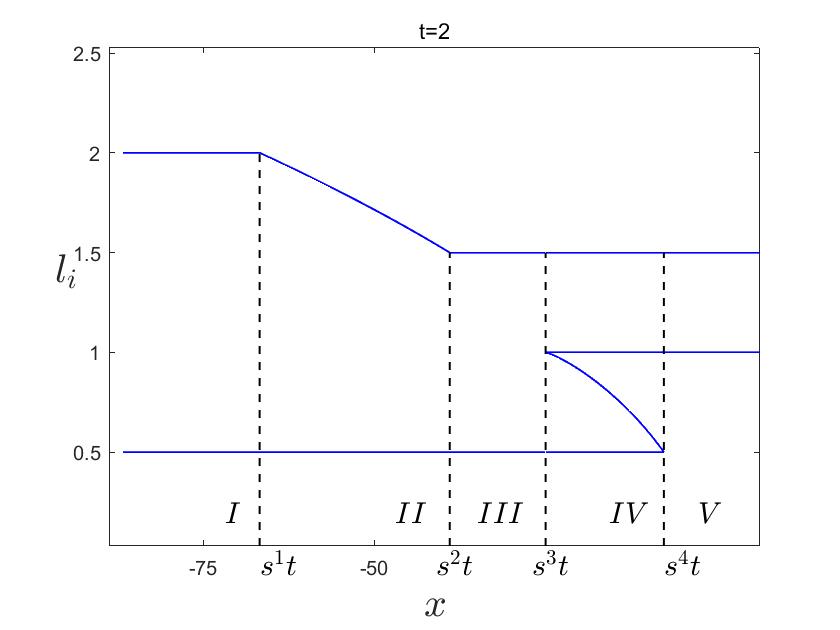}}
\quad
\subfigure[]{\includegraphics[width=0.31\linewidth]{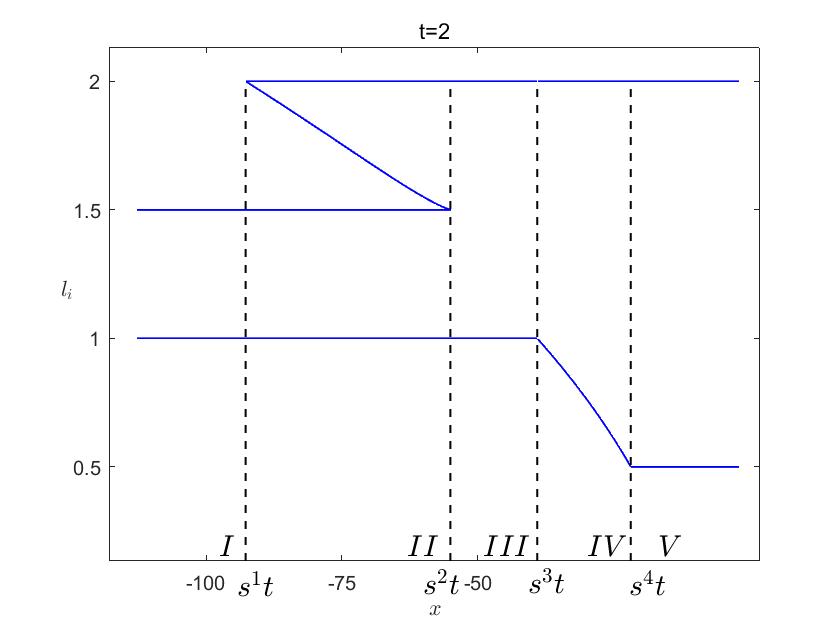}}
\quad
\subfigure[]{\includegraphics[width=0.31\linewidth]{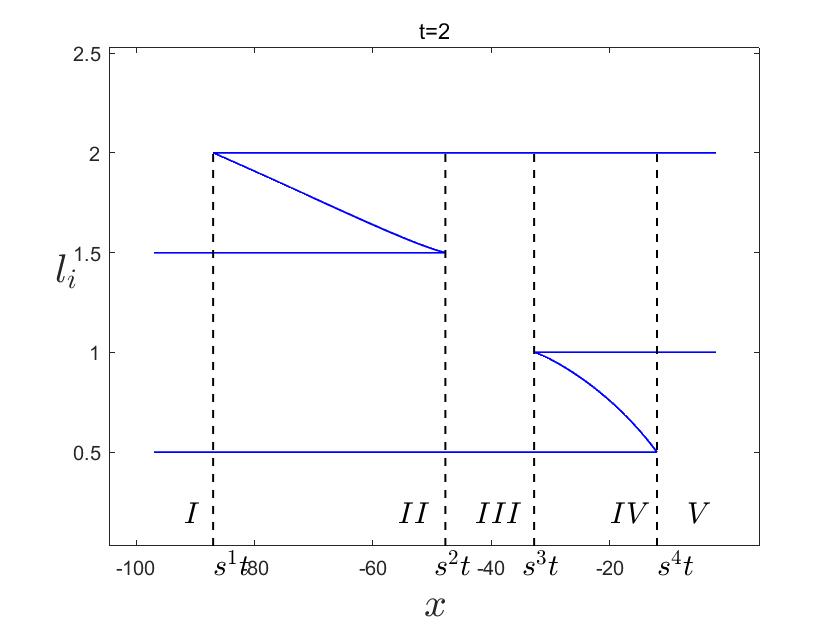}}
\quad
\subfigure[]{\includegraphics[width=0.31\linewidth]{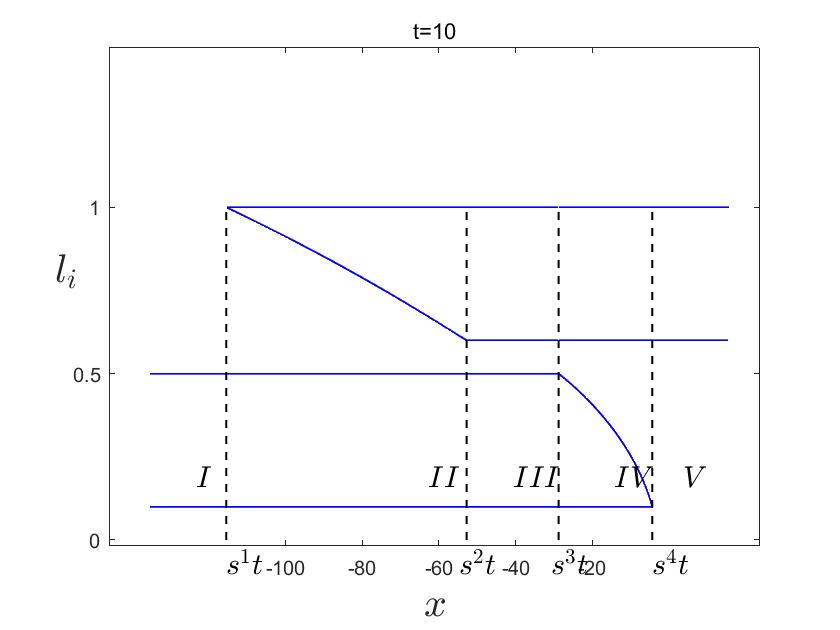}}
\quad
\flushleft{\footnotesize
\textbf{Fig.~$\bm{9}$.} Structures of Riemann invariants under different initial discontinuous conditions corresponding to the six categories from the case when both boundary values (corresponding to the left and right states) lie on the same side of the axis $\rho = 2\nu$.}
\end{figure}

\begin{figure}[htbp]
\centering
\setcounter{subfigure}{0}
\subfigure[]{\includegraphics[width=0.31\linewidth]{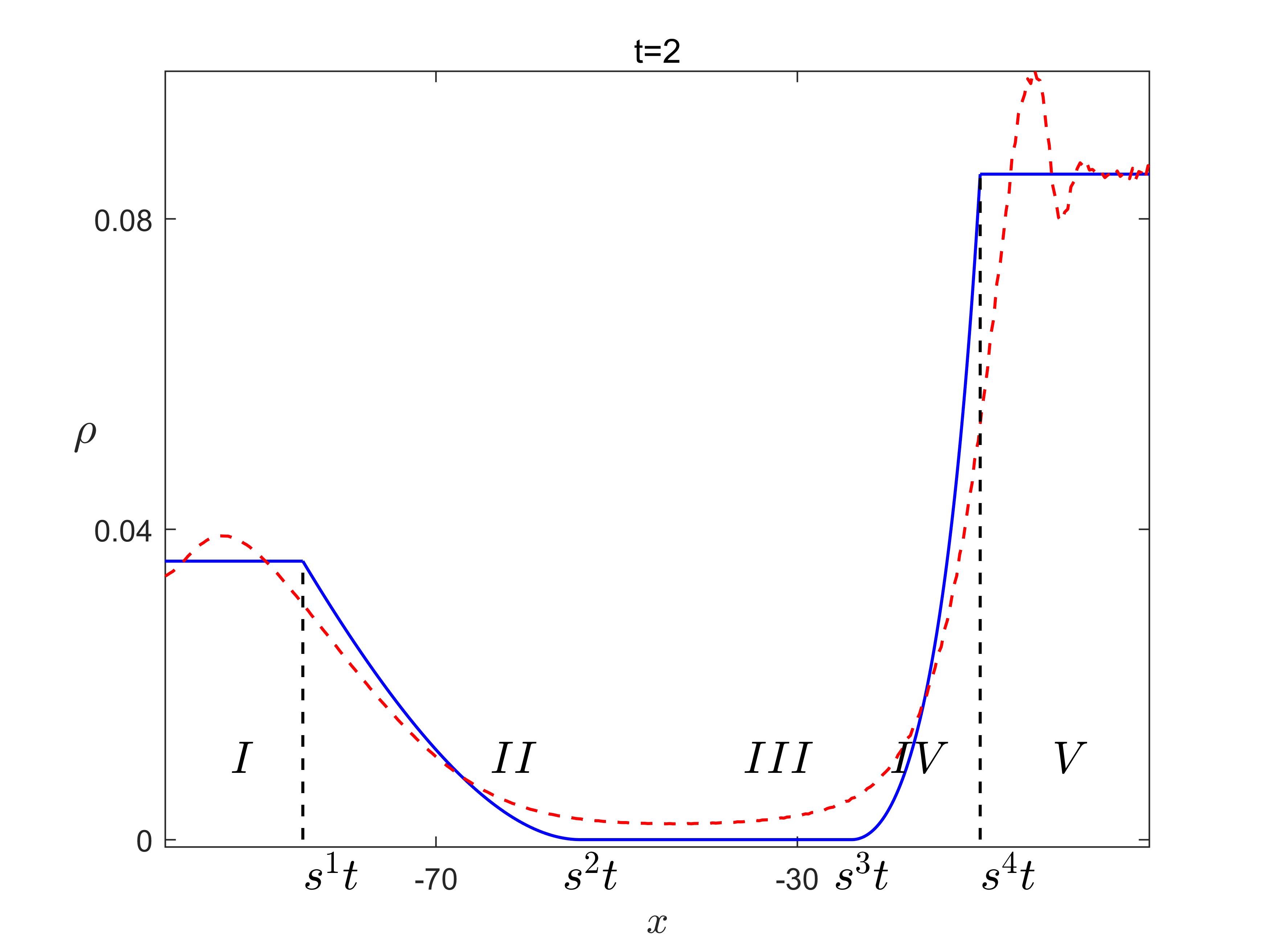}}
\quad
\subfigure[]{\includegraphics[width=0.31\linewidth]{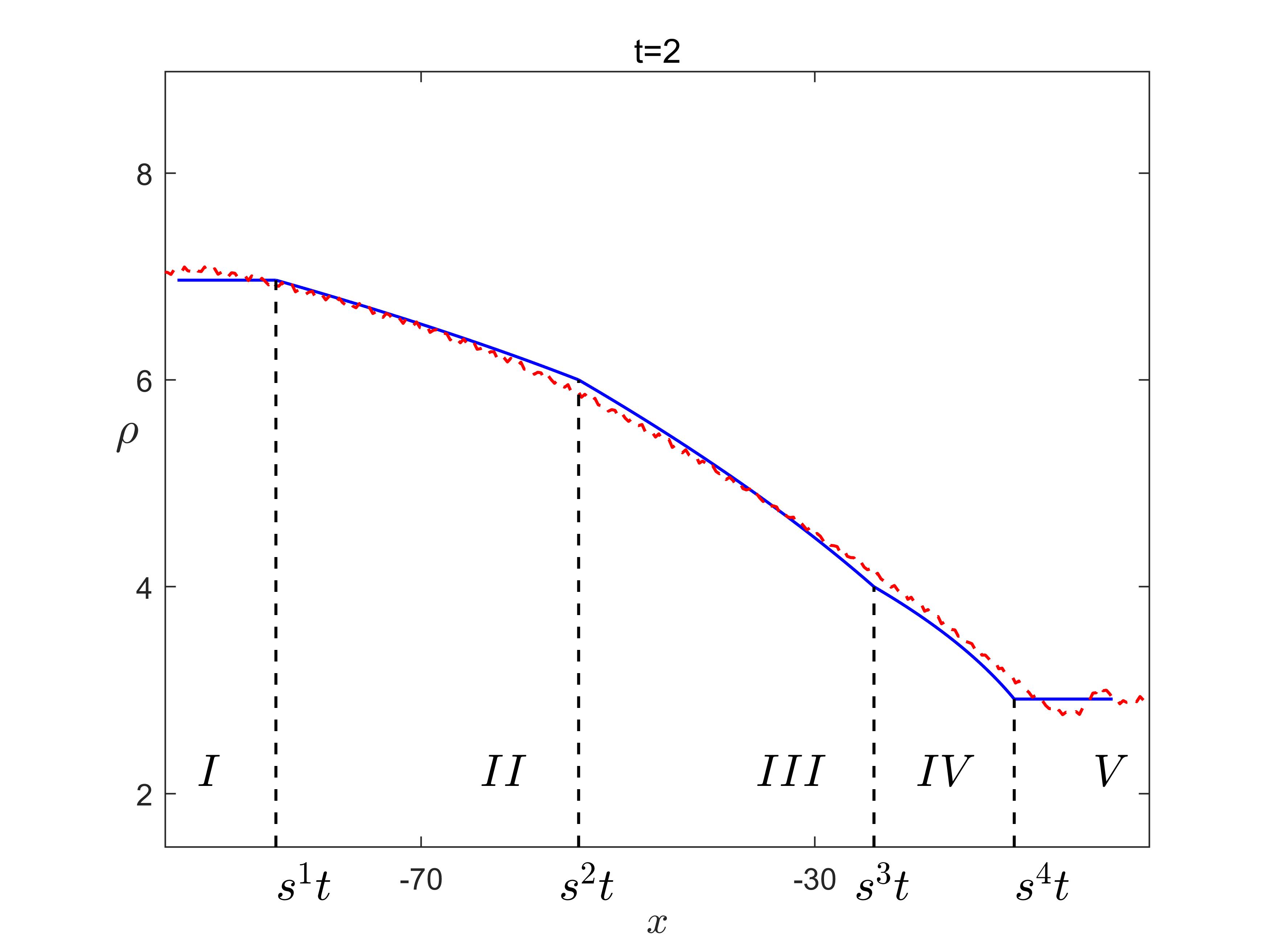}}
\quad
\subfigure[]{\includegraphics[width=0.31\linewidth]{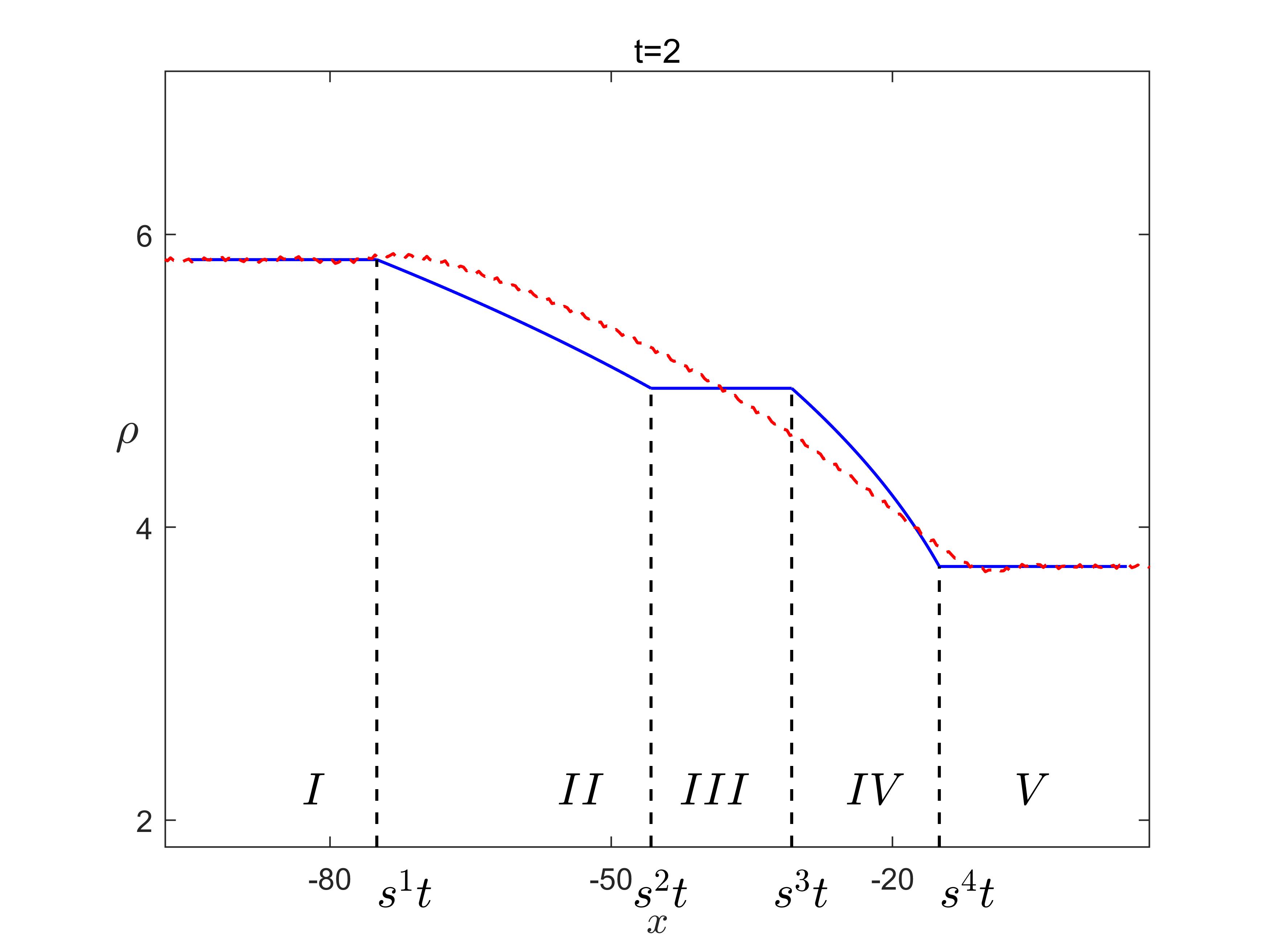}}
\quad
\subfigure[]{\includegraphics[width=0.31\linewidth]{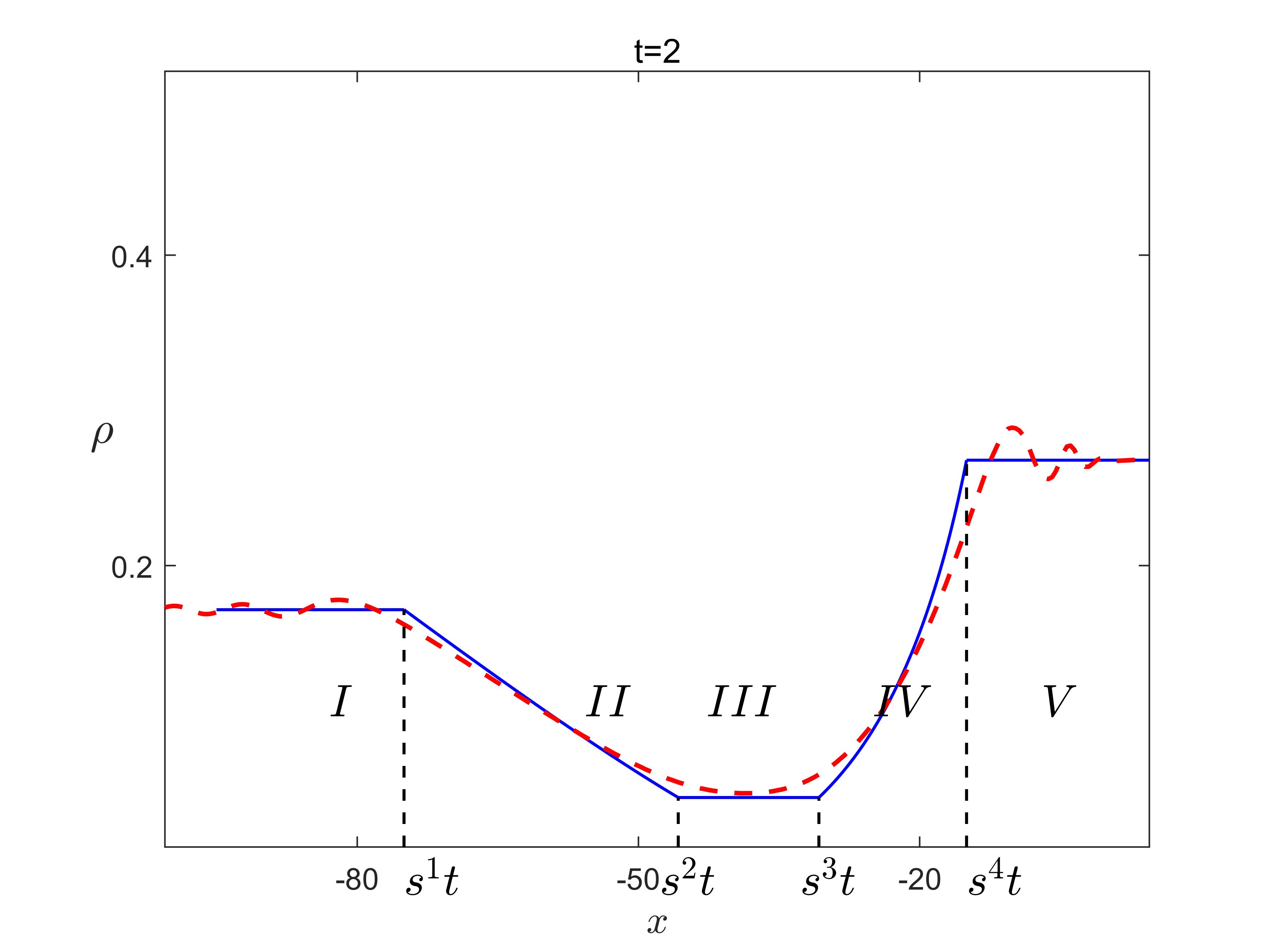}}
\quad
\subfigure[]{\includegraphics[width=0.31\linewidth]{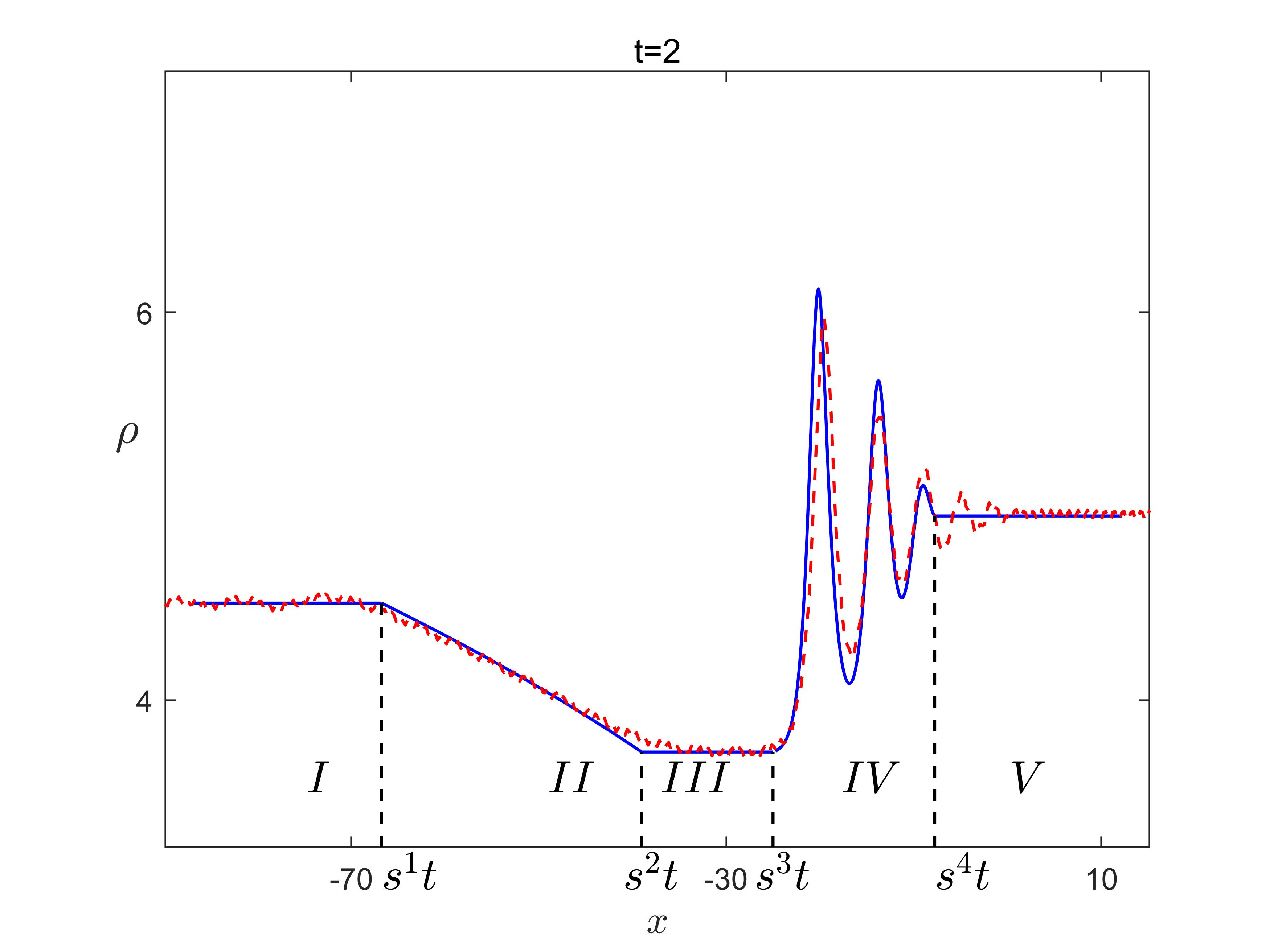}}
\quad
\subfigure[]{\includegraphics[width=0.31\linewidth]{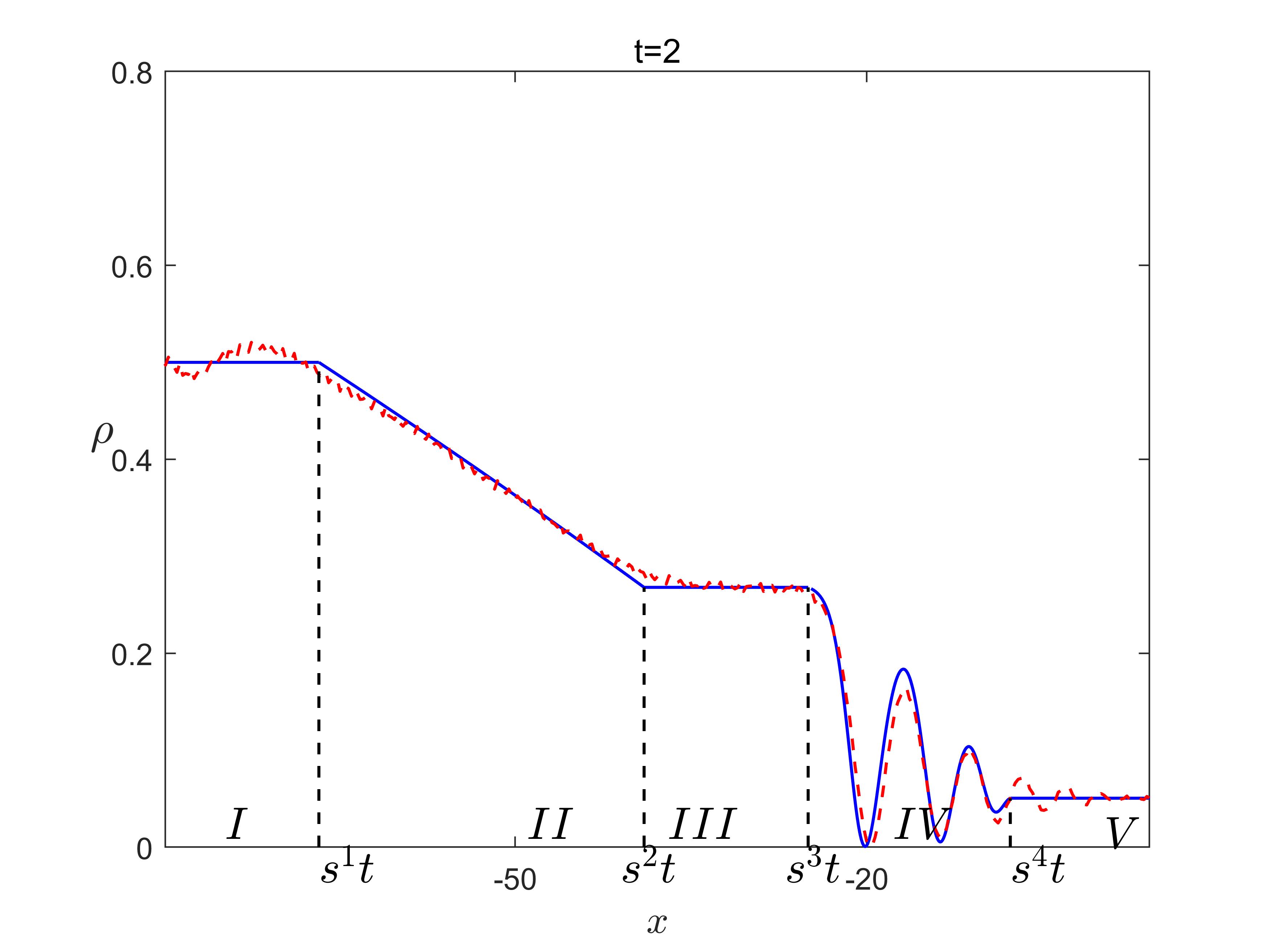}}
\quad
\subfigure[]{\includegraphics[width=0.31\linewidth]{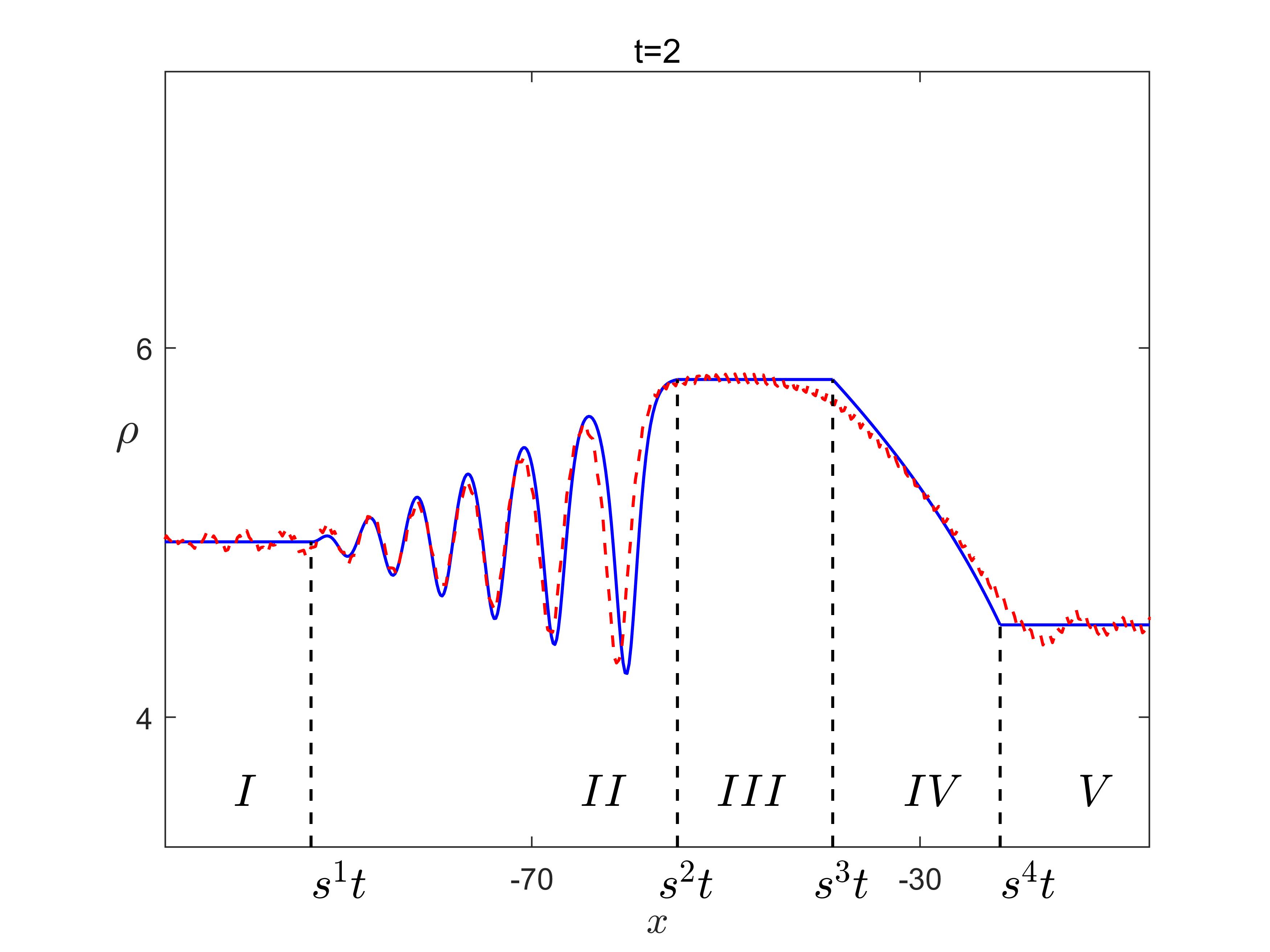}}
\quad
\subfigure[]{\includegraphics[width=0.31\linewidth]{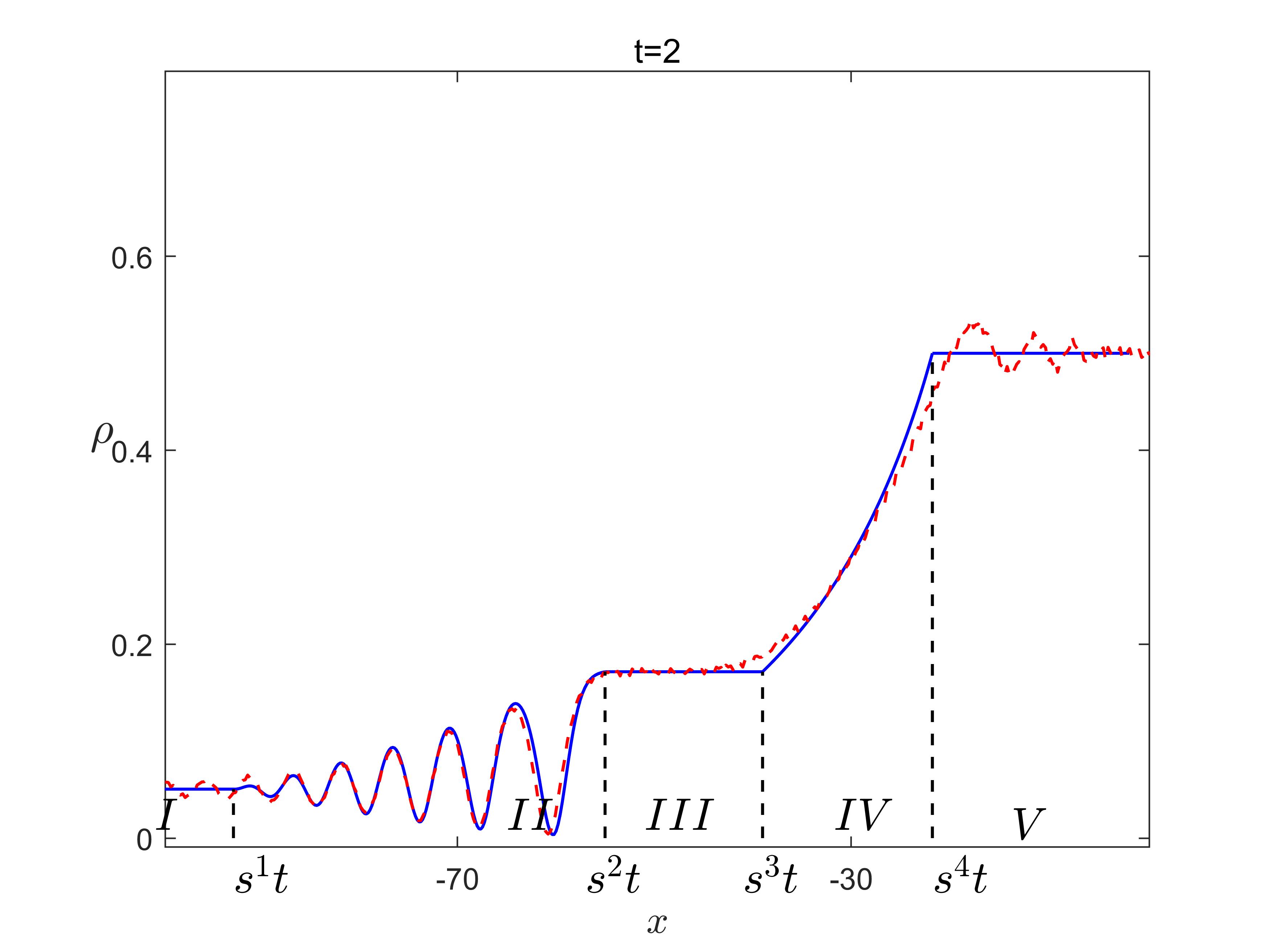}}
\quad
\subfigure[]{\includegraphics[width=0.31\linewidth]{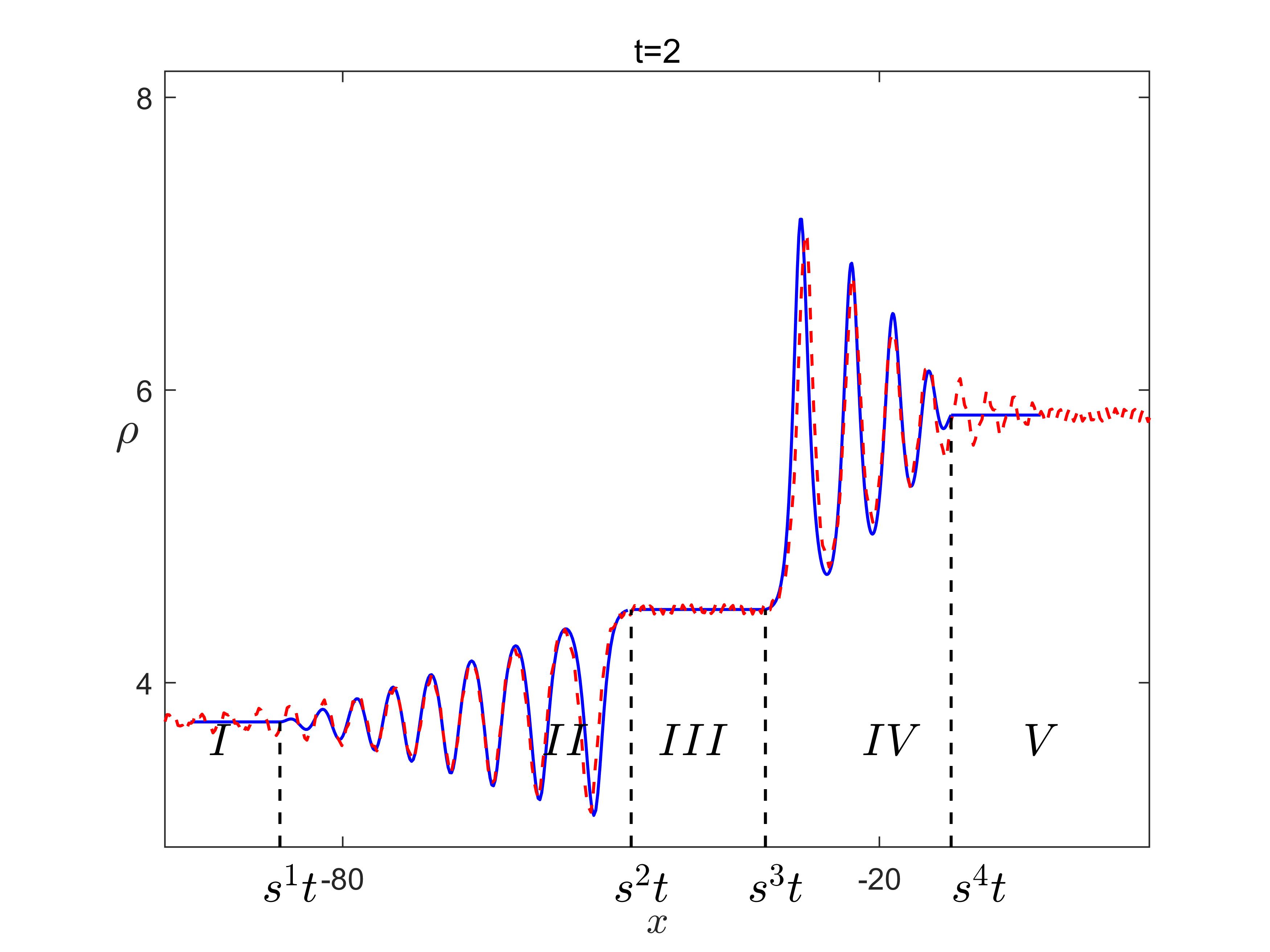}}
\quad
\subfigure[]{\includegraphics[width=0.31\linewidth]{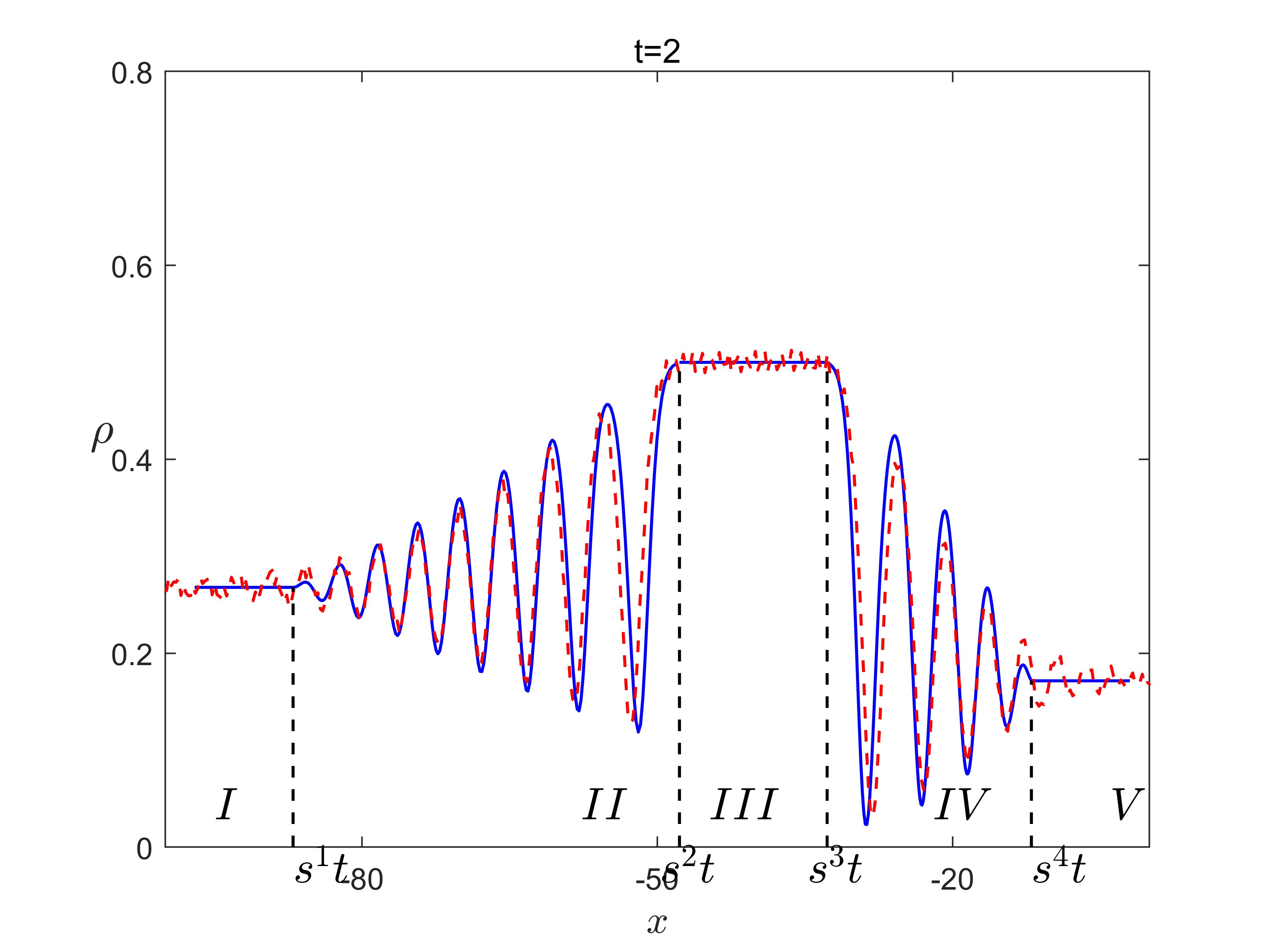}}
\quad
\subfigure[]{\includegraphics[width=0.31\linewidth]{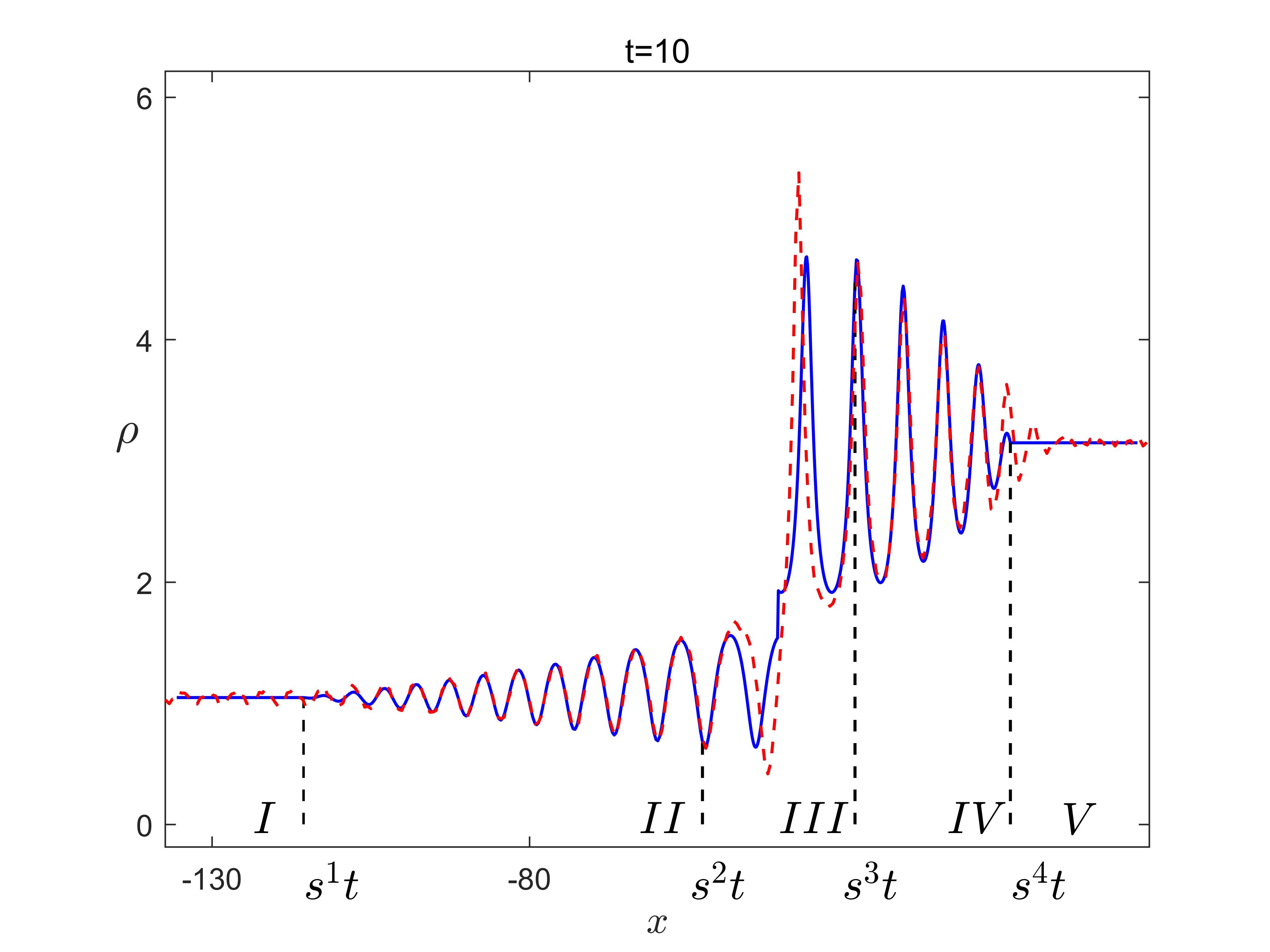}}
\quad
\subfigure[]{\includegraphics[width=0.31\linewidth]{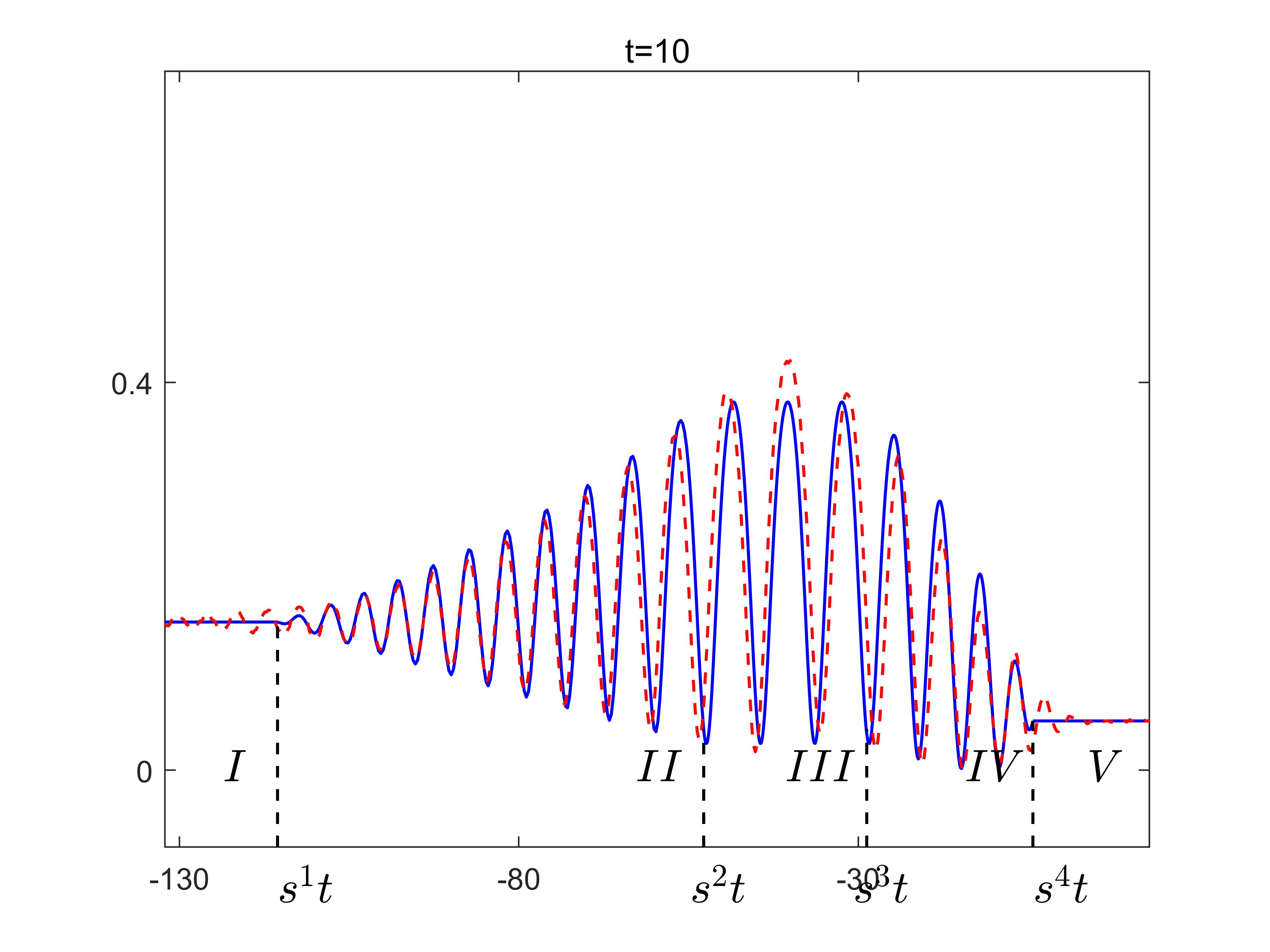}}
\quad
\flushleft{\footnotesize
\textbf{Fig.~$\bm{10}$.} Evolution of the twelve density waves with respect to $x$ corresponding to the six Riemann invariant structures in Fig. 7, where one Riemann invariant structure corresponds to two density waves via Eq. (2.30) and (2.31). The red dashed lines represent the results of numerical simulation.}
\end{figure}

Fig.\ 9 illustrates the characteristics of six types of structures of Riemann invariants, which intuitively show the structures of waves in the flow field. The Riemann invariants correspond to two density mappings so that these six cases yield twelve wave structures evolving from the initial discontinuity. As depicted in Fig.\ 10, the flow field is clearly divided into five distinct regions with distinct characteristics. Among them, the left (Region I) and right (Region V) uniform platform regions represent the initial boundary states of the flow. Within these two regions, the flow velocity and density remain constant, and connecting these two uniform regions is the intermediate evolution zone.

The distributions of Riemann invariants corresponding to Case A are shown in Fig.\ 9(a), and the mapping density profiles are presented in Fig.\ 10(a) and Fig.\ 10(b). Fig.\ 10(a) consists of two rarefaction waves (Regions II and IV) and a special vacuum region (Region III). Different from the structure in Fig.\ 10(a), the middle third region of the flow field in Fig.\ 10(b) is not a vacuum but another RW. The boundary velocities in each region are given by
\begin{equation}
\begin{aligned}
s^1=-\frac{3}{2}\left( 5l_{+}^{L2}+2l_{+}^{L}l_{-}^{L}+l_{-}^{L2} \right) ,\ s^2=-12l_{-}^{L2},\\
s^3=-12l_{+}^{R2},\ s^4=-\frac{3}{2}\left( 5l_{-}^{R2}+2l_{+}^{R}l_{-}^{R}+l_{+}^{R2} \right) .
\end{aligned}
\end{equation}

In case B, the structures of its Riemann invariants are clearly shown in Fig.\ 9(b). Notably, the diagram exhibits that the two rarefaction waves (Regions II and IV) are connected by a platform region (Regions III), which can be shown in Fig.\ 10(c) and Fig.\ 10(d). Each velocities are obtained by
\begin{equation}
\begin{aligned}
s^1=-\frac{3}{2}\left( 5l_{+}^{L2}+2l_{+}^{L}l_{-}^{L}+l_{-}^{L2} \right) ,\ s^2=-\frac{3}{2}\left( 5l_{+}^{R2}+2l_{+}^{R}l_{-}^{L}+l_{-}^{L2} \right) ,\\
s^3=-\frac{3}{2}\left( 5l_{-}^{L2}+2l_{+}^{R}l_{-}^{L}+l_{+}^{R2} \right) ,\ s^4=-\frac{3}{2}\left( 5l_{-}^{R2}+2l_{+}^{R}l_{-}^{R}+l_{+}^{R2} \right) .
\end{aligned}
\end{equation}

Case C presents a DSW structure as shown in Fig.\ 9(c). Here, the DSW (Region IV) is connected to the RW (Region II) via a platform (Region III). The corresponding density wave in Fig.\ 10(e) and Fig.\ 10(f) can more clearly reveal the details of Region IV: the left boundary of this region is a soliton edge, manifesting as a bright soliton or a dark soliton respectively; while the right boundary is a small-amplitude harmonic edge. The velocities of the edges are obtained by Whitham equations
\begin{equation}
\begin{aligned}
&s^1=-\frac{3}{2}\left( 5l_{+}^{L2}+2l_{+}^{L}l_{-}^{L}+l_{-}^{L2} \right) ,\ s^2=-\frac{3}{2}\left( 5l_{+}^{R2}+2l_{+}^{R}l_{-}^{L}+l_{-}^{L2} \right) ,\\
&s^3=-\frac{1}{2}\left( -3l_{-}^{L2}-8l_{-}^{R2}-4l_{-}^{R}l_{+}^{R}-3l_{+}^{R2}-2l_{-}^{L}\left( 2l_{-}^{R}+l_{+}^{R} \right) \right) ,\\
s^4=&-\frac{48l_{-}^{L3}-6l_{-}^{L}\left( l_{-}^{R}-l_{+}^{R} \right) ^2-24l_{-}^{L2}\left( l_{-}^{R}+l_{+}^{R} \right) -3\left( l_{-}^{R}-l_{+}^{R} \right) ^2\left( l_{-}^{R}+l_{+}^{R} \right)}{4l_{-}^{L}-2\left( l_{-}^{R}+l_{+}^{R} \right)}.
\end{aligned}
\end{equation}

Case D is quite similar to Case C and the structures of Riemann invariants are shown in Fig.\ 9(d), with the main difference being that the positions of the DSW and the RW are swapped. Specifically, in Case D, the left boundary of the DSW (corresponding to Region II) is the edge of the small-amplitude harmonic, while its right boundary is the edge of the soliton, which presents a dark soliton. The characteristics can be referred to the demonstrations in Fig.\ 10(g) and Fig.\ 10(h). The velocities are respectively
\begin{equation}
\begin{aligned}
\small s^1=-\frac{1}{2}&\left( 3l_{-}^{L2}+3l_{+}^{L2}+4l_{+}^{L}l_{+}^{R}+8l_{+}^{R2}+2l_{-}^{L}\left( 2l_{+}^{L}+l_{+}^{R} \right) \right) +\frac{4\left( l_{-}^{L}-l_{+}^{R} \right) \left( l_{+}^{R}-l_{+}^{L} \right) \left( l_{-}^{L}+l_{+}^{L}+4l_{+}^{R} \right)}{l_{-}^{L}+l_{+}^{L}-2l_{+}^{R}},\\
&s^2=\frac{1}{2}\left( -3l_{-}^{L2}-8l_{+}^{L2}-4l_{+}^{L}l_{+}^{R}-3l_{+}^{R2}-2l_{-}^{L}\left( 2l_{+}^{L}+l_{+}^{R} \right) \right) ,\\
&s^3=-\frac{3}{2}\left( 5l_{-}^{L2}+2l_{+}^{R}l_{-}^{L}+l_{+}^{R2} \right) ,\ s^4=-\frac{3}{2}\left( 5l_{-}^{R2}+2l_{+}^{R}l_{-}^{R}+l_{+}^{R2} \right) .
\end{aligned}
\end{equation}
\begin{figure}[htbp]
\centering
\setcounter{subfigure}{0}
\subfigure[]{\includegraphics[width=0.31\linewidth]{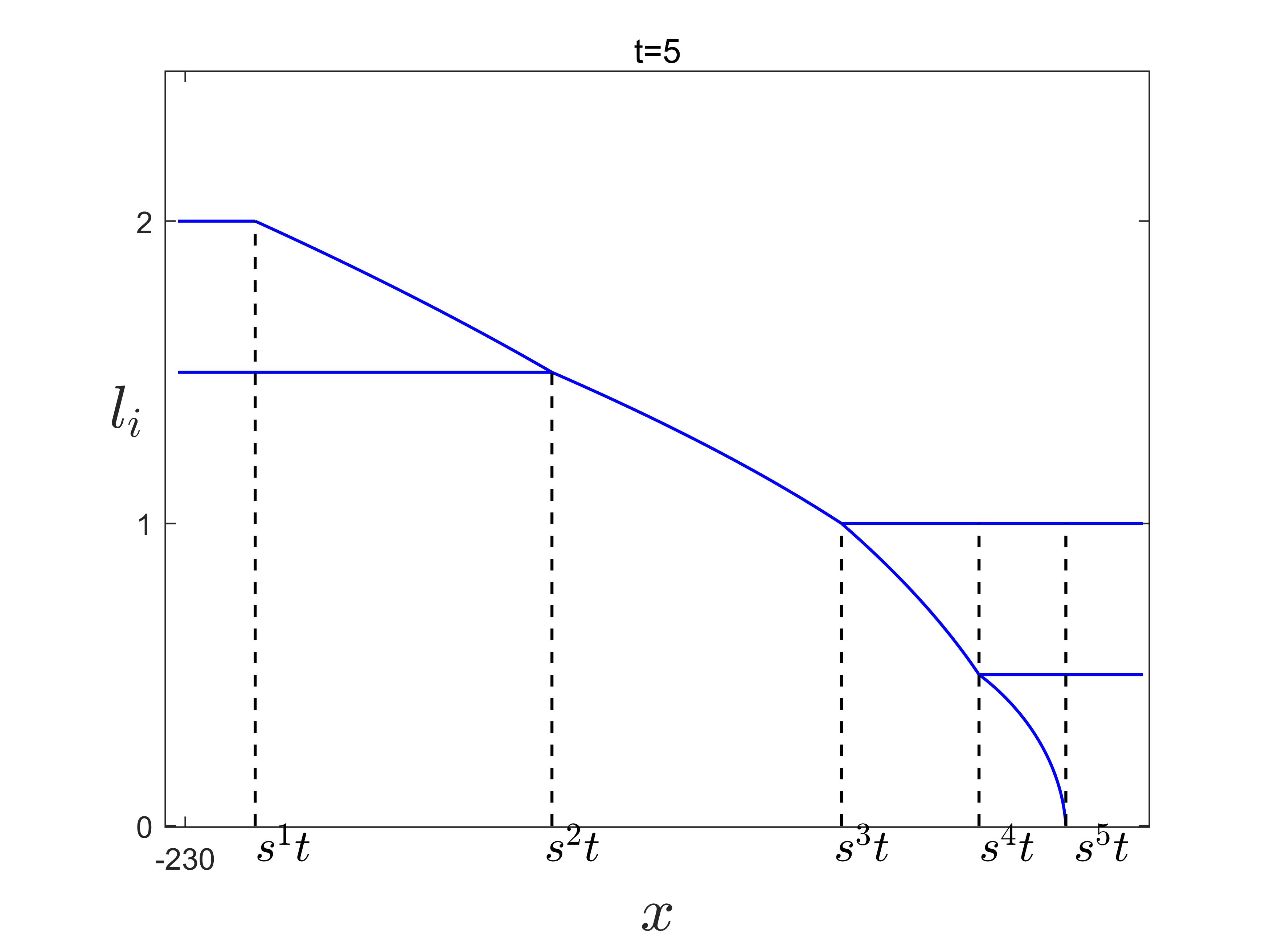}}
\quad
\subfigure[]{\includegraphics[width=0.31\linewidth]{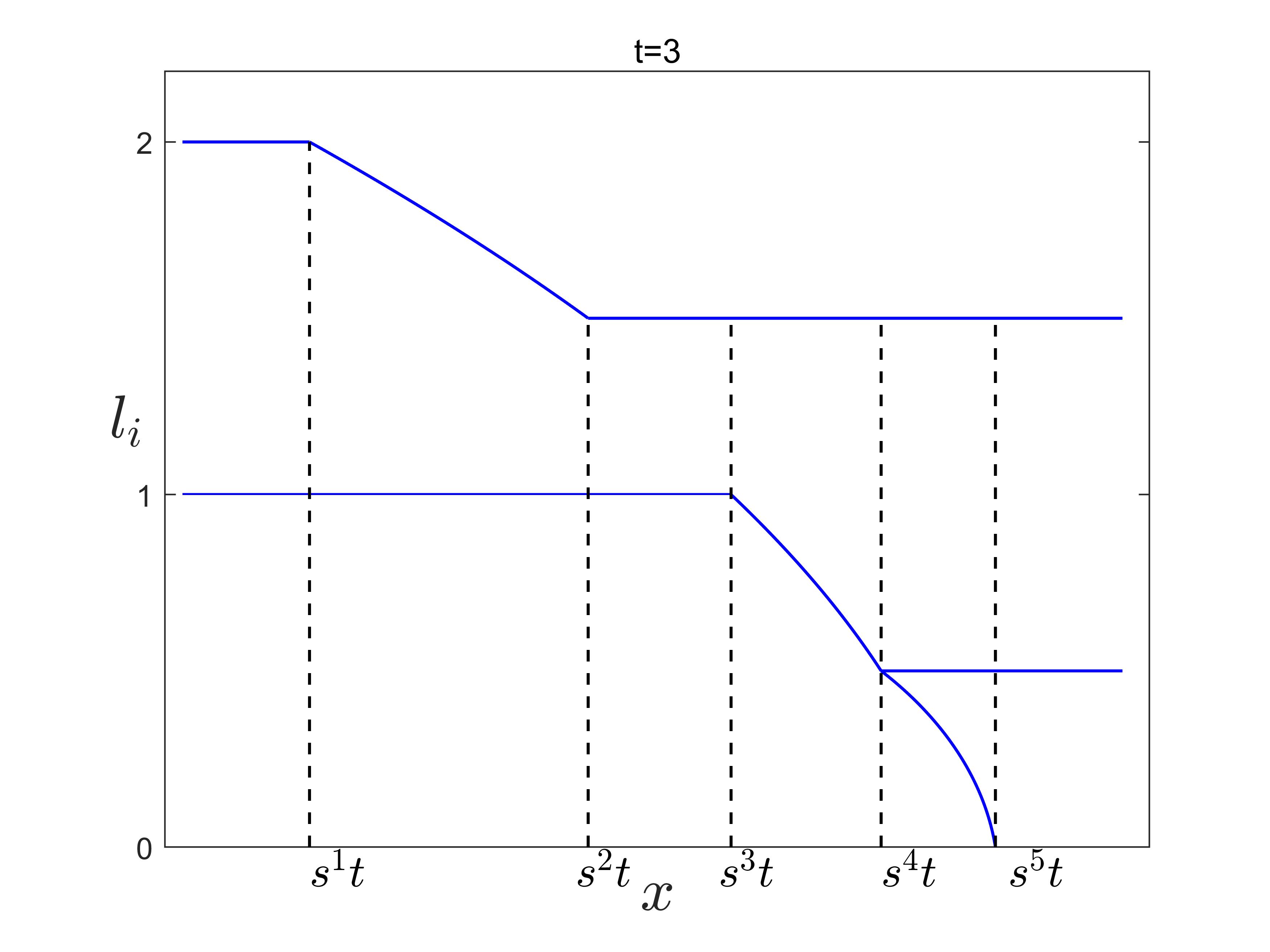}}
\quad
\subfigure[]{\includegraphics[width=0.31\linewidth]{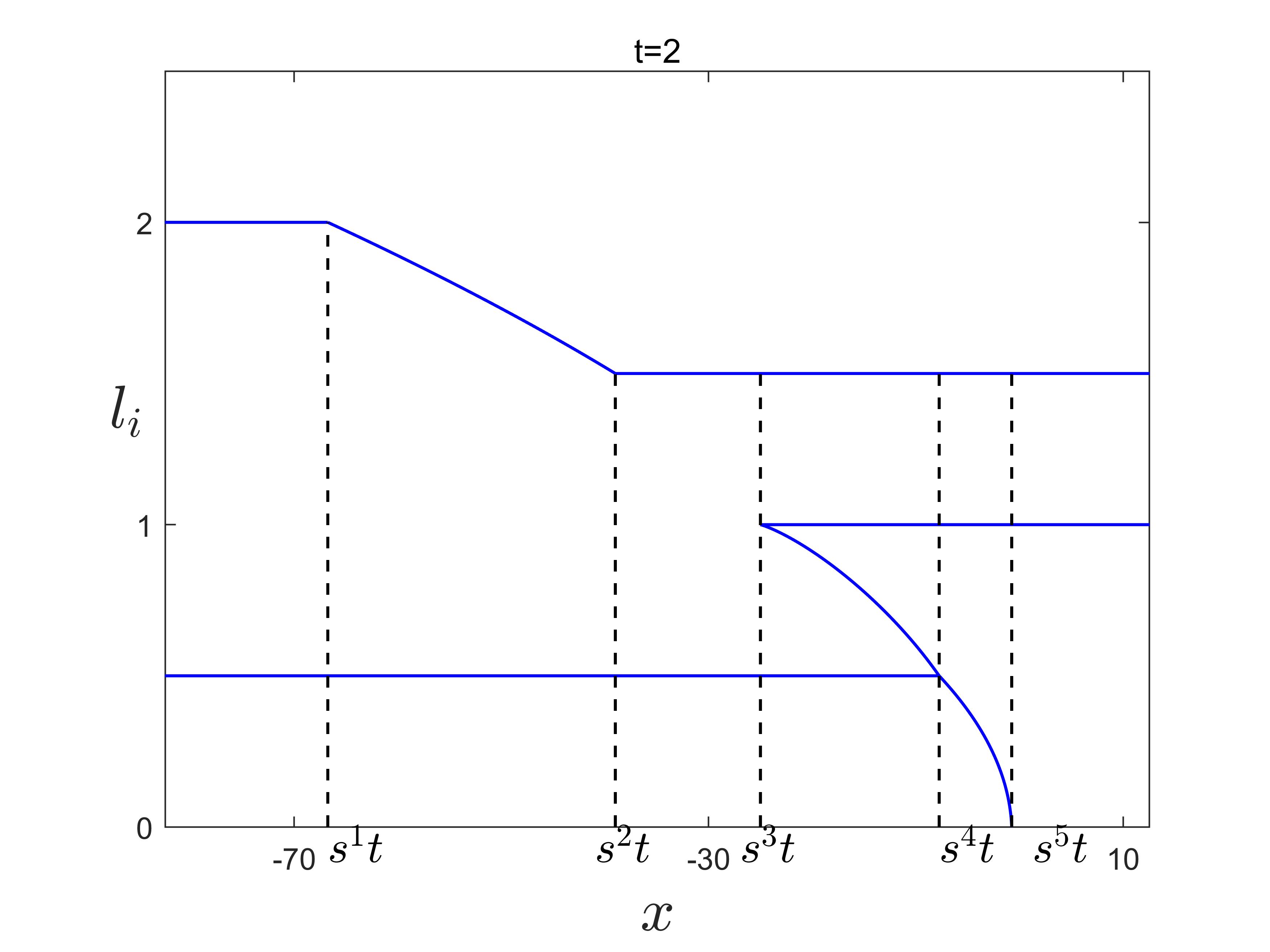}}
\quad
\subfigure[]{\includegraphics[width=0.31\linewidth]{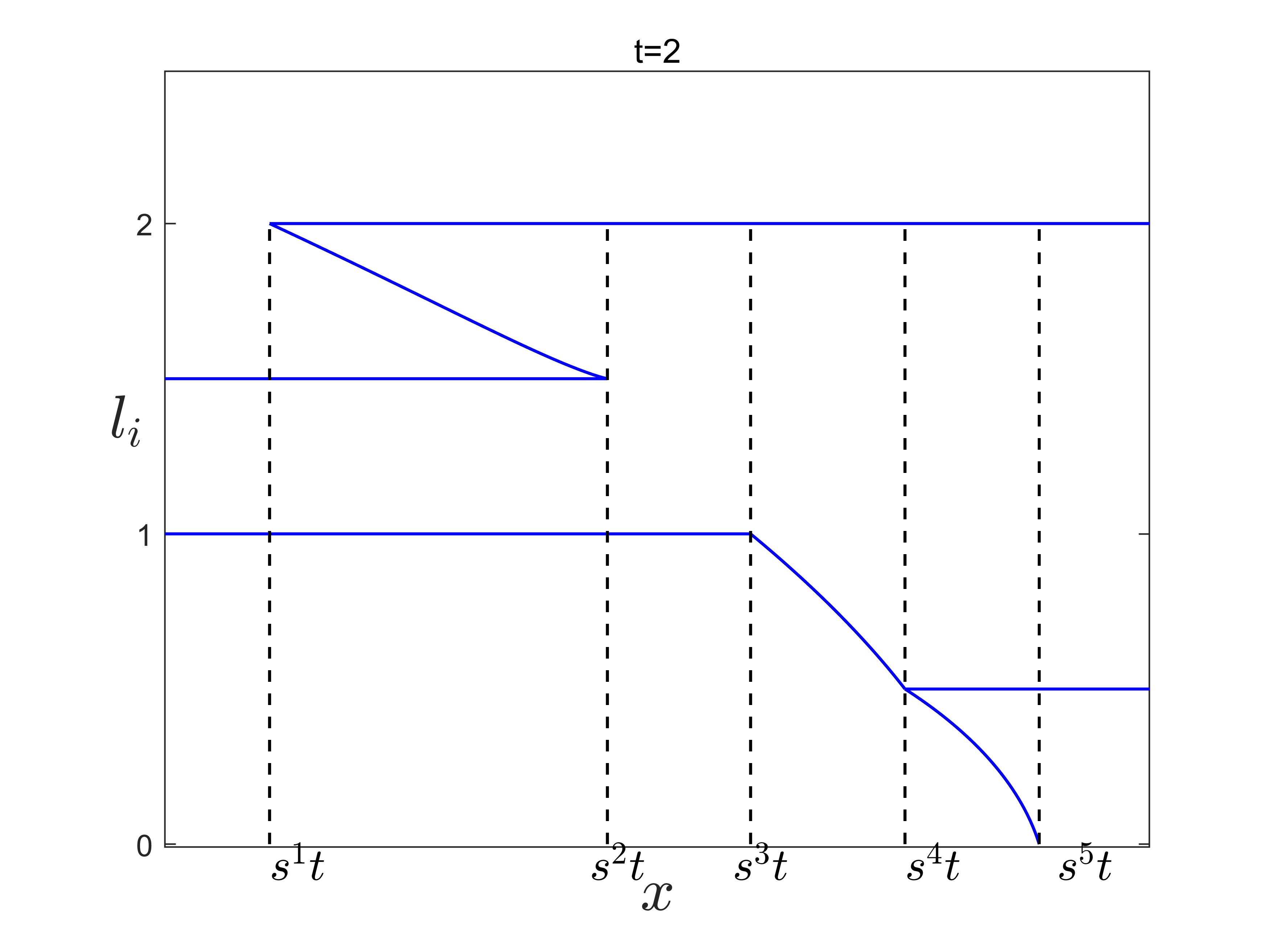}}
\quad
\subfigure[]{\includegraphics[width=0.31\linewidth]{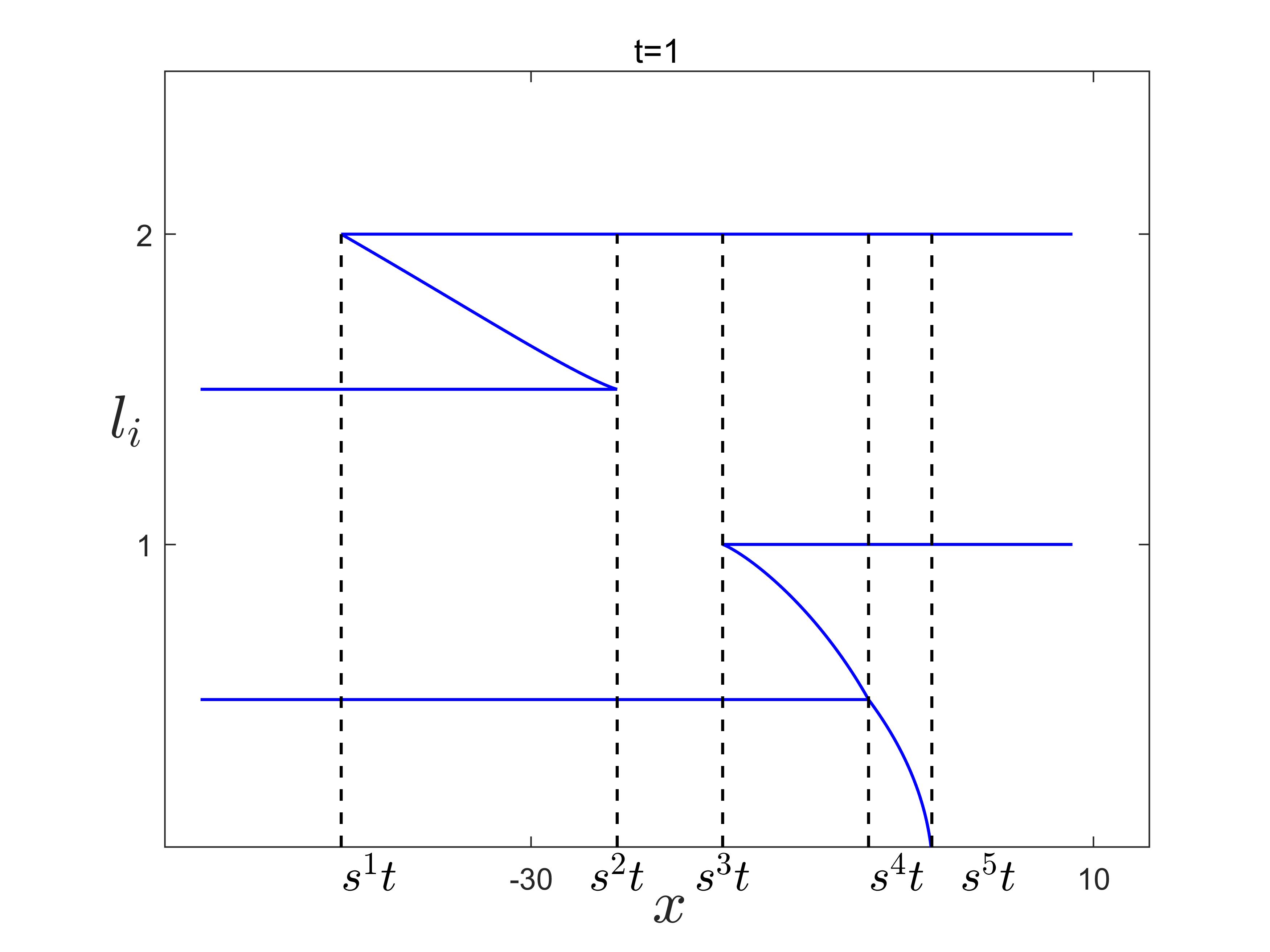}}
\quad
\subfigure[]{\includegraphics[width=0.31\linewidth]{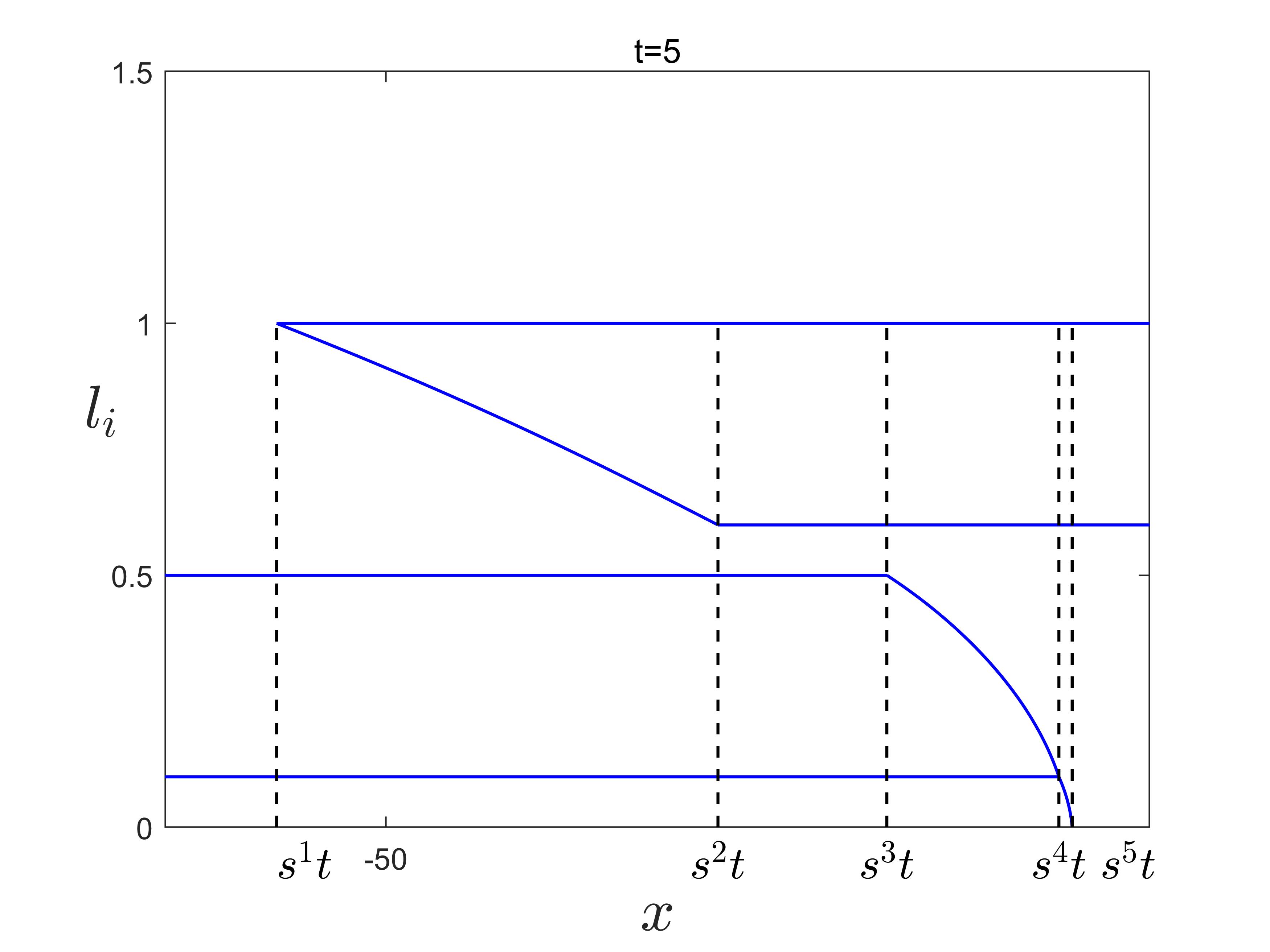}}
\quad
\flushleft{\footnotesize
\textbf{Fig.~$\bm{11}$.} Distributions of Riemann invariants under different initial discontinuous conditions corresponding to the six categories from the case when boundary points (corresponding to the left and right states) lie on different side of the axis $\rho = 2\nu$.}
\end{figure}
\begin{figure}[htbp]
\centering
\setcounter{subfigure}{0}
\subfigure[]{\includegraphics[width=0.31\linewidth]{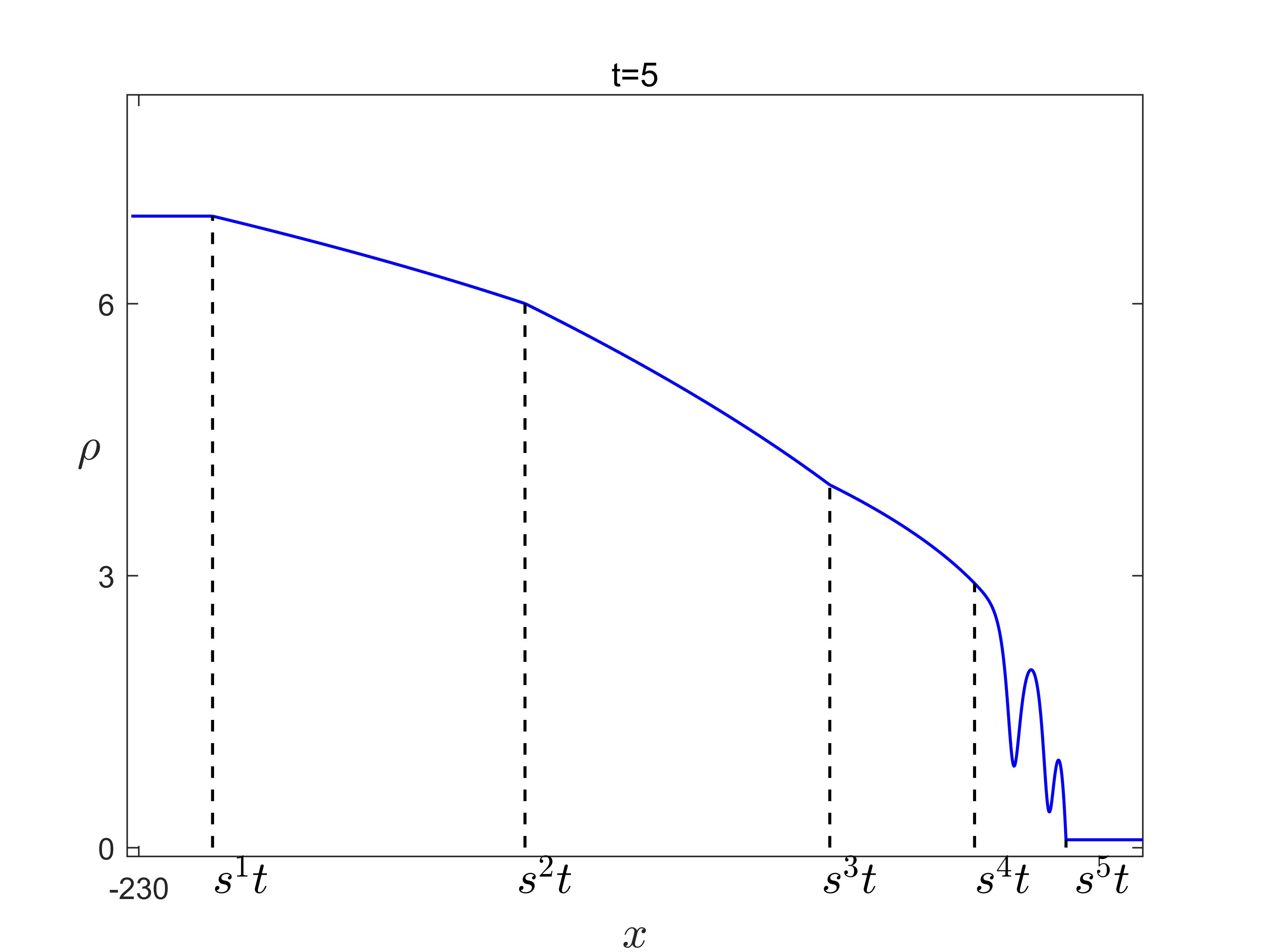}}
\quad
\subfigure[]{\includegraphics[width=0.31\linewidth]{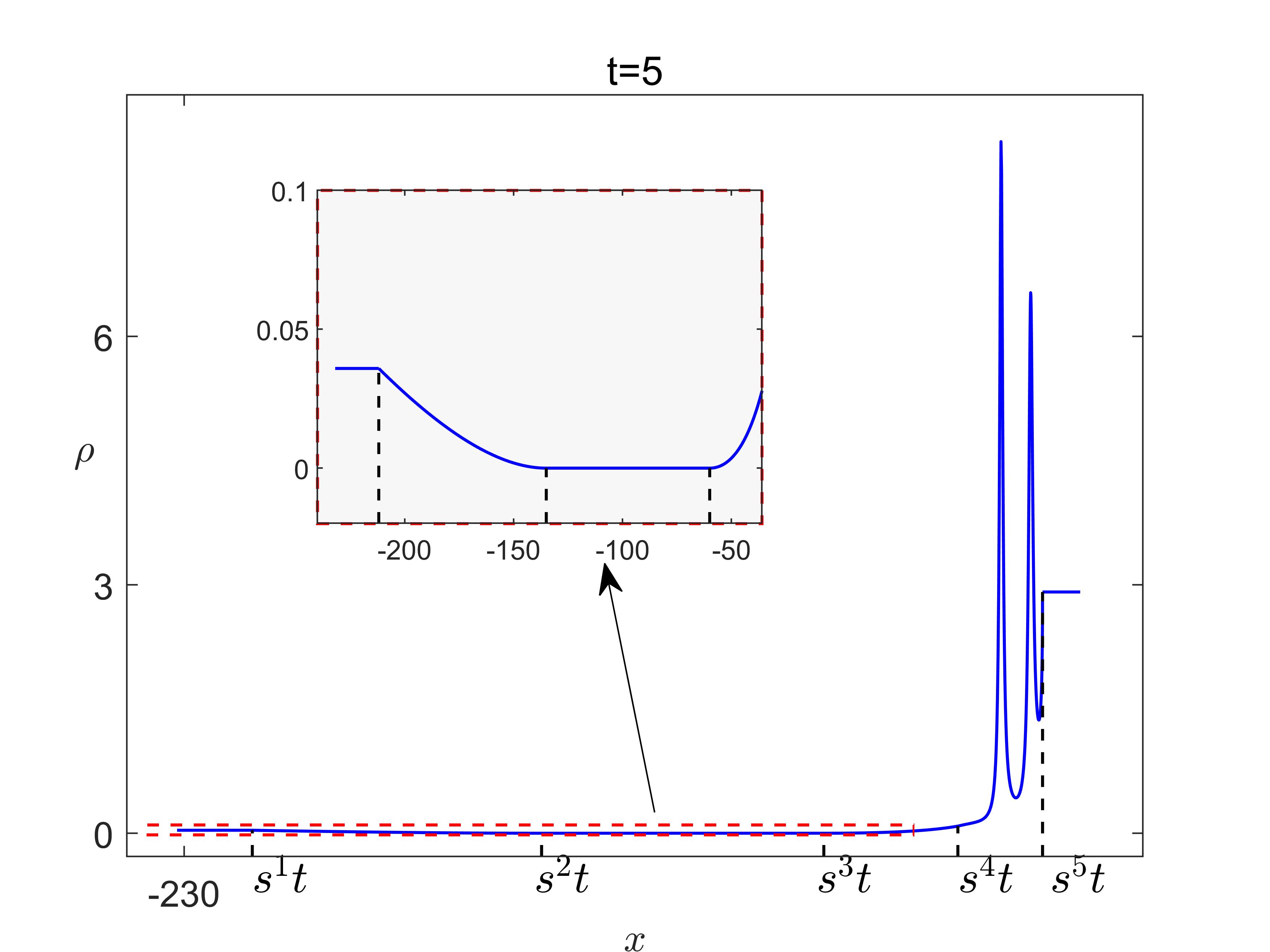}}
\quad
\subfigure[]{\includegraphics[width=0.31\linewidth]{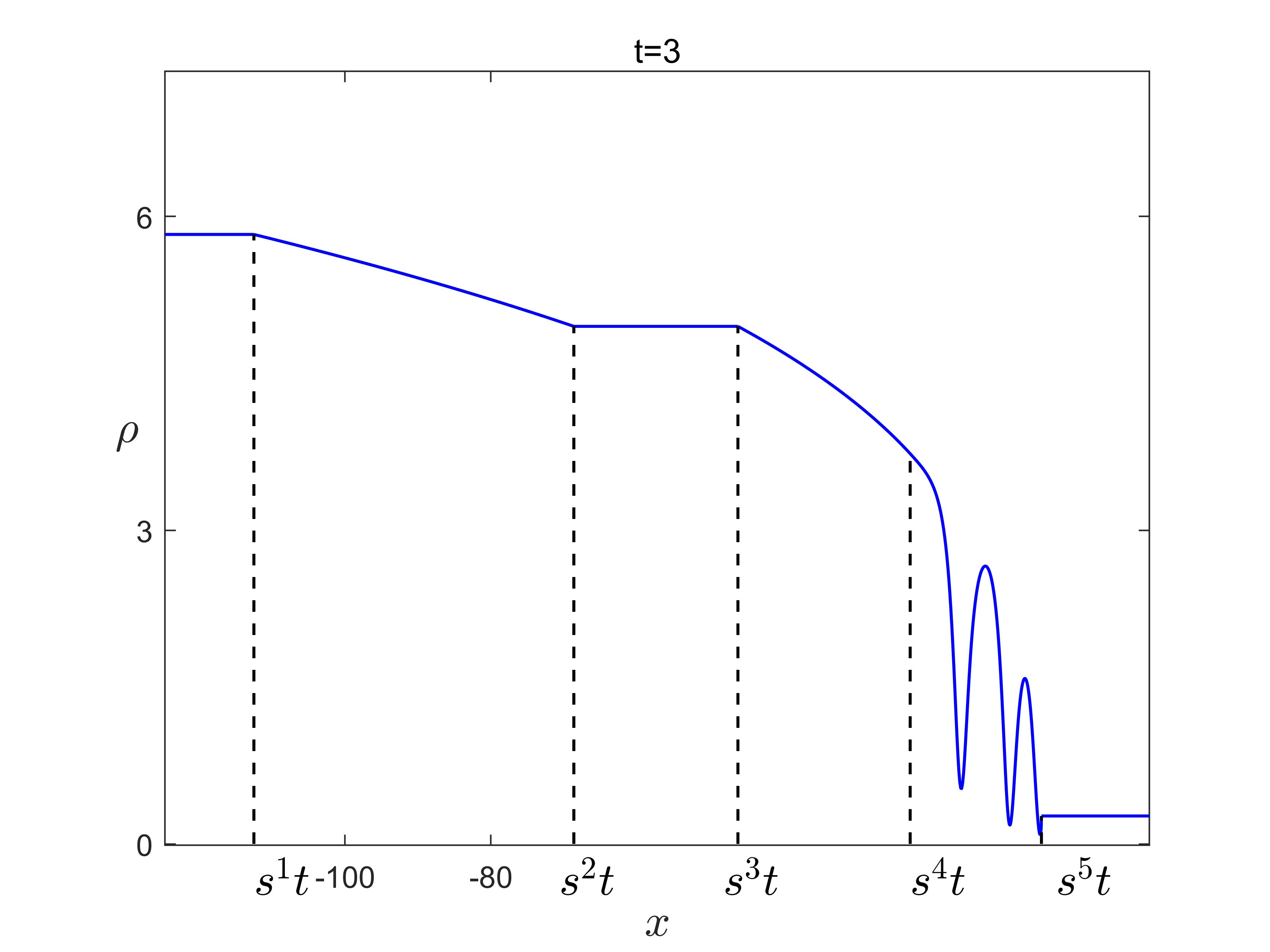}}
\quad
\subfigure[]{\includegraphics[width=0.31\linewidth]{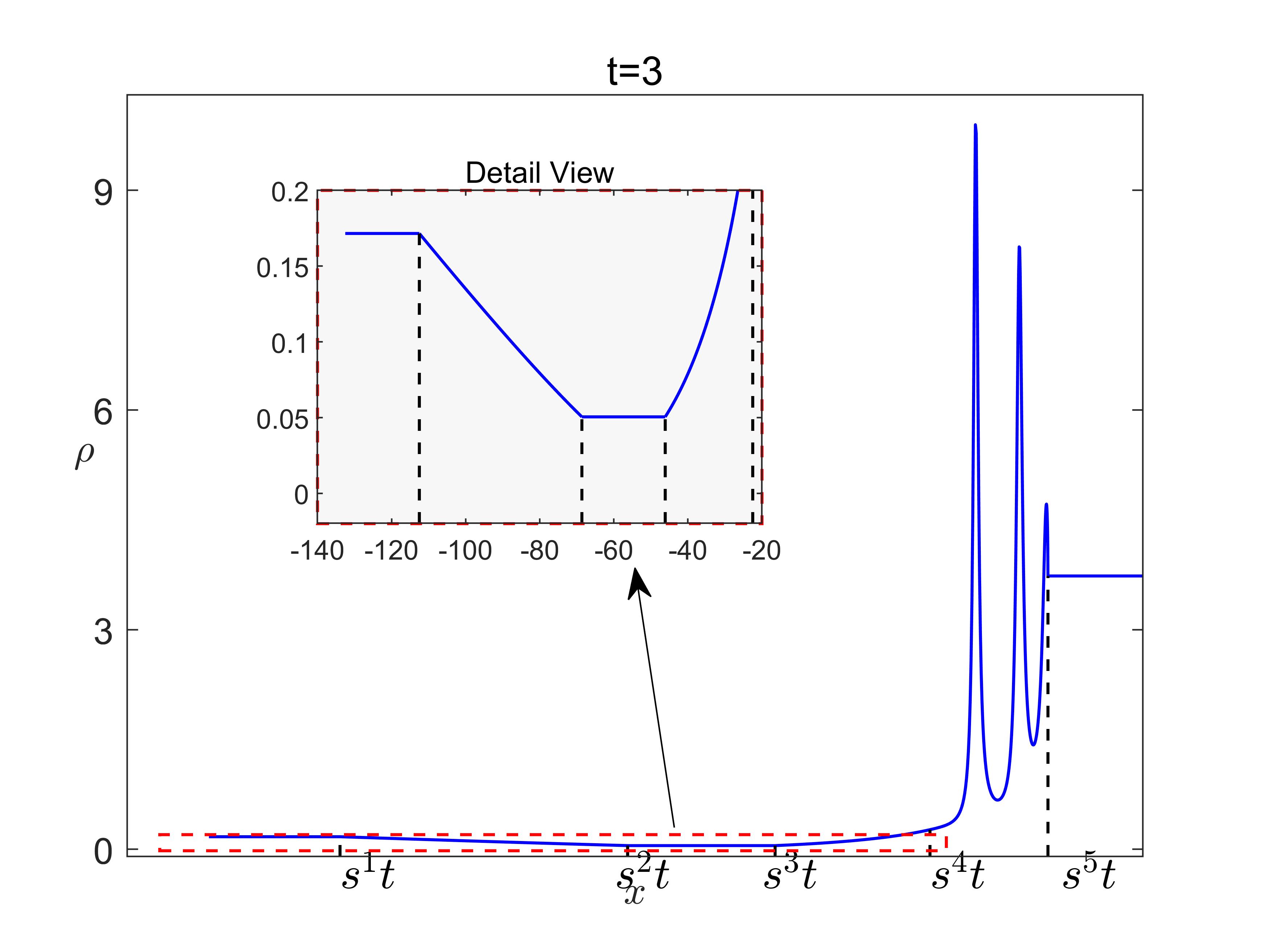}}
\quad
\subfigure[]{\includegraphics[width=0.31\linewidth]{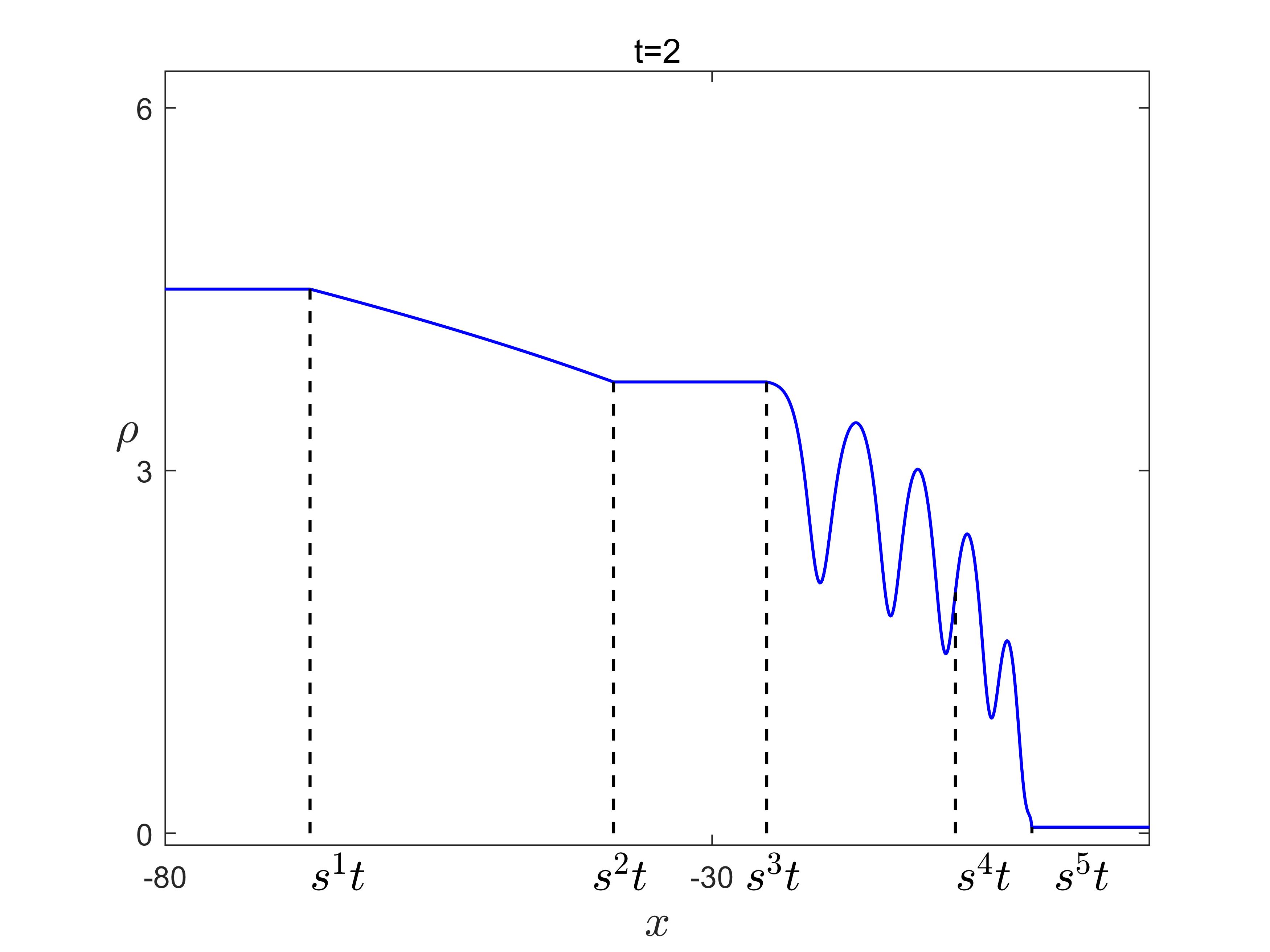}}
\quad
\subfigure[]{\includegraphics[width=0.31\linewidth]{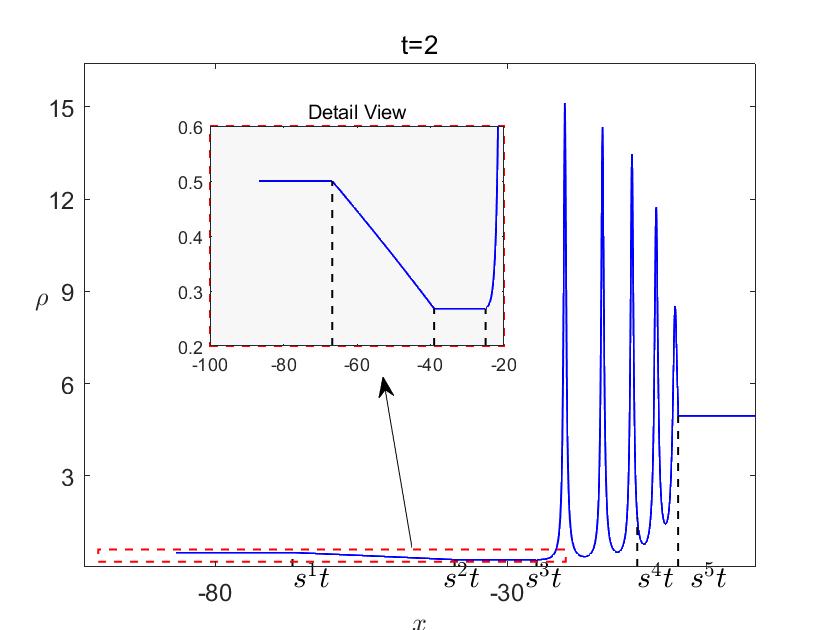}}
\quad
\subfigure[]{\includegraphics[width=0.31\linewidth]{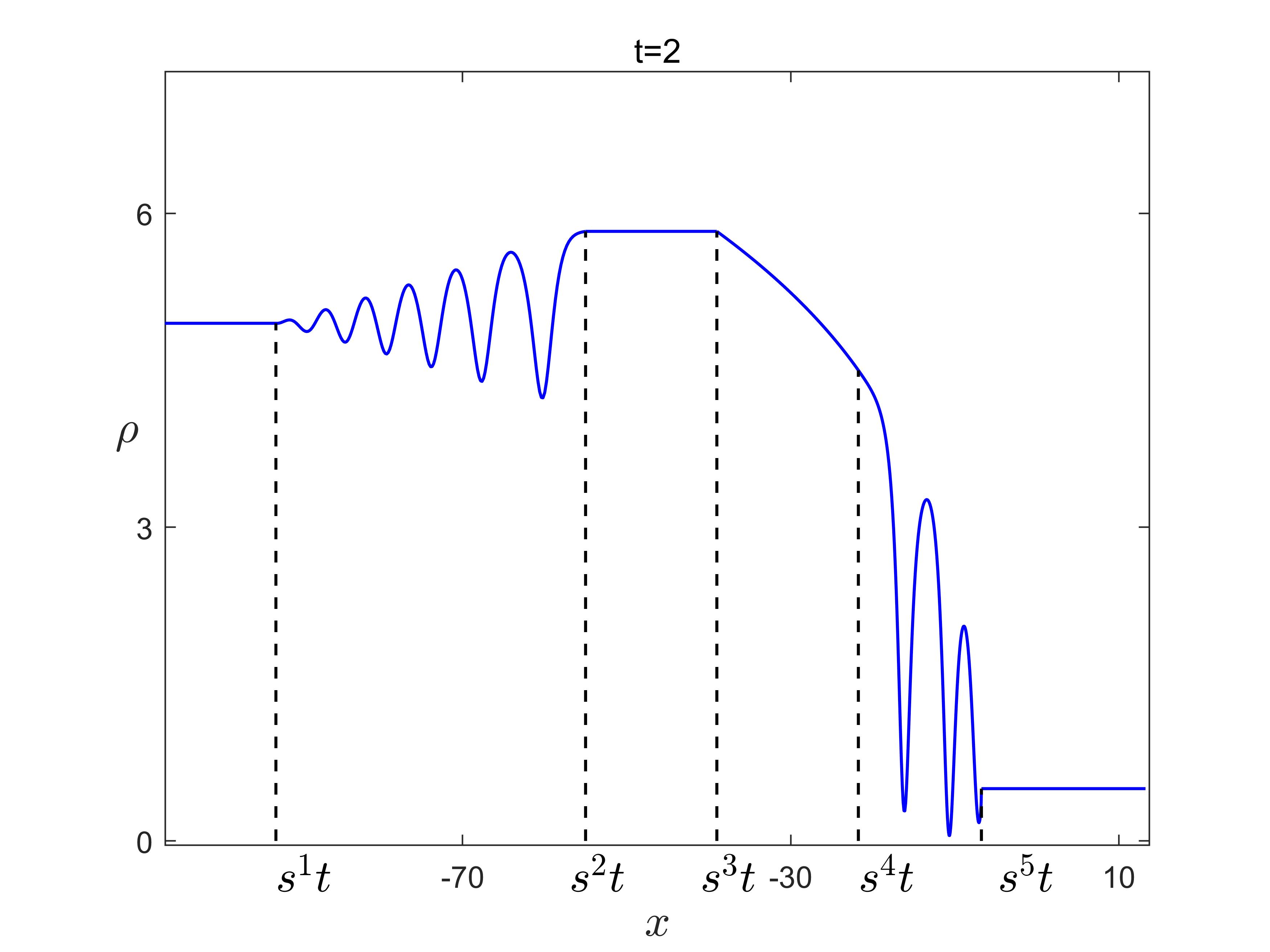}}
\quad
\subfigure[]{\includegraphics[width=0.31\linewidth]{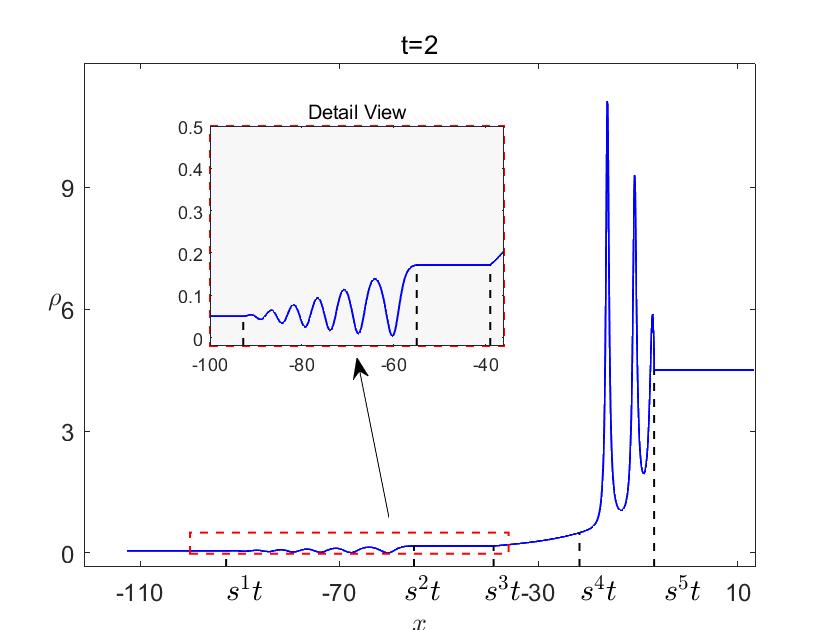}}
\quad
\subfigure[]{\includegraphics[width=0.31\linewidth]{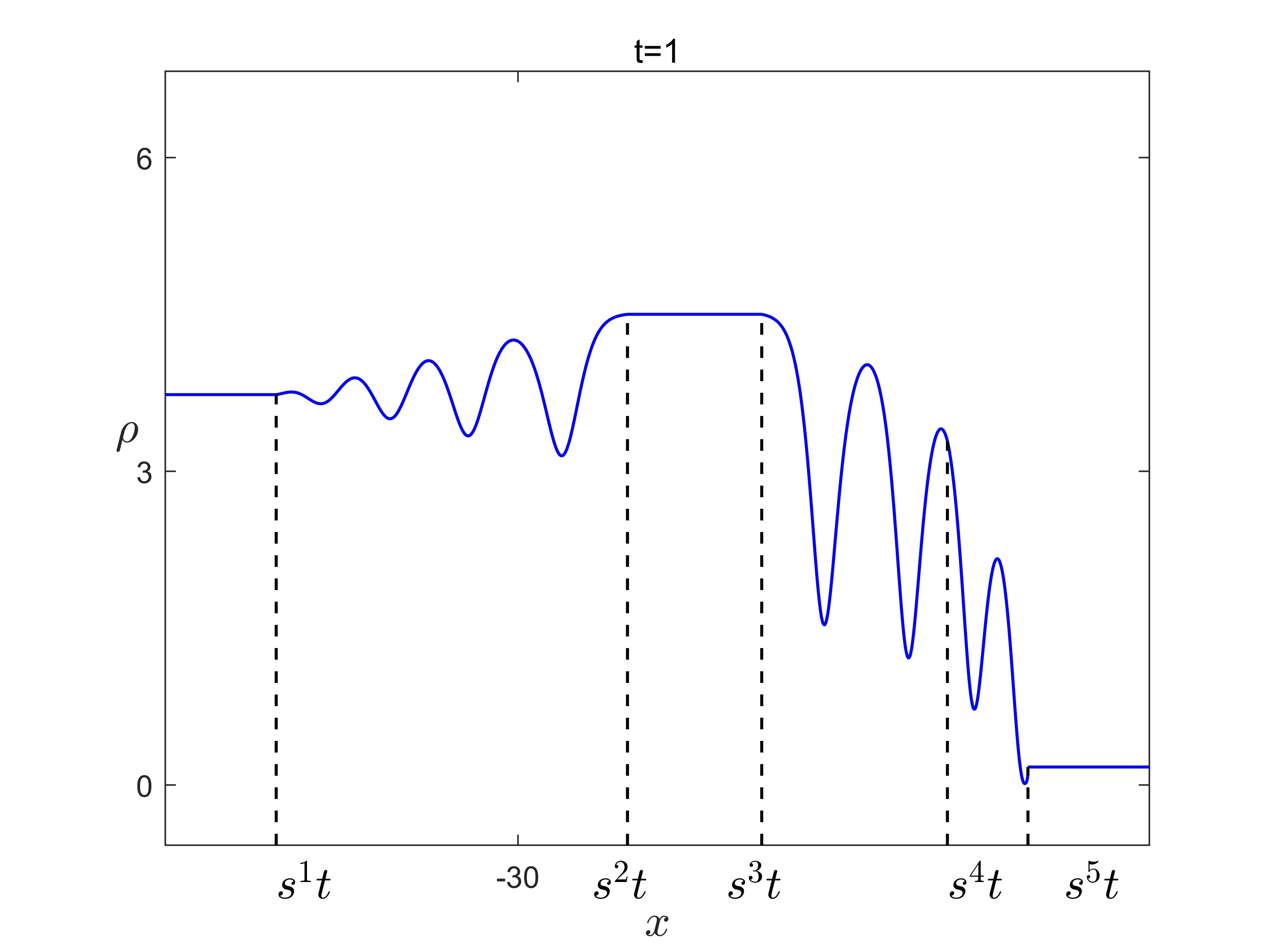}}
\quad
\subfigure[]{\includegraphics[width=0.31\linewidth]{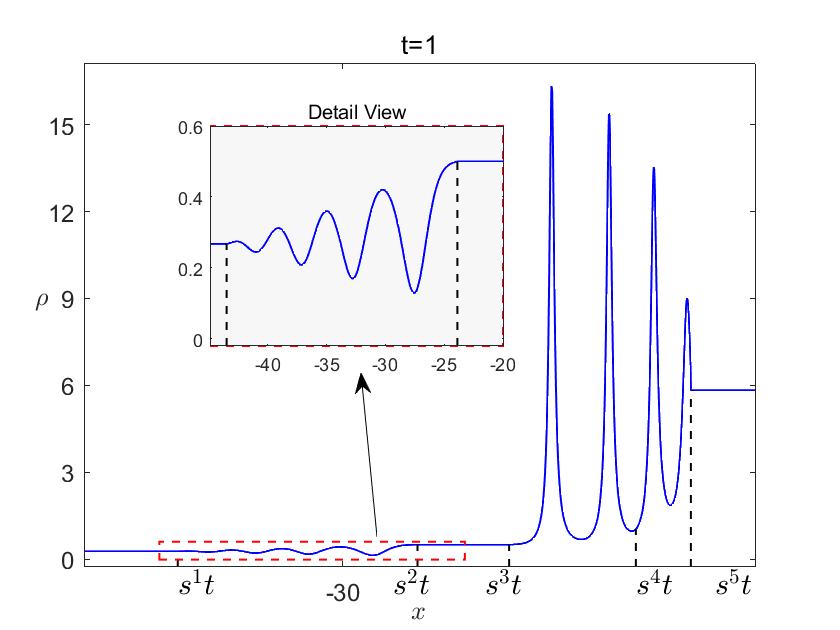}}
\quad
\subfigure[]{\includegraphics[width=0.31\linewidth]{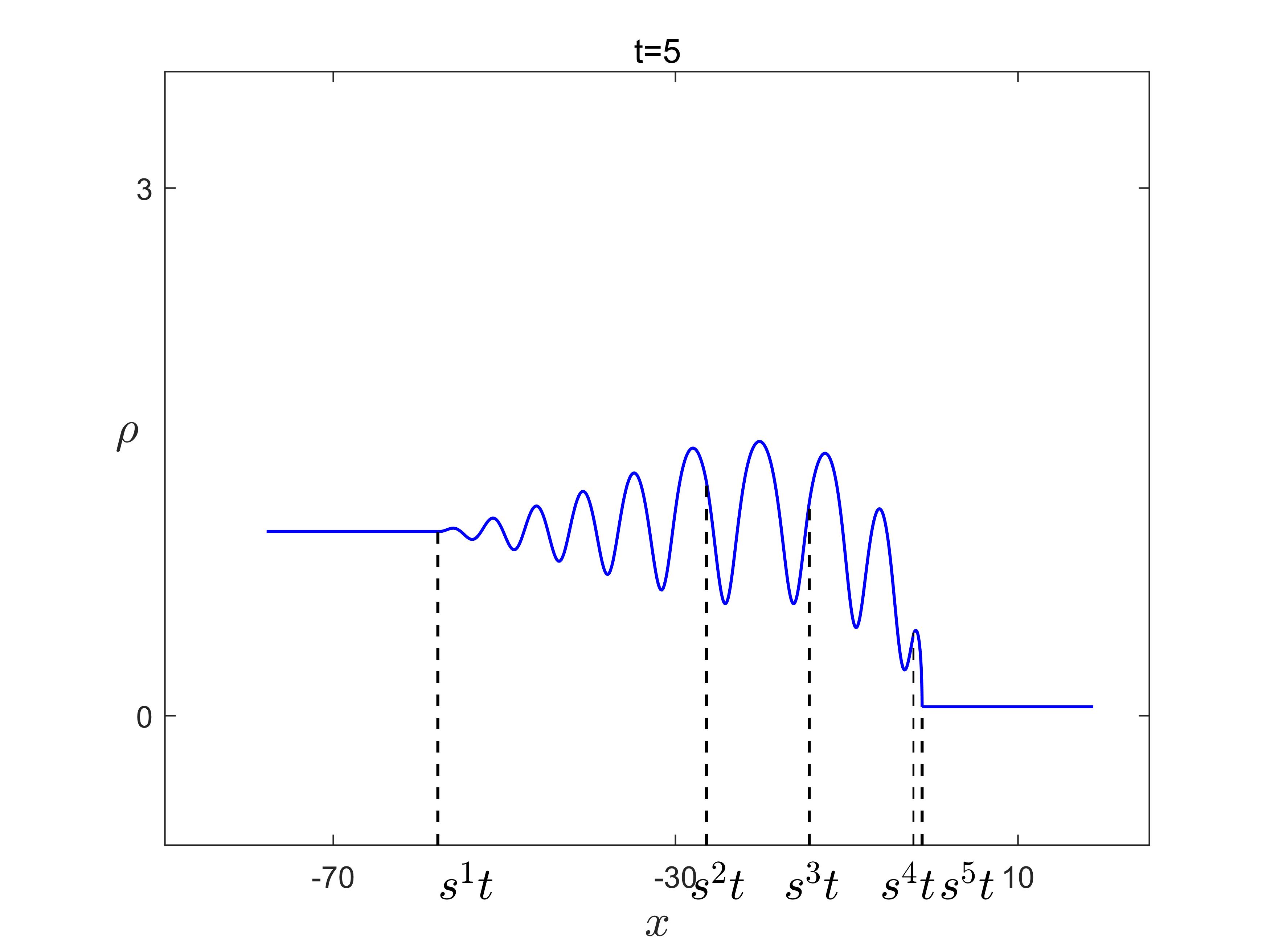}}
\quad
\subfigure[]{\includegraphics[width=0.31\linewidth]{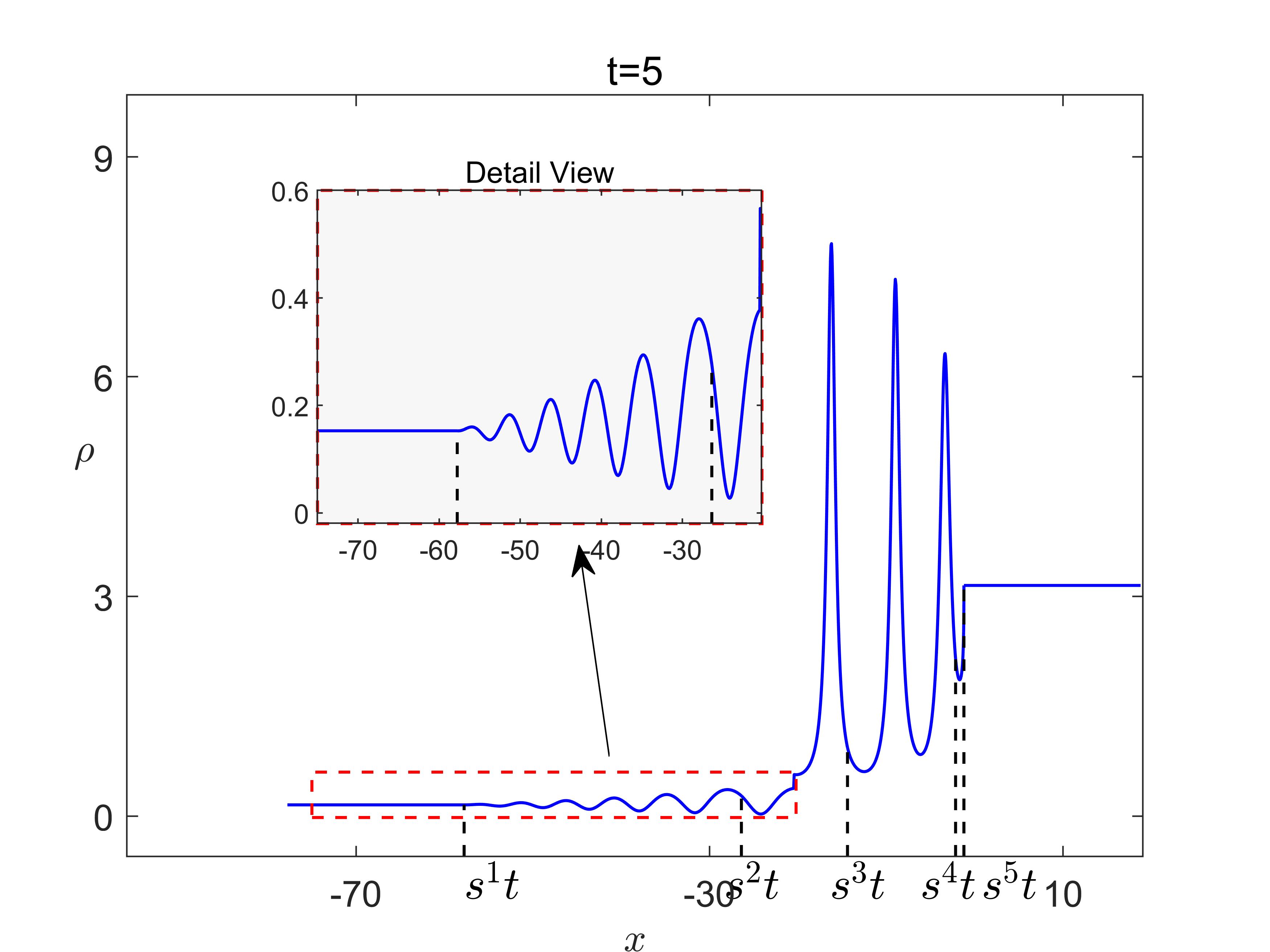}}
\quad
\flushleft{\footnotesize
\textbf{Fig.~$\bm{12}$.} Evolution of the twelve density waves with respect to $x$ corresponding to the six Riemann invariant structures in Fig.\ 9, where one Riemann invariant structure corresponds to two density waves via Eq. (2.30) and (2.31). The embedded sub-figures represent the magnified views enclosed by red boxes.}
\end{figure}

Case E shows that two DSWs are connected by a platform (as shown in Fig.\ 9(e)), which correspond to two density cases. Region II and Region IV are the DSW regions, and the remaining parts are platform regions. In Fig.\ 10(i), the left boundary of Region II and the right boundary of Region IV are the edges of small-amplitude harmonics, while the other side of them are the edges of soliton. In Fig.\ 10(j), the edges of DSW are the same as those in Fig.\ 10(i), but the bright soliton in Region IV becomes a dark soliton. The velocities corresponding to the respective boundaries are
\begin{equation}
\begin{aligned}
\small s^1=-\frac{1}{2}&\left( 3l_{-}^{L2}+3l_{+}^{L2}+4l_{+}^{L}l_{+}^{R}+8l_{+}^{R2}+2l_{-}^{L}\left( 2l_{+}^{L}+l_{+}^{R} \right) \right) +\frac{4\left( l_{-}^{L}-l_{+}^{R} \right) \left( l_{+}^{R}-l_{+}^{L} \right) \left( l_{-}^{L}+l_{+}^{L}+4l_{+}^{R} \right)}{l_{-}^{L}+l_{+}^{L}-2l_{+}^{R}},\\
&s^2=\frac{1}{2}\left( -3l_{-}^{L2}-8l_{+}^{L2}-4l_{+}^{L}l_{+}^{R}-3l_{+}^{R2}-2l_{-}^{L}\left( 2l_{+}^{L}+l_{+}^{R} \right) \right) ,\\
&s^3=-\frac{1}{2}\left( -3l_{-}^{L2}-8l_{-}^{R2}-4l_{-}^{R}l_{+}^{R}-3l_{+}^{R2}-2l_{-}^{L}\left( 2l_{-}^{R}+l_{+}^{R} \right) \right) ,\\
s^4&=-\frac{48l_{-}^{L3}-6l_{-}^{L}\left( l_{-}^{R}-l_{+}^{R} \right) ^2-24l_{-}^{L2}\left( l_{-}^{R}+l_{+}^{R} \right) -3\left( l_{-}^{R}-l_{+}^{R} \right) ^2\left( l_{-}^{R}+l_{+}^{R} \right)}{4l_{-}^{L}-2\left( l_{-}^{R}+l_{+}^{R} \right)}.
\end{aligned}
\end{equation}

In Case F, as shown in Fig.\ 9(f), there are still two DSWs. The corresponding density profiles are shown in Fig.\ 10(i) and Fig.\ 10(j). It is worth noting that the platform region (Region III) that existed in Case E has disappeared and is replaced by a nonlinear wave that can be represented by an elliptic cosine wave. For each boundary, the velocities are
\begin{equation}
\begin{aligned}
\small s^1=-\frac{1}{2}&\left( 3l_{-}^{R2}+3l_{+}^{L2}+4l_{+}^{L}l_{+}^{R}+8l_{+}^{R2}+2l_{-}^{R}\left( 2l_{+}^{L}+l_{+}^{R} \right) \right) +\frac{4\left( l_{-}^{R}-l_{+}^{R} \right) \left( l_{+}^{R}-l_{+}^{L} \right) \left( l_{-}^{R}+l_{+}^{L}+4l_{+}^{R} \right)}{l_{-}^{R}+l_{+}^{L}-2l_{+}^{R}},\\
&s^2=v_3\left( l_{-}^{L},\ l_{+}^{L},\ l_{-}^{R},\ l_{+}^{R} \right) ,\ s^3=v_2\left( l_{-}^{L},\ l_{+}^{L},\ l_{-}^{R},\ l_{+}^{R} \right) ,\\
s^4=&-\frac{48l_{-}^{L3}-6l_{-}^{L}\left( l_{-}^{R}-l_{+}^{R} \right) ^2-24l_{-}^{L2}\left( l_{-}^{R}+l_{+}^{R} \right) -3\left( l_{-}^{R}-l_{+}^{R} \right) ^2\left( l_{-}^{R}+l_{+}^{R} \right)}{4l_{-}^{L}-2\left( l_{-}^{R}+l_{+}^{R} \right)}.
\end{aligned}
\end{equation}

For the six cases mentioned above, we have conducted numerical simulations after assigning specific values to the initial conditions. As can be seen in Fig.\ 10, the results of the numerical simulation are basically consistent with those of the theoretical analysis. The slight oscillations at the edges mainly stem from the limitations of numerical methods near the discontinuity surfaces.

Finally, we consider the scenarios where the boundary points lie on either side of the line $\rho = 2\nu$. As we discussed earlier, when the boundary points are located in different monotonicity regions, a combined shock wave structure will emerge. The classification method for these cases is the same as that for the same monotonicity region (as shown in Fig.\ 11), with the exception that a contact DSW is added. The velocities corresponding to six various cases are consistent with the previous discussion, however, the additional velocity of $s^5$ is expressed as $s^5=-\frac{3\left( l_{-}^{R}-l_{+}^{R} \right) ^2}{2}.$ Fig.\ 12 shows the corresponding density evolution under these six classifications.

\vspace{7mm}\noindent\textbf{6 breaking of the wave with a cubic profile}
\hspace*{\parindent}
\renewcommand{\theequation}{6.\arabic{equation}}\setcounter{equation}{0}\\

In the study of nonlinear waves, the occurrence of wave breaking is often closely related to the local behavior of the initial profile near the breaking point. This section, we further consider another typical case of wave breaking when the local behavior of the initial profile at the breaking point can be approximated by a power function $x^{1/n}$ with the exponent $n$ taking positive integers. And we can use the GP approach to analyze DSW as modulated periodic solution where Whitham equations govern the evolution laws of these modulation parameters. 

From the content in the preceding Section 3, we have derived the Riemann invariants. For simple waves, one of their Riemann invariants must remain constant. Based on this, we choose the Riemann invariant $l_-$ as a constant, which then naturally satisfies the second equation of Eqs. (3.4). The evolution equation of the other Riemann invariant $l_+$ transforms to the Hopf equation, which has a solution of the following form
\begin{equation}
\begin{aligned}
x+\frac{3}{2}\left( 5l_{+}^{2}+2l_+l_-+l_{-}^{2} \right) t=w\left( l_+ \right),
\end{aligned}
\end{equation}
where $w\left( l_+ \right)$ is the inverse function corresponding to the initial profile $l_+$.

We will consider the case that the profile of the Riemann invariants $l_+$ at breaking instant is represented by a cubic root curve. Without loss of generality, establishing a coordinate system such that the wave breaking point occurs at $x=0$ and $t=0$, we thus obtain the solution of Eqs. (3.4)
\begin{equation}
\begin{aligned}
&l_-=l_{-}^{L}=const,\ l_{+}^{L}=const,\\
&x+\frac{3}{2}\left( 5l_{+}^{2}+2l_+l_-+l_{-}^{2} \right) t=\left( l_+-l_{+}^{L} \right) ^3.
\end{aligned}
\end{equation}
The functions of the Riemann invariants $l_\pm$ and the functions of two sets of corresponding densities $\rho$ at different times are illustrated in Fig.\ 13. Consequently, there is no singularity when $t < 0$ while a singularity emerges at $t = 0$ and then a multi-valued region comes into existence.
\begin{figure}[htbp]
\centering
\setcounter{subfigure}{0}
\subfigure[]{\includegraphics[width=0.311\linewidth]{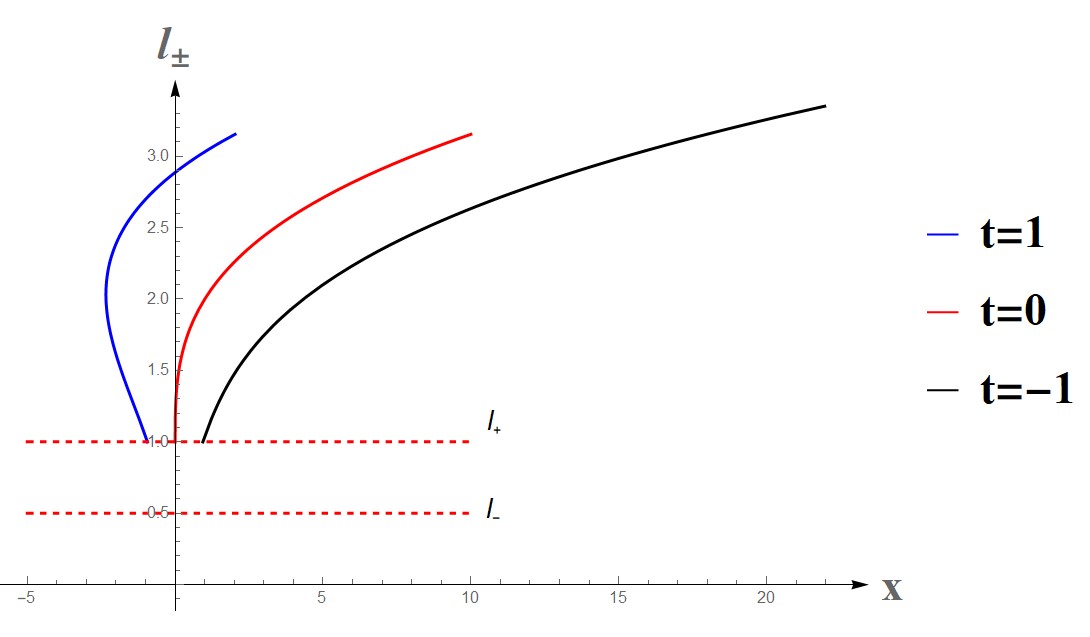}}
\quad
\subfigure[]{\includegraphics[width=0.312\linewidth]{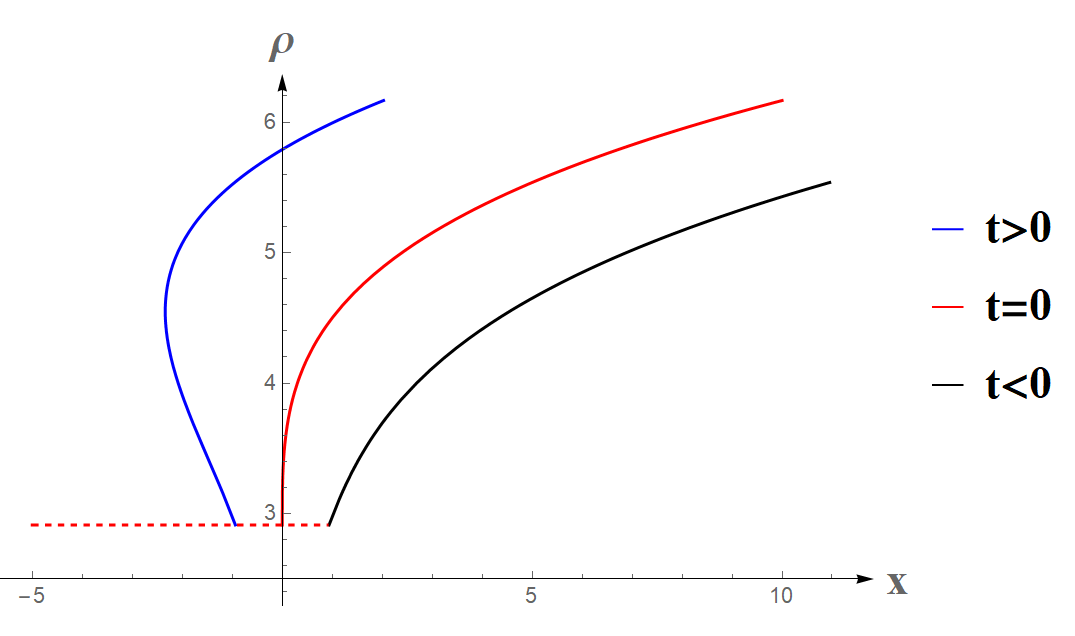}}
\quad
\subfigure[]{\includegraphics[width=0.311\linewidth]{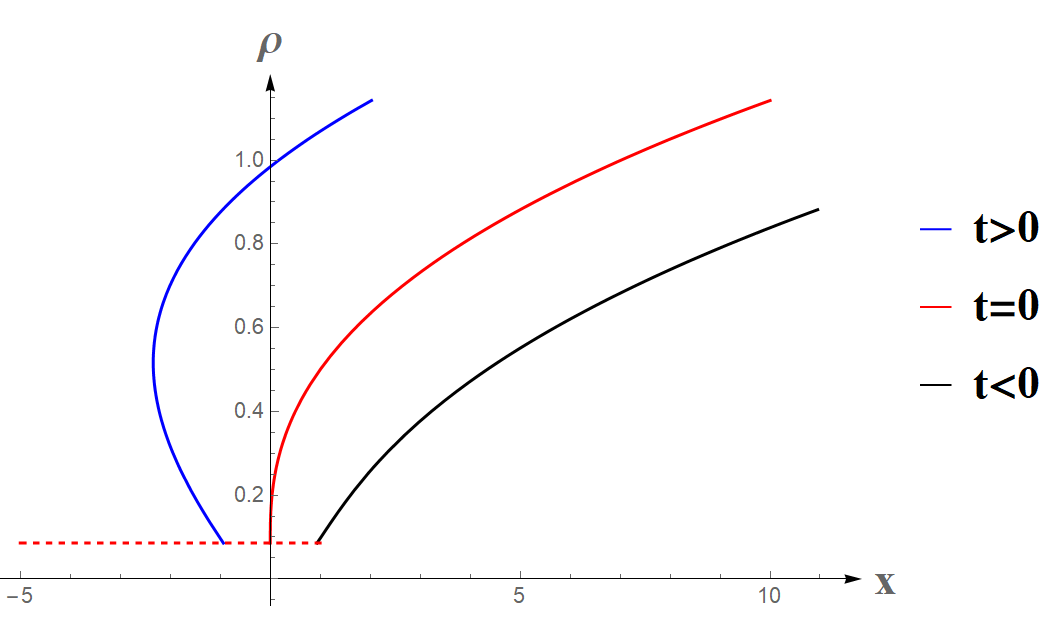}}
\quad
\flushleft{\footnotesize
\textbf{Fig.~$\bm{13}$.} The distributions of the Riemann invariants $l_{\pm}$ and the functions of two sets of corresponding densities $\rho$ under cubic initial case at $t = -1$, $t = 0$, and $t = 1$.}
\end{figure}

Next, we employ the generalized hodograph method to solve the Whitham equations (2.32) and obtain the general form of the solutions
\begin{equation}
x-v_i\left( l \right) t=\omega_i \left( l \right),~i = 1, 2, 3, 4,
\end{equation}
where $v_i(l)$ are Whitham velocities as specified in Eqs. (2.33). By differentiating Eqs. (6.3) with respect to $l_j$ $(i\ne j)$ and eliminating the variable $t$ yields
\begin{equation}
\frac{1}{\omega _i-\omega _j}\frac{\partial \omega _i}{\partial l_j}=\frac{1}{v_i-v_j}\frac{\partial v_i}{\partial l_j},
\end{equation}
as one can see, $\omega _i$ and $v_i$ are symmetric and the expression of  $\omega_i \left( l \right)$
\begin{equation}
\omega _i\left( l \right) =\left( 1-\frac{L}{\partial _iL}\partial _i \right) W.
\end{equation}
Then Eqs. (6.4) are transformed into $(i\ne j)$
\begin{equation}
\frac{\partial ^2W}{\partial l_i\partial l_j}=\frac{1}{2\left( l_i-l_j \right)}\left( \frac{\partial W}{\partial l_j}-\frac{\partial W}{\partial l_j} \right) .
\end{equation}
We can obtain the generating function
\begin{equation}
W=\frac{l^2}{\sqrt{\prod_{i=1}^4{\left( l-l_i \right)}}}=\sum_{k=0}^{\infty}{\frac{W^{\left( k \right)}}{l^k}},
\end{equation}
which the coefficients obtained from expanding $W$ in terms of the inverse powers of $l$ 
\begin{equation}
\begin{aligned}
W^{\left( 0 \right)}&=1,~W^{\left( 1 \right)}=\frac{\eta _1}{2},~W^{\left( 2 \right)}=\frac{3\eta _1^2}{8}-\frac{\eta _2}{2},\\
&W^{\left( 3 \right)}=\frac{\eta _3}{2}+\frac{5\eta _1^3}{16}-\frac{3\eta _1\eta _2}{4},
\end{aligned}
\end{equation}
where,
\begin{equation}
\begin{aligned}
\eta _1=l_1+l_2~+~& l_3+l_4,~\eta _2=l_3~l_4+l_2\left( l_3+l_4 \right) +l_1\left( l_2+l_3+l_4 \right) ,\\
&\eta _3=\left( l_1+l_2 \right) l_3~l_4+l_1~l_2\left( l_3+l_4 \right) .
\end{aligned}
\end{equation}
Therefore, $\omega _i\left( l \right) $ take the forms
\begin{equation}
\omega _i\left( l \right) =\sum_{k=1}^n{C^{\left( k \right)}\omega _{i}^{\left( k \right)}}\left( l \right),
\end{equation}
where
$$
\omega _{i}^{\left( k \right)}\left( l \right) =\left( 1-\frac{L}{\partial _iL}\partial _i \right) W^{\left( k \right)}.
$$
For matching the DSW edges with  dispersionless region, the following conditions must be satisfied:
\begin{equation}
\begin{aligned}
l_2=l_+,\ l_1=l_-,\ at\ m=0,\\
l_4=l_+,\ l_1=l_-,\ at\ m=1.
\end{aligned}
\end{equation}
Choosing the coefficients $C^{(k)}$ to match $l_4$ with $l_+$ at $m=1$ so that $\omega _4=\left( l_{-}^{L}-l_4 \right) ^3$, after a simple calculation yields
\begin{equation}
\begin{aligned}
C^{\left( 0 \right)}&=-\frac{1}{35}\left( 35l_{+}^{L3}+35l_{+}^{L2}l_{-}^{0}-7l_{+}^{L}l_{-}^{02}+l_{-}^{03} \right) ,\\
&C^{\left( 1 \right)}=\frac{2}{35}\left( 35l_{+}^{L2}+14l_{+}^{L}l_{-}^{0}-l_{-}^{02} \right) ,\\
&C^{\left( 2 \right)}=-\frac{8}{35}\left( 7l_{+}^{L}+l_{-}^{0} \right) ,C^{\left( 3 \right)}=\frac{16}{35}.
\end{aligned}
\end{equation}
therefore, the solutions of Whitham equation are as follows
\begin{equation}
x-v_i\left( l \right) t=C^{\left( 3 \right)}\omega _{i}^{\left( 3 \right)}\left( l \right) +C^{\left( 2 \right)}\omega _{i}^{\left( 2 \right)}\left( l \right) +C^{\left( 1 \right)}\omega _{i}^{\left( 1 \right)}\left( l \right) +C^{\left( 0 \right)}\omega _{i}^{\left( 0 \right)}\left( l \right),
\end{equation}
where $i=3, 4$.

Eqs. (6.13) indicate the solutions of $l_3(x, t)$ and $l_4(x, t)$. The dependencies of Riemann invariants $l_+$ and $l_-$ in relation to $x$ are depicted in Fig.\ 14(a). Fig.\ 14(b) and Fig.\ 14(d) demonstrate the behaviors of two parameter sets $\rho_i$ of periodic solutions with regard to $x$. Substituting the four Riemann invariants into periodic solutions of the HOCLL equation, the DSWs are shown in Fig.\ 14(c) and Fig.\ 14(e) and the numerical results are consistent with theoretical solutions.

\begin{figure}[H]
\centering
\setcounter{subfigure}{0}
\subfigure[]{\includegraphics[width=0.31\linewidth]{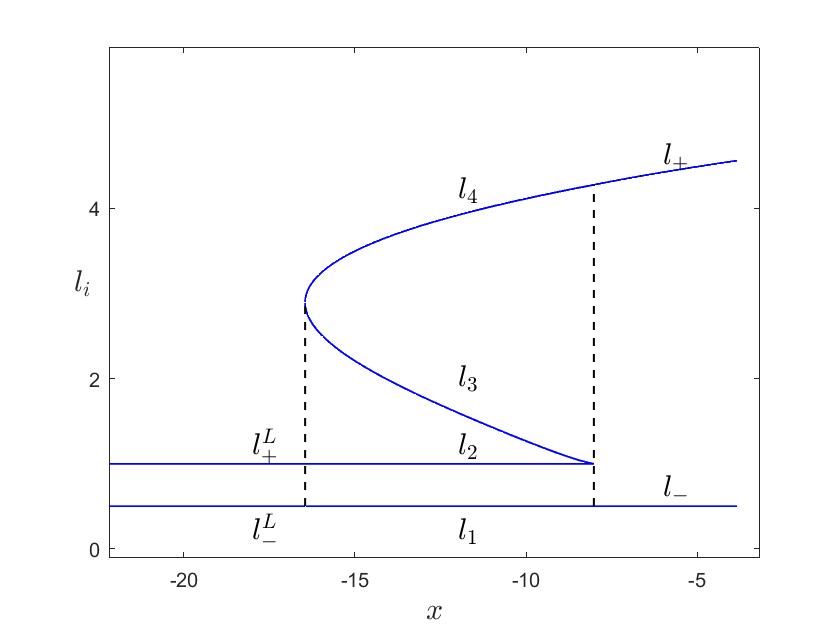}}
\quad
\subfigure[]{\includegraphics[width=0.31\linewidth]{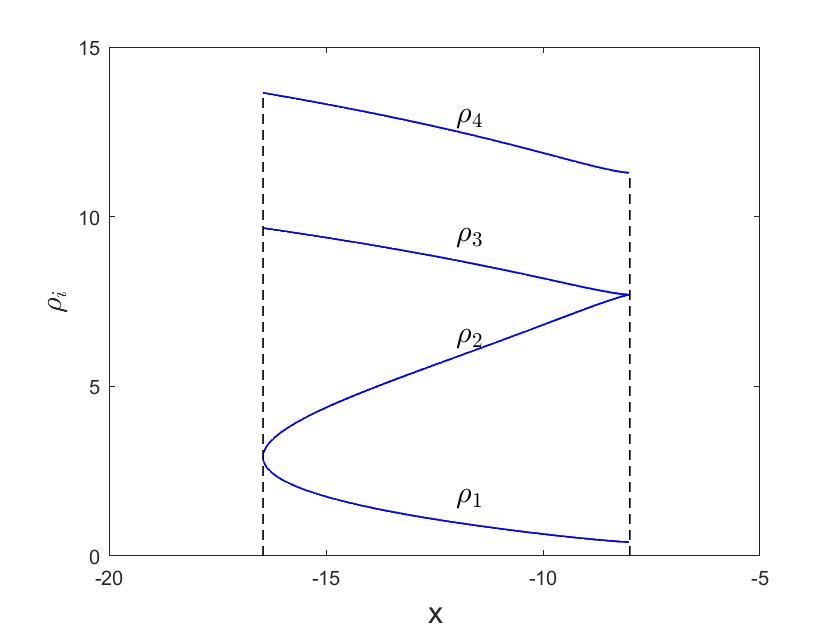}}
\quad
\subfigure[]{\includegraphics[width=0.31\linewidth]{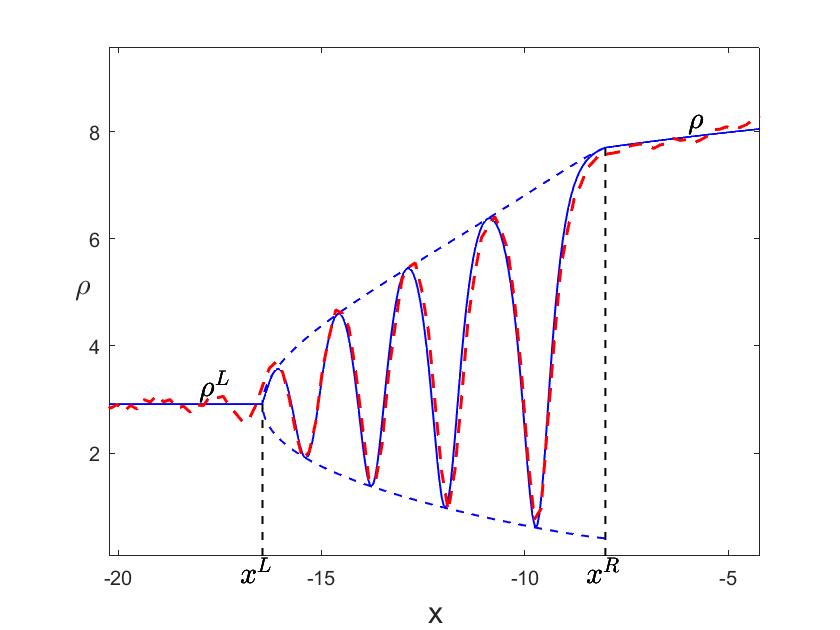}}
\quad
\subfigure[]{\includegraphics[width=0.31\linewidth]{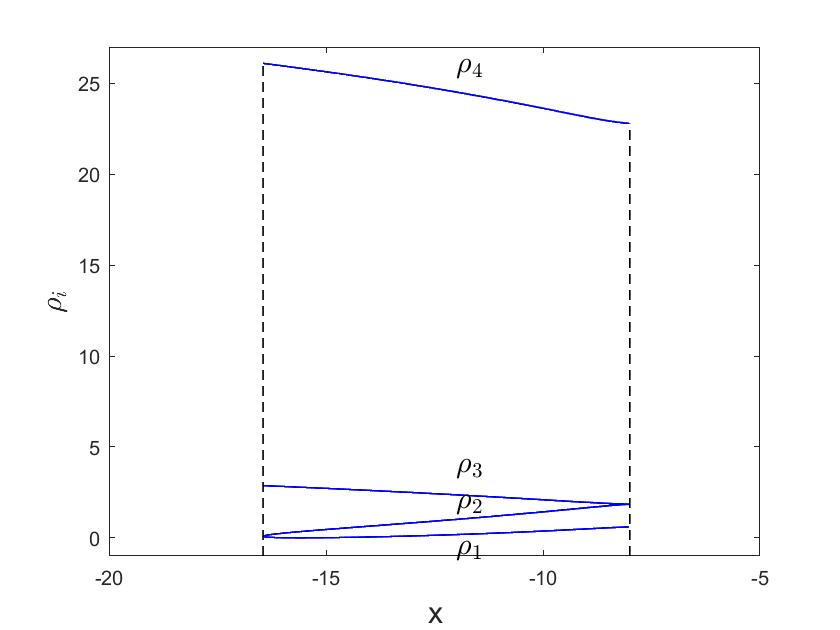}}
\quad
\subfigure[]{\includegraphics[width=0.31\linewidth]{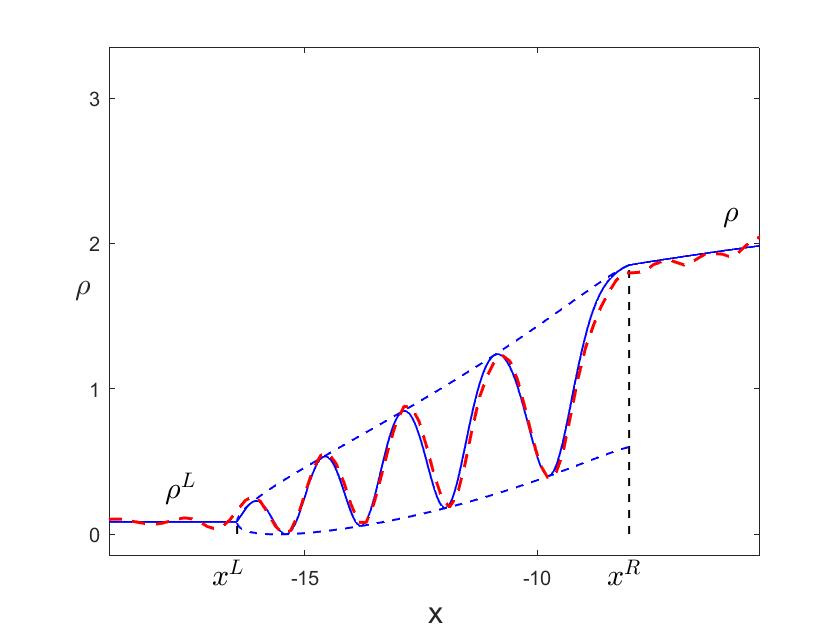}}
\flushleft{\footnotesize
\textbf{Fig.~$\bm{14}$.} The DSW solutions of initial cubic profile at $t = 0.3$. (a) The distributions of the Riemann invariants $l_{\pm}$. (b) (d) Dependence of the two sets of parameters $\rho_i$ on $x$ in periodic solutions corresponding to Riemann invariants. (c) (e) Two sets of analytical solutions (blue solid line) and the numerical simulation (red dashed line) results for the DSW.}
\end{figure}

Further, in the limit of $l_3 \rightarrow  l_2$ in Eqs. (6.13), the expression of $l_4$ can be represented as a function of $t$
\begin{equation}
l_4\left( t \right) =l_{+}^{L}+\frac{7t}{2}+\frac{1}{2}\sqrt{\frac{7}{3}}\sqrt{t\left( 4l_{-}^{L}+20l_{+}^{L}+21t \right)},
\end{equation}
then we can obtain the law at the soliton edge of the DSW
\begin{equation}
x^R\left( t \right) =-\frac{49t^3}{4}-t\left( \frac{3l_{-}^{L2}}{2}+3l_{-}^{L}l_{+}^{L}+\frac{15l_{+}^{L2}}{2}+\frac{7l_{-}^{L}t+35l_{+}^{L}t}{2} \right) -\frac{1}{12}\sqrt{\frac{7}{3}}\left( t\left( 4l_{-}^{L}+20l_{+}^{L}+21t \right) \right) ^{\frac{3}{2}}.
\end{equation}

In the small-amplitude limit, i.e. $l_3 \rightarrow l_4$, the law of $x = x^L(t)$ 
\begin{equation}
x+(12l_{4}^{2}-\frac{2}{3}\left( l_{-}^{L}-l_{+}^{L} \right) ^2)t=-\frac{16\left( l_4-l_{+}^{L} \right) ^3\left( 8l_4-7l_{-}^{L}-l_{+}^{L} \right)}{35\left( -2l_4+l_{-}^{L}+l_{+}^{L} \right)},
\end{equation}
calculating the maximum of the function $x(l_4)$, the time $t$ depending on $l_4$ presented as
\begin{equation}
t=-\frac{8\left( l_4-l_{+}^{L} \right) ^2\left( 16l_{4}^{2}+7l_{-}^{L2}+6l_{-}^{L}l_{+}^{L}+3l_{+}^{L2}-4l_4\left( 5l_{-}^{L}+3l_{+}^{L} \right) \right)}{35\left( 16l_{4}^{3}-16l_{4}^{2}\left( l_{-}^{L}+l_{+}^{L} \right) +\left( l_{-}^{L}-l_{+}^{L} \right) ^2\left( l_{-}^{L}+l_{+}^{L} \right) +4l_4\left( l_{-}^{L}+l_{+}^{L} \right) ^2 \right)},
\end{equation}
therefore, the law of motion of the small-amplitude edge is
\begin{equation}
\begin{aligned}
\small &x^L=-\frac{4\left( l_4-l_{+}^{L} \right) ^2}{35\left( 16l_{4}^{3}-16l_{4}^{2}\left( l_{-}^{L}+l_{+}^{L} \right) +\left( l_{-}^{L}-l_{+}^{L} \right) ^2\left( l_{-}^{L}+l_{+}^{L} \right) +4l_4\left( l_{-}^{L}+l_{+}^{L} \right) ^2 \right)}\\
&\left( 128l_{4}^{3}\left( l_4-l_{-}^{L}+l_{+}^{L} \right) +8l_{4}^{2}\left( l_{-}^{L2}-30l_{+}^{L}l_{-}^{L}-19l_{+}^{L2} \right) \right. +\left( l_{-}^{L}-l_{+}^{L} \right) ^2\left( 21l_{-}^{L2}+46l_{-}^{L}l_{+}^{L}+13l_{+}^{L2} \right) \\
&\left. -4l_4\left( l_{-}^{L3}-23l_{-}^{L2}l_{+}^{L}-45l_{-}^{L}l_{+}^{L2}+3l_{+}^{L3} \right) \right),
\end{aligned}
\end{equation}
where, $l_4$ is a function of t from Eq. (6.16). Of course, the solution to the breaking problem for a power function $x^{1/n}$ at the instant of breaking can be similarly found.\\

\noindent\textbf{Conclusion}\\
\hspace*{\parindent}

This paper investigates the evolutions of waves under two initial conditions for the higher-order Chen-Lee-Liu (HOCLL) equation by utilizing Whitham modulation theory.

Firstly, we have obtained the single-phase periodic solutions of the HOCLL equation via the finite-band integration method, and then averaged the conservation laws of the equation using the averaging method to derive the Whitham modulation equations.

Secondly, we have acquired the hydrodynamics representation of the equation by means of the Madelung transformation, with consideration given to the dispersionless limit.

Thirdly, we have constructed a full description of the Riemann problem-a crucial problem in the field of dispersive dynamics for HOCLL equation. Since the HOCLL equation is a non-convex equation, the wave structures generated by the evolution of an initial discontinuity in Riemann problem are more complex than those of the classical Riemann problem. Based on the magnitudes of the Riemann invariants corresponding to the initial conditions at the two boundaries, as well as the monotonic regions where the Riemann invariants lie at different boundaries, we have classified the initial discontinuity problems into twelve categories and obtained twenty-four types of potentially generated wave structures, while also accounting for the propagation velocities of wave boundaries. In addition, we have performed numerical simulation solutions on the evolution of the equation under the initial conditions, and the theoretical analysis results are consistent with the numerical simulation results.

Finally, we have considered the evolution laws of waves under the condition of cubic initial condition formed at the wave breaking moment. Using the generalized hodograph method, we have derived the structures of the Riemann invariants and calculated the motion laws of the DSWs at the soliton edge and small-amplitude harmonic edge.\\

\noindent\textbf{Acknowledgments}\\
\hspace*{\parindent}

We express our sincere thanks to each member of our discussion group for their suggestions. This work has been supported by the Fund Program for the National Natural Science Foundation of China under Grant No. 12575005, the Scientific Activities of Selected Returned Overseas Scholars in Shanxi Province under Grant No. 20220008, and the Shanxi Province Science Foundation under Grant No. 202303021221031.\\

\noindent\textbf{Conflict of interest}\\
\hspace*{\parindent}

The authors have not disclosed any competing interests.

\end{document}